\documentclass[12pt]{article}
\usepackage{geometry}
\usepackage{graphicx,psfrag,epsf, eufrak}
\usepackage{enumerate}
\usepackage{amsmath}
\usepackage{amssymb}
\usepackage{amsthm}

\usepackage[authoryear]{natbib}
\usepackage{appendix}
\usepackage[dvipsnames]{xcolor}

\usepackage{amsfonts}
\usepackage{mathtools}
\usepackage{mathrsfs}
\usepackage[linesnumbered, ruled]{algorithm2e}

\usepackage[unicode=true,
bookmarks=false,bookmarksnumbered=true,bookmarksopen=true,
 breaklinks=true,pdfborder={0 0 1},backref=page,colorlinks=true, urlcolor=blue]
{hyperref}

\newcommand{\blind}{0}

\newcommand{\fix}[1]{\ifmmode{#1}\else{$#1$}\xspace\fi}
\newcommand{\bb}[1]{\mathbb{#1}}
\newcommand{\cl}[1]{\mathcal{#1}}
\newcommand{\Q}{\fix{\mathbb{Q}}}

\newcommand{\x}{\times}

\newtheorem{assumption}{Assumption}

\newtheorem{remark}{Remark}

\begin{document}

\def\spacingset#1{\renewcommand{\baselinestretch}%
{#1}\small\normalsize} \spacingset{1}

%%%%%%%%%%%%%%%%%%%%%%%%%%%%%%%%%%%%%%%%%%%%%%%%%%%%%%%%%%%%%%%%%%%%%%%%%%%%%%

\if0\blind
{
  %\title{\bf A framework for Sequential Monte Carlo methods in Continuous-Discrete State Space Models}
  \title{\bf Particle-Based Inference for Continuous-Discrete State Space Models}
  \author{Christopher Stanton%\thanks{}
  \hspace{.2cm}\\
    Department of Statistical Science, University College London\\
    and \\
    Alexandros Beskos \\
    Department of Statistical Science, University College London}
  \maketitle
} \fi

\if1\blind
{
  \bigskip
  \bigskip
  \bigskip
  \begin{center}
    {\LARGE\bf Title}
\end{center}
  \medskip
} \fi

\bigskip
\begin{abstract}
This article develops a methodology allowing application of the complete machinery of particle-based inference methods upon the class of \emph{continuous-discrete State Space Models} (CD-SSMs). Such models correspond to a latent continuous-time It\^o diffusion process which is observed with noise at discrete time instances. Due to the continuous-time nature of the hidden signal, standard Feynman-Kac formulations and their accompanying particle-based approximations have to overcome several challenges, arising mainly due to the following considerations: (i) finite-time transition densities of the signal are typically intractable; 
(ii) ancestors of sampled signals are determined w.p.~1, thus cannot be resampled; (iii) diffusivity parameters given a sampled signal yield Dirac distributions.  
We overcome all above issues by introducing a framework based on carefully designed path proposals and reparameterisations thereof.
That is, we obtain new expressions for the Feynman-Kac model that accommodate the effects of a continuous-time signal and overcome induced degeneracies.  
The constructed formulations enable use of the full range of particle-based algorithms for 
CD-SSMs: for filtering/smoothing and parameter inference, whether online or offline.
Our framework is compatible with guided proposals in the filtering steps that are essential for efficient algorithmic performance in the presence of informative observations or in higher dimensions, and is applicable for a very general class of CD-SSMs, including the case when the signal is modelled as a hypo-elliptic diffusion. We incorporate our methods into an established probabilistic programming package and present several numerical examples.
 %We illustrate the new methodology through several numerical applications. 
\end{abstract}

\noindent%

\begingroup
\renewcommand\thefootnote{}\footnotetext{
\textit{Corresponding author:} Christopher Stanton: \texttt{christopher.stanton.20@ucl.ac.uk}
}
\addtocounter{footnote}{0}
\endgroup

\newpage

\spacingset{1.45}
\section{Introduction}
\label{sec:intro}
This work considers the development of general particle-based Monte Carlo methods for statistical inference within the broad class of \emph{continuous-discrete State Space Models} (CD-SSMs) \citep{sark:19}. 
For such a CD-SSM, the  latent process is a continuous-time diffusion process \citep{okse:13} defined as the solution of a Stochastic Differential Equation (SDE):
\begin{equation}
\label{eq:CD-X}
    dX(s) = b(s, X(s))ds + \sigma(s, X(s)) dB(s), \qquad X(0) = x(0)\in \mathbb{R}^{d},
\end{equation} 
for a $d_w$-dimensional Brownian motion $B=\{B(s)\}_{s\ge 0}$, 
a drift function $b:[0,\infty)\times \mathbb{R}^{d}\to\mathbb{R}^{d}$ and a diffusion coefficient function $\sigma:[0,\infty)\times \mathbb{R}^{d}\to \mathbb{R}^{d\times d_w}$,  with $d\ge 1$, $1\le d_w\le d$. The signal process $X=\{X(s)\}_{s\ge 0}$ is assumed to be indirectly observed at $T\ge 1$ discrete times $0=s_0<s_1< \dots < s_T$, with data $Y_{1:T}=(Y_{1},\ldots,Y_{T})$ such that:
%$s_t=\frac{S}{T}$, with non-linear noise. 
%Letting $V_t = V(s_t)$ for each $t=1, \dots, T$, we observe:
%
\begin{equation}
\label{eq:CD-Y}
Y_t\,\big|\,\{X(s)\}_{s\ge 0},Y_{1:t-1},Y_{t+1:T} \sim  f_t(Y_t|X(s_t)), \qquad t \in \{1, \dots, T\},
\end{equation} 
independently over $t=1,\ldots,T$, for a conditional probability density function (pdf) $f_t(\cdot|\cdot)$.
% where we have set $V_t=V(s_t)$,  for simplicity.
%Where it is assumed that $Y_t | V_t$ has a known, tractable density $f_t(y_t|v_t)$. 
%The above description gives rise to what we call the class of \emph{continuous-discrete SSMs}.
%Without loss of generality, we assume equidistant observation times: 
Models of this nature -- and the associated inference procedures -- are of interest in many fields, including ecology \citep{knap:12}, epidemiology \citep{dure:12}, neuroscience \citep{sams:25} and finance \citep{ait:05}. We define 
$\mathcal{T}=\{1,\ldots, T\}$ and $\Delta_t = s_t-s_{t-1}$, $t\in \mathcal{T}$.
For convenience, %we use the notation 
%$V_t = \{X(s)\}_{s\in[s_{t-1},s_t]}$, $t\in\mathcal{T}$, and  we also 
we often write $X([u,v])$ for $\{X(s)\}_{s\in[u,v]}$, with $0\le u<v\le s_T$. % and $\bX_t(s)$, $s\in[s_{t-1},s_t]$. 
We will take under consideration the full range of particle-based algorithms related with SSMs developed in the literature, including ones that 
involve filtering, smoothing and parameter estimation \citep{chop:20}. For instance, the filtering problem, expressed within the CD-SSM setting, considers inference about $X(s_t) | Y_{1:t}$, for each $t\ge 1$, via an online approach. The smoothing problem can be considered as either the offline problem of inferring the joint distribution $X([0,s_T]) | Y_{1:T}$, or as the sequential problem of inferring  $X([0,s_{t}]) | Y_{1:t}$ for each $t\in \mathcal{T}$. 
%We note that the name `continuous-discrete SSM' we use here resembles the terminology in \cite{mide:21}. 
Importantly, we assume a low-frequency observation setting with $\Delta_t$ sufficiently large so that use of a single $\Delta_t$-step of an approximation scheme as proxy of the signal dynamics over $\Delta_t$ will introduce considerable bias into the model.

%SSMs can be represented as Feynman-Kac models \citep{del:04}. Particle-based methods can then be developed as Monte Carlo approximations of such models. In our setting, an initial expression for a Feynman-Kac model corresponding to the CD-SSM
%(\ref{eq:diffusion})-(\ref{eq:noise_function}) 
%is as follows, for a test function $\varphi$:
%
%\begin{align}
%\label{eq:FK}
%\mathbb{E}\big[\varphi(\bx_{1:n})|Y_{1:n}\big] = \frac{\mathbb{E}\big[\varphi(\bx_{1:n})\,\mathsf{G}_{1}(\bx_{1})\prod_{i=2}^{n}\mathsf{G}_{i}(X(s_{i-1}),\bx_{i})\big]}
%{\mathbb{E}\big[\mathsf{G}_{1}(\bx_{1})\prod_{i=2}^{n}\mathsf{G}_{i}(X(s_{i-1}),\bx_{i})\big]}, 
%\end{align}
%
%where we have defined the weights:
%
%\begin{align}
%\label{eq:FKW}
% &\mathsf{G}_1(\bx_{1}) = \frac{p(\bx_{1})\,p(Y_1|X(s_1))}{q(\bx_{1})}, \nonumber  \\[0.1cm]   &\mathsf{G}_{i}(\bx_{i-1},\bx_{i})= \frac{p(\bx_{i}|
% X(s_{i-1}))p(Y_i|X(s_i))}{q(\bx_{i}|\bx(s_{i-1}))}, \quad i\ge 2.  
%\end{align}
%
%
%Here, expectations are taken with respect to (w.r.t.) some
%chosen `guided' proposal kernel $q(\cdot|\cdot)$.

A standard Sequential Monte Carlo (SMC) algorithm is immediately available for the approximation of the smoothing distribution $X([0,s_T]) | Y_{1:T}$. That is, iteratively for $t\in\mathcal{T}$:
\begin{itemize}
\item[(i)] particles are sampled in-between observation instances $s_{t-1}$, $s_t$, according to some path proposal dynamics $M\big(dX([s_{t-1},s_t])\big|X(s_{t-1})\big)$; 
\item[(ii)]
sampled particles representing $X([s_{t-1},s_t])$ are assigned un-normalised importance weights according to the potential:
\begin{align}
\label{eq:Girsanov}
\frac{dP(\cdot |X(s_{t-1}))}{dM(\cdot|X(s_{t-1}))}\big(X([s_{t-1},s_t])\big)\cdot f_t(Y_t|X(s_t));
\end{align}
\item[(iii)] the weighted particles are resampled. 
\end{itemize}
Here $P(\cdot|\cdot)$ is the probability law corresponding to the dynamics of the signal SDE (\ref{eq:CD-X}).
Approximate samples from the smoothing distribution $P(dX([0,s_T])|Y_{1:T})$ can be generated by selecting particles at $s_T$ according to the final set of weights and tracing their  ancestral path \citep{kita:96}.
Improved SMC algorithms provably overcome path degeneracy characterising this standard approach by  updating the ancestors of obtained particles, see e.g.~\cite{gods:04}. 
Separate particle-based algorithms are used in the setting when a parameter vector $\theta\in \Theta\subseteq \mathbb{R}^{p}$, $p\ge 1$, must also be inferred.

A number of challenges arise due to the continuous-time nature of the signal in the CD-SSM model class of interest in this work. The main ones are summarised below:
\begin{itemize}
\item[(C.i)] 
First, sampling from $X(s_t) | X(s_{t-1})$ is possible by generating imputed points at high frequency from approximate numerical schemes, immediately enabling the use of `blind' proposals in a setting that ignores the continuous-time nature of the signal and only considers its values at the discrete-time observation instances. Use of observation-informed proposals in this context does require access to the signal transition density $P(X(s_t)\in dx'|X(s_{t-1})=x)$.
However, for CD-SSMs, this quantity is typically intractable. As already implied by the particle filter briefly described via steps (i)-(iii) in the previous paragraph, a solution to this problem is to augment the state space by considering the full continuous-time path of the SDE in-between observations. 
Notice that expression~(\ref{eq:Girsanov}) does not require a (finite-time) transition density, and is analytically available via Girsanov's theorem \citep{okse:13}. 

\item[(C.ii)] 
Then, a number of particle-based methods developed for smoothing and parameter inference involve updating the ancestors of particles to overcome path degeneracy, e.g.~the 
Forward-Filtering, Backward-Sampling (FFBS) algorithm  of \cite{gods:04}. However, owning to the continuous-time nature of the signal, the ancestor of path $X([s_{t-1},s_{t}])$, $t\ge 2$,
can only involve a particle $X([s_{t-2},s_{t-1}])$
%say $X([s_{t-2},s_{t-1}])^{j}$, 
%$X([s_{t-1},s_{t}])^{k}$, $1\le j,k\le N$, with $N\ge 1$ denoting the number of particles,
for which:
\begin{align}
\label{eq:issue}
\texttt{end}(X([s_{t-2},s_{t-1}]))
\equiv \texttt{start}(X([s_{t-1},s_{t}])). 
\end{align}
Thus, the overall method will be degenerate.
From a probabilistic perspective, this issue arises because the dominating measure of the path-valued proposal for $X([s_{t-1},s_{t}])$ depends intrinsically on $X(s_{t-1})$. For the algorithm to be allowed to select any ancestral particle $X([s_{t-2},s_{t-1}])$, the path proposal for $X([s_{t-1},s_{t}])$ -- or, more precisely, of a transform of thereof -- must have a density w.r.t.~a reference measure that will not depend on the previous state via $X(s_{t-1})$. This is infeasible when working with 
$X([s_{t-1},s_{t}])$, but can be made possible via careful transforms of the signal process. 
Indeed, such a reparameterisation gives rise to a new CD-SSM formulation, for which both target and proposals have densities w.r.t.~the common reference measure $ \mathrm{Leb} \otimes \mathbb{W}_t$, 
with $\mathrm{Leb}(\cdot)$ denoting the Lebesgue measure of dimension $d$ and $\mathbb{W}_t[\cdot]$  the distribution of the standard $d_w$-dimensional Wiener process (to be fully specified in the sequel). 
%that can be evaluated up to discretisation. 
The construction has connections with the transforms of \cite{chib:04, goli:08} used in the context of MCMC algorithms for SDE models.

\item[(C.iii)] The above two challenges are also present when one considers parameter inference for CD-SSMs via application of particle-based MCMC methods (e.g.~particle Gibbs \citep{andr:10}). 
Advanced versions of particle Gibbs based on backward or ancestral sampling
improve mixing and have provably robust properties w.r.t.\@ $T$, see e.g.~\cite{lind:13, lind:14}. Implementation of such sampling approaches in the CD-SSM setting immediately connects with the challenge raised in point (ii) above.
In addition, in the CD-SSM setting, the full conditional of 
$\theta|X([0,s_T])$ will be a Dirac measure for components of $\theta$ appearing in the diffusion coefficient $\sigma(s,X(s))$. This is a standard concern when data augmentation Bayesian methods are applied upon discretely observed SDEs \citep{robe:01}.

\end{itemize}

In this contribution we take under consideration the full spectrum of particle-based algorithms developed for standard SSMs, and produce a methodology that overcomes challenges (C.i)-(C.iii) summarised above, thus making such algorithms available for a large class of CD-SSMs. Our work achieves that by developing careful reparameterisations of the signal and corresponding re-expressions of the Feynman-Kac formulae for CD-SSMs.% so that challenges (i)-(iii) listed above can be overcome, and the full toolbox of particle-based methodology can be made available for CD-SSMs. 

We briefly summarise our main contributions as follows.

\begin{itemize}

\item 
We consider effective path proposals $M(dX([s_{t-1}, s_t])| X(s_{t-1}))$ which take observation~$y_t$ into account, but allow for explicit expressions for the importance weights, thus bypassing (C.i). In brief, these proposals are one of two types, `forward' and `backward', and give rise to corresponding \emph{guided} Feynman-Kac formulae for the class of CD-SSMs. 
The deduced particle-based Monte Carlo methods will typically be much more efficient than ones build upon blind proposals.
\item 
 We carefully define two transformations of the latent path signal, one for each of the derived (forward and backward) guided approaches, thus developing corresponding reparameterised Feynman-Kac formulae. The Monte Carlo approximations of these constructions via particle-based methods lead to algorithms that overcome challenges (C.i)-(C.iii). Although the two transforms are seemingly closely connected, algorithms based on the backward reparameterisation are applicable to the class of hypo-elliptic SDEs, whereas this is not the case for methods based on the forward approach.
\end{itemize}

We present our approach by providing corresponding expressions of Feynman-Kac formulae, thus
directly connecting our methodological contributions to practical particle-based algorithms developed in the literature for the approximation of such formulae.
The informed path proposals improve performance versus vanilla blind ones, typically for any SMC algorithm under consideration. Our path transforms allow 
for the first time (to the best of our knowledge) for the full range of SMC algorithms based upon sampling new ancestors to be applicable in the context of CD-SSMs. 
%SMC algorithms to sample new ancestors, which (to the best of our knowledge) allow for the first time all SMC algorithms 
%and hence require access to the transition density of the proposal are now made available. 
Such algorithms include: offline FFBS-type smoothing methods \citep{gods:04, bunc:13}; online smoothing methods for additive functionals of quadratic cost in the number of particles $N\ge 1$ \citep{del:10, poyi:11}, along with their extensions for parameter estimation via (a form of) stochastic gradient descent within an offline or online approach; in the context of particle MCMC \citep{andr:10}, backward and ancestral variants \citep{whit:10, lind:13, lind:14} of the iterated conditional SMC (iCSMC) algorithm, where incorporation of changes in the genealogy is well-known to improve performance both empirically and theoretically \citep{lee:20, karj:23}.
 %Our constructions also imply the applicability of the corresponding algorithms when extended to joint estimation of the parameter and the latent signal via Particle Gibbs. 
 Our path transforms can also be applied in the context of more recent particle MCMC developments involving extensions to higher dimensions via local proposals about the reference path \citep{fink:23} or methods that use the iCSMC proposal to update the latent signal but enable joint updates of signal and parameter \citep{core:25}, circumventing the need for Gibbs updates.

 We note that very recent research employing particle-based methods for CD-SSMs focuses on reducing the bias that necessarily occurs when using finite discretisations of continuous-time paths in numerical applications. E.g.~the arxiv preprint \cite{alva:25} uses one of the methods that also appears in our work (\cite{alva:25} cite the arxiv version \citep{stan:24} of our paper) as part of a  
larger intricate algorithm that utilises random truncation and coupling techniques to obtain unbiased estimates of the MLE of (elliptic) CD-SSMs within a stochastic-gradient-type approach. We stress that improvements in our algorithms can be directly incorporated in such methodologies.  

We mention that for the purposes of offline inference, %alongside particle-based methods, 
standard (i.e., not particle-based) MCMC algorithms have been successfully applied for the class of CD-SSMs as defined herein, see e.g.~\cite{chib:04, goli:08} for some of the first developments. If the logarithm of the posterior of parameters and latent variables is differentiable, one can carry out inference for CD-SSMs with standard MCMC, e.g.~via use of available probabilistic programming software, such as \texttt{Stan} \citep{carp:17}. %that uses efficient implementatiosn of Hamoltonian Monte Carlo.
We make the following remarks on the importance of the particle-based direction for treating CD-SSMs. First, particle-based algorithms can provide fast results in online settings, for general SSMs. Then, advanced MCMC approaches (e.g., HMC, NUTS) can be ineffective for target posteriors with complex curvature structures -- and such settings easily arise, e.g.~for SSMs with small observation noise, when leapfrog schemes can require excessively small step-sizes thus leading to defective algoritms (see, e.g., the numerical results in \cite{grah:22}). Lastly, the signal SDEs (elliptic and hypo-elliptic) upon our methodology develops in this work serve as a blueprint for the treatment of more general continuous-time model classes, including ones with discontinuous paths, 
i.e.~SDEs with jumps or discrete-valued pure jump-processes, when standard MCMC methods are either inappropriate (e.g., derivative-driven algorithms) or require complex model-specific development of proposals over `blocks' of signal paths, see e.g.~\cite{joha:10} for the case of SDEs with jumps and the references therein. Indicatively, one of the authors has been involved in \cite{yone:22}, where indeed elliptic SDEs with jumps are considered, and a particle-based algorithm for online parameter inference is proposed. Though that work lacks the general scope of the present one, it does serve to illustrate the potential for the full machinery of particle-based methods to be made available for general continuous-time signals with discontinuities. 

Another key property of particle-based methods is their scalability w.r.t.~the time index~$T$. \cite{karj:23} prove that the backward variant of iCSMC requires a number of particles that scales as $\mathcal{O}(1)$ w.r.t.\@~$T$, and provides mixing time of $\mathcal{O}(\log T)$. Thus, the overall cost of the algorithm is $\mathcal{O}(T\log T)$ w.r.t.\@ $T$ (as~$\mathcal{O}(T)$ is the computing cost per step), with the number of particles selected independently of $T$. The above contrast with reported results for standard MCMC, which point at ideal costs of $\mathcal{O}(T^2)$, $\mathcal{O}(T^{4/3})$ and $\mathcal{O}(T^{5/4})$ for RWM, MALA and HMC, respectively \citep{gelm:97, robe:98, besk:13}.  
Such robustness of particle-based algorithms is 
achieved precisely due to the fact that careful reselection of ancestors of particles within FFBS or similar methods overcomes the path degeneracy of standard particle filters.
Our contribution in this work makes such ancestor alteration possible for the class CD-SSMs.

\begin{figure}[!h]
  \includegraphics[width=\linewidth]{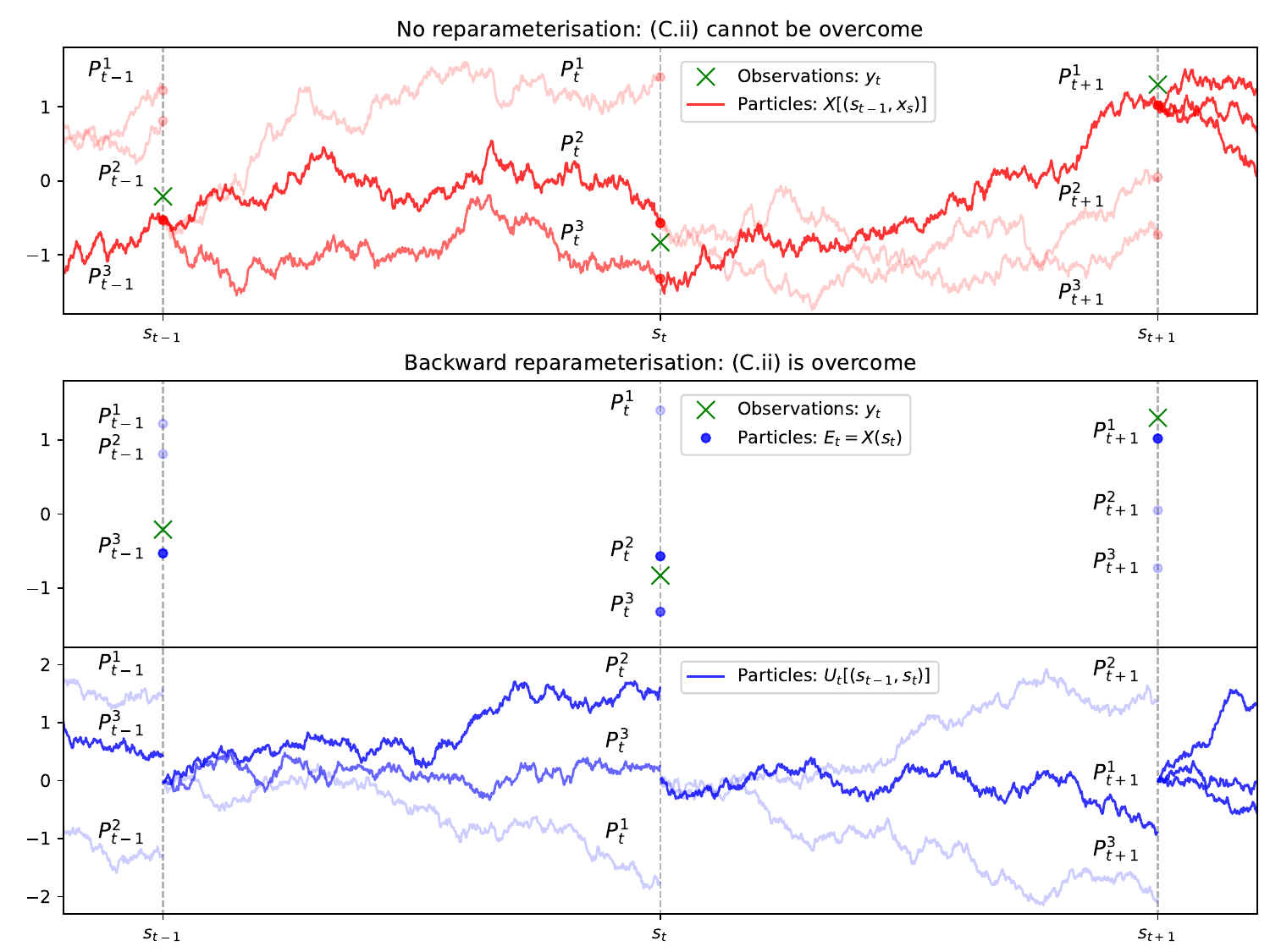}
  \caption{Visual representation of particle filter output for a 1D CD-SSM with $N=3$ particles. Signal: $dX(s) = -X(s)ds + dB(s)$; Observation density: $f_t(y_t|x(s_t)) = \mathcal{N}(y_t; x_t, 0.1^2)$. Top Panel: No reparameterisation applied. Bottom Panel: Backward reparameterisation applied.}
  \label{fig:intro_plot}
\end{figure}

To aid the communication of the developments in this work, Figure~\ref{fig:intro_plot} provides a visual illustration of the output of the particle filter 
 across multiple timesteps on an 1-dimensional CD-SSM with $N=3$ particles, both without any transformations of the latent continuous-time signal and with the (backward) 
transformation developed in this work.
The (path-valued) particles are labelled $P_t^i$, $i \in \{1, 2, 3\}$. First, the top panel shows the particles run without use of our reparameterisation. Consider the weighting and resampling steps that occur at steps $t-1$ and $t$. 
Particle $P_{t-1}^3$ ends up close to the observation at time $s_{t-1}$ and is assigned the largest weight (indicated by the strength of shading of the paths). As a result, when particles are resampled, only particle $P_{t-1}^3$ survives, and is the sole ancestor of particles $P_t^{1}, P_t^2, P_t^3$.
Due to the continuous-time nature of the latent signal, the start of each of the 3 particles at step $t$ is equal to the end point of its ancestor particle $P_{t-1}^3$. 
Similarly at step $t$, particles $P_t^2$ and $P_t^3$ are the ones found close to the observation, thus are given the largest weights. After resampling at step $t$, only $P_t^2$ and $P_t^3$ survive.
Now, consider an implementation of FFBS that aims to generate a sample from the smoothing distribution. 
Assume that the currently selected particle at step $t+1$ is $P_{t+1}^1$ and one would like to choose its ancestor, via an FFBS update, from one of the particles at step $t$. As discussed in (C.ii), $P_t^{1}$, $P_t^2$, cannot be chosen, as their end-points do not match the start point of $P_{t+1}^1$. 
One can only pick $P_t^3$ as the ancestor of $P_{t+1}^1$ following the (forward) simulation of the particle filter. Thus, the backward step is pointless and the FFBS algorithm reduces to trivial genealogy tracking.
Now, we consider running the particle filter under the backward reparameterisation developed in this paper.
As we will see later, running such a modified particle filter involves simulating, for each particle, first an end-point $E_t = X(s_t)$ of the path at step $t$,  and then, separately the driving (Brownian) noise $U_t([s_{t-1}, s_t])$ of the path. 
A particle $P_t$ then consists of both these components. Now, consider the implementation of FFBS, where we want to select an ancestor at step $t$ for particle $P_{t+1}^1$. 
Under our transform, it \emph{is} now possible for $P_t^2$ to be selected as an ancestor of $P_{t+1}^1$.
For each particle $P_t^i$ at step $t$, a corresponding weight that determines the probability of selecting that particle 
as an ancestor can be calculated and used to reselect the ancestor of $P_{t+1}^i$.

The remainder of the paper is structured as follows. In Section \ref{sec:FK}, we recall the Feynman-Kac formalism of a SSM and provide pseudocode for some of the deduced particle-based methods. In Section \ref{sec:CD-SSM} we embed the CD-SSM within the Feynman-Kac framework, in its original form, before consideration of path proposals or reparameterisations. In preparation of what follows, we describe forward and backward decompositions of SDE paths.
In Section \ref{sec:guided} we present our constructions for forward/backward path proposals, and develop the corresponding guided Feynman-Kac formulae.
In Section \ref{sec:transform} we present the reparameterised forward and backward Feynman-Kac formulae, which imply the feasibility of implementation of particle-based Monte Carlo methods that reselect ancestors for the class of CD-SSMs. In Section \ref{sec:numerics}, we present a collection of numerical experiments that demonstrate the effectiveness of our Feynman-Kac constructions on both filtering and smoothing problems.
%and the corresponding re-expressions of the Feynman-Kac model. 
%so that the deduced particle-based methods are effective for 
%CD-SSMs.
%
%approaches for guiding path proposals in the presence of informative observations: the existing approach of an unconditional path proposal using Girsanov's theorem, and an alternative path proposal constructed from conditioning on the end point that as we will see is necessary for smoothing in the hypoelliptic case. In Section 4, we present the reparameterisation of the Feynman-Kac model, under which the proposal kernels have densities w.r.t $\mathbb{W} \otimes dx$, immediately implying that one can use particle-based smoothers. 
%
%In Section %\ref{sec:algorithm} we %present examples of %particle-based algorithms %that emerge from the %framework, covering both %online and offline %approaches, and algorithms %that extend to estimation %of the parameter. 
%Section \ref{sec:numerics} %shows a collection of %numerical experiments. 
We give concluding remarks and outline further research directions in Section \ref{sec:conc}.

\section{Feynman-Kac Model}
\label{sec:FK}

Given measurable spaces $(\mathsf{X}_t,\mathscr{X}_t)$, $t\in \mathcal{T}$, a Feynman-Kac model \citep{del:04, del:13} can be defined through an initial law $M_1[dx_1]$, the Markov probability kernels $M_t[x_{t-1}, dx_t]$, $t\ge 2$, and the potential functions $G_1: \mathsf{X}_1 \rightarrow [0, \infty)$, $G_t: \mathsf{X}_{t-1}\times \mathsf{X}_{t} \rightarrow [0, \infty)$, $t \ge 2$. One then obtains, via a change of measure, the following probability law on 
$\big(\mathsf{X}_1\times \cdots \times \mathsf{X}_t, \mathscr{X}_1\otimes\cdots \otimes \mathscr{X}_t  \big)$: 
\begin{align}
\label{eq:FK-model}
    \mathbb{Q}_{t}[dx_{1:t}] = \frac{1}{L_t} \Big\{G_1(x_1)\prod_{i=2}^t G_i(x_{i-1}, x_i)\Big\}\times \mathbb{M}_{t}[dx_{1:t}], \qquad t \in \mathcal{T},
\end{align}
where we have set: 
\begin{align*}
\mathbb{M}_{t}[dx_{1:t}] = M_1[dx_1]\bigotimes_{i=2}^{t} M_{i}[x_{i-1},dx_i], %\qquad t \in \mathcal{T}.
\end{align*}
and we have assumed that the normalising constant $L_t>0$ is well-defined, that is:
\begin{equation*}
    L_t = \bb{E}_{\bb{M}_t}\Big[G_1(X_1)\prod_{i=2}^t G_i(X_{i-1}, X_i)\Big] < \infty.% \qquad t \in \cl{T}.
\end{equation*}
It is of interest to approximate integrals w.r.t.~the distribution $\bb{Q}_t[dx_{1:t}]$, %$t \in \cl{T}$. 
In particular, 
$\bb{Q}_t[dx_{1:t}]$ corresponds to the \emph{smoothing distribution} of an SSM at time $t\in \cl{T}$. We will also work with the \emph{filtering distribution} of an SSM at $t\in\cl{T}$, i.e.~the marginal of $\bb{Q}_t$ at $x_t$, which we denote $\Q_t[dx_t]$. 
We use the notation $\Q_t[dx_{i:j}]$, $1\le i\le  j\le t$, to refer to general marginal laws of $\bb{Q}_t[dx_{1:t}]$.
%
%\begin{itemize}
%    \item This sequence of target distributions: $\bb{Q}_t[dx_{1:t}]$ arise most commonly as the target distributions of state space models. 
%\end{itemize}
%
%\subsection{Filtering \& Smoothing}
%Where $\M_t$ is the Markov measure consisting of $\M_1$ and kernels $M_i[dx'_{i-1}, dx'_i]$ for $i=2, \dots, t$. 

Successive filtering laws $\Q_t[dx_t]$ are related via the following recursion, for $t\ge 2$:
\begin{align}
     \bb{Q}_{t}[dx_{t-1:t}] &= const. \times  \big(\bb{Q}_{t-1}[dx_{t-1}]\otimes M_t[x_{t-1}, dx_t]\big) \times  G_t(x_{t-1}, x_t).
     \label{eq:FK-filter}%\\ 
    %\mathbb{Q}_{t}[dx_{t-1:t}] &= 
    %G_t(x_{t-1}, x'_t)\,\mathrm{Q}_{t-1}[dx'_{t-1:t}] \label{eq:filtering-change-of-measure} \\
    %\Q_t[dx_t] &= \int_{x_{t-1} \in \cl{X}}\Q_t[dx_{t-1:t}] \label{eq:filtering-marginalisation} 
\end{align}
That is, starting from $\bb{Q}_{t-1}[dx_{t-1}]$, one  obtains the one-step-ahead predictive distribution, and  then applies a change of measure, to arrive at the next filtering law $\bb{Q}_{t}[dx_{t}]$.
In what follows, we make use of the convention $M_1[x_0,dx_1]\equiv M_{1}[dx_1]$ and $G_1(x_0,x_1)\equiv G_1(x_1)$. The same convention is also applied  for other transition kernels, potentials and related quantities we use in the sequel.
%
%\begin{itemize}
%    \item Extension \eqref{eq:filtering-extension}
%    \item Change of Measure \eqref{eq:filtering-change-of-measure}
%    \item Marginalisation \eqref{eq:filtering-marginalisation}
%\end{itemize}
%
Recursion (\ref{eq:FK-filter}) is analytically intractable except for the case of finite state spaces, or when the Feynman-Kac model corresponds to a linear, Gaussian SSM (LG-SSM). The particle filter \citep{gord:93} provides an effective Monte Carlo approximation of the recursion and, in particular, of the filtering laws $\Q_t[dx_t]$, $t\in \mathcal{T}$. One can also approximate the full smoothing law $\mathbb{Q}_T[dx_{1:T}]$ in an offline approach, from the particle filter output via genealogy tracking. That is, we select particles that represent $\mathbb{Q}_T[dx_T]$ according to the final set of sampled weights, then trace the ancestral path of the particles \citep{kita:96}. This algorithm suffers from path degeneracy: for $t << T$, the representation of the smoothing marginal is reduced to a single particle.

The Feynman-Kac distribution $\Q_T[dx_{1:T}]$ is a Markov measure, as it gives rise to the Markov chain $X_{1:T}\sim \Q_T[dx_{1:T}]$. Assume that (with some abuse of notation) $M_t[x_{t-1}, dx_t] = M_t(x_t|x_{t-1})\nu{(dx_t)}$, %$t\in \mathcal{T}$, 
where $M_t(\cdot|\cdot)$ is a transition density and $\nu(dx_t)$ is a $\sigma$-finite dominating measure, common for all $t\in\mathcal{T}$.
%we then have the following expression for the backward kernel of $\Q_T[dx_{1:T}]$:
%
%\begin{equation}\label{eq:fk-backward-kernel}
%   Q^\leftarrow_{t-1}[x_{t}, dx_{t-1}] \propto G_{t}(x_{t-1}, x_{t})\,m_{t}(x_{t}|x_{t-1})\,\bb{Q}_{t-1}[dx_{t-1}].
%\end{equation}
%
%Above, and in the sequel, we make use of the standard conventions $M_1[x_0,dx_1]\equiv M_{1}[dx_1]$ and $G_1(x_0,x_1)\equiv G_1(x_1)$.
FFBS  \citep{gods:04} is an offline smoothing algorithm, used in this work as representative of particle-based methods of improved efficiency due to well-designed reselection of ancestors of particles during  execution. Starting with a forward pass of the particle filter that delivers Monte Carlo approximations of the filtering laws $\bb{Q}_t[dx_t]$, $t \in \cl{T}$, FFBS then also carries out a backward pass to generate an approximate sample from the joint smoothing distribution $\bb{Q}_T[dx_{1:T}]$. In brief, given current sample $x_{t:T}\sim \bb{Q}_T[dx_{t:T}]$, $t\ge 2$, FFBS provides 
$x_{t-1}|x_{t:T}\sim \bb{Q}_T[dx_{t-1}|x_{t:T}]$
via, first, use of the formula:
\begin{align*}
\bb{Q}_T[dx_{t-1}|x_{t:T}] &\propto M_{t}(x_{t}|x_{t-1})\,G_t(x_{t-1},x_{t})\times \bb{Q}_{t-1}[dx_{t-1}].
\end{align*}
Then, the filtering distribution $\bb{Q}_{t-1}[dx_{t-1}]$ is replaced by its Monte Carlo approximation provided during the forward pass, thus a proxy for $\bb{Q}_T[dx_{t-1}|x_{t:T}]$ is obtained by re-weighting the particles representing the filtering law $\bb{Q}_{t-1}[dx_{t-1}]$. 

For concreteness, Algorithms \ref{alg:FK-PF}--\ref{alg:FK-FFBS} provide pseudo-code for standard versions of the particle filter, genealogy tracking and FFBS.
We provide the following explanations for the pseudo-codes: 
$N\ge 1$ is the number of particles; operations over $j$ are repeated independently for $j=1,\ldots,N$;
operation $\texttt{Resample}(W^{1:N})$, with $W^{j}\ge 0$ and $\sum_{j=1}^{N}W^{j}=1$, applies multinomial (or some  other type of) resampling; operation 
$\mathcal{M}(W^{1:N})$ samples an index in $1,\ldots, N$ according to the probabilities $W^{1:N}$.

\medskip

\begin{algorithm}[H] 
    \caption{Particle Filter for Feynman-Kac model in (\ref{eq:FK-model}).} %\citep{chopin2020introduction}}
    \label{alg:FK-PF}
    \SetKwInOut{Input}{Input}
    \SetKwInOut{Output}{Output}

    \KwIn{$M_1[dx_1]$, $M_t[x_{t-1},dx_{t}]$ and  $G_1(x_1)$, $G_t(x_{t-1},x_t)$, $t=2, \ldots, T$; no of particles $N$.}
    \KwOut{Particles $X^{1:N}_{1:T}$, ancestors $A^{1:N}_{2:T}$, weights $W_{1:T}^{1:N}$.}
    Sample: $X^j_1\sim M_1[dx_1]$\;
    Assign weights: $w^j_1 \gets G_1(X^j_1)$\;
    Normalise weights: $W^j_1 \gets \frac{w^j_1}{\sum_{k=1}^{N}w^k_1}$\;
    \For{$t=2,...,T$}{
        Sample: $A_t^{j}\sim \texttt{Resample}(W^{1:N}_{t-1})$\; %User defined Resampling Scheme 
        Sample: $X_t^j \sim M_t[X_{t-1}^{A_t^j},dx_t]$\;
        Assign weights: $w_t^j \gets G_t(X_{t-1}^{A_t^j}, X_t^{j})$\;
        Normalise weights: $W_t^j \gets \frac{w_t^j}{\sum_{k=1}^N w_t^k}$;
    }
\end{algorithm}
\medskip

\begin{algorithm}[H]  
    \caption{Genealogy Tracking for Feynman-Kac model in (\ref{eq:FK-model}).} %\cite{godsill2004monte}}
    \label{alg:FK-GT}
    \SetKwInOut{Input}{Input}
    \SetKwInOut{Output}{Output}
    \KwIn{Particle output, $X_{1:T}^{1:N}$, $A_{2:T}^{1:N}$, $W_{T}^{1:N}$, from Algorithm \ref{alg:FK-PF}.}
    \KwOut{Approximate sample $(X_1^{B_1}, \dots, X_T^{B_T})$ from $\bb{Q}_T[dx_{1:T}]$.}
    Sample: $B_T \sim \mathcal{M}(W_T^{1:N})$\;
    \For{$t=T,...,2$}{
        Set: $B_{t-1} \gets 
        A_{t}^{B_t}$;
        }
\end{algorithm}

\medskip

\begin{algorithm}[H] 
    \caption{FFBS for Feynman-Kac model in (\ref{eq:FK-model}).} %\cite{godsill2004monte}}
    \label{alg:FK-FFBS}
    \SetKwInOut{Input}{Input}
    \SetKwInOut{Output}{Output}
    \KwIn{Full particle output, $X_{1:T}^{1:N}$, $A_{2:T}^{1:N}$, $W_{1:T}^{1:N}$, from Algorithm \ref{alg:FK-PF}.}
    \KwOut{Approximate sample $(X_1^{B_1}, \dots, X_T^{B_T})$ from $\bb{Q}_T[dx_{1:T}]$.}
    Sample: $B_T \sim \mathcal{M}(W_T^{1:N})$\;
    \For{$t=T,...,2$}{
        Assign weights: $\hat{w}_{t-1}^j \gets M_{t}(X_{t}^{B_{t}}|X_{t-1}^j)\,G_t(X_{t-1}^{j}, X_t^{B_{t}})\cdot W_{t-1}^j$\;
        Normalise weights: $\hat{W}_{t-1}^j \gets  \frac{\hat{w}_{t-1}^j}{\sum_{k=1}^N \hat{w}_{t-1}^k}$\;
        Sample: $B_{t-1} \sim \mathcal{M}(\hat{W}_{t-1}^{1:N})$;
        }
\end{algorithm}

\medskip
We stress that implementation of particle filter and genealogy tracking (Algorithms \ref{alg:FK-PF} and~\ref{alg:FK-GT}) involves only simulating from $M_t[x_{t-1}, dx_t]$ and evaluating $G_t(x_{t-1}, x_t)$,  $t \in \mathcal{T}$. However, FFBS (Algorithm \ref{alg:FK-FFBS}) requires the additional assumption that $M_t[x_{t-1}, dx_t]$ has a density, for all $x_{t-1}\in\mathsf{X}_{t-1}$,  w.r.t.~a reference $\sigma$-finite measure $\nu(dx_t)$ that  \emph{cannot depend on $x_{t-1}$}. This condition is required to implement the efficient particle-based methods that reselect ancestors (further examples of which are given in Appendix \ref{sec:further_methods}) and is typically not satisfied when dealing with the class of CD-SSMs via standard approaches. A main contribution of this work is to develop reparameterised Feynman-Kac models for CD-SSMs that will indeed  satisfy the aforementioned  condition, and will enable the implementation of effective particle-based methods for CD-SSMs that overcome challenges (C.i)-(C.iii) stated in the Introduction.

\section{Continuous-Discrete SSMs}
\label{sec:CD-SSM}

\subsection{Preliminaries}

We return to the CD-SSM specified via
(\ref{eq:CD-X})-(\ref{eq:CD-Y}).
We introduce the path-valued random variables $V_t = \{V_{t}(s)\}_{s\in[0,\Delta_t]}$, 
via use of the time-shift $V_t(s)= X(s+s_{t-1})$. 
%where we have used the time-shift operator $\mathcal{C}_{-s}$ such that 
%$\mathcal{C}_{-s}(X(t))=X_{t-s}$, $0\le s\le t$. Thus, $V_t = \{V_t(s)\}_{s_{\in[0,\Delta_t]}}$.
%taking values 
%on $[0,\Delta_t]$ after application of a time-shift. 
In addition, we consider the end point $E_t=V_t(\Delta_t)\equiv\texttt{end}(V_t)$. We define the full latent state at step $t\in\mathcal{T}$ as: 
\begin{align}
\label{eq:CD-latent}
X_t := (V_t, E_t); \qquad  X_t\in\mathsf{X}_t := C([0,\Delta_t],\bb{R}^d) \times \bb{R}^{d}.
\end{align}
We write realisations of $X_t$ as $x_t=(v_t, e_t)$. 

The random variables $(X_{1:T}, Y_{1:T})$ form a discrete-time SSM, with joint distribution:
\begin{equation}
\label{eq:CD-joint}
P[X_{1:T} \in dx_{1:T}, Y_{1:T} \in dy_{1:T}] =\Big[ \bigotimes_{t=1}^T f_t(y_t|e_t)\,dy_{1:T} \Big]\otimes  \mathbb{P}_T[dx_{1:T}],
\end{equation}
where $\bb{P}_T[dx_{1:T}]$ is the Markov measure on $\mathsf{X}_1\times \cdots \times \mathsf{X}_T$ corresponding to the distribution of time-shifted segments of the latent SDE model in (\ref{eq:CD-X}) on the time intervals $[0,\Delta_t]$, $t\in\mathcal{T}$. 
That is, we write: 
\begin{align*}
\bb{P}_T[dx_{1:T}] =  P_1[dx_1]\bigotimes_{t=2}^{T} P_{t}[x_{t-1},dx_t],
\end{align*}
for accordingly defined kernels $P_t[x_{t-1}, dx_t]$, $t\in\mathcal{T}$,  such that $P_t[x_{t-1}, dx_t]\equiv P_t[e_{t-1}, dx_t]$, since the distribution of $X_t$ depends on $X_{t-1}$ only through the end-point $E_{t-1}$. As done here, we will often replace argument $x_t=(v_t,e_t)$ with $e_t$ in probability expressions when these only depend on  $e_t$ without further mention. 
From \eqref{eq:CD-joint}, we obtain the conditional law:
\begin{equation}
\label{eq:CD-conditional}
P[X_{1:T} \in dx_{1:T} | Y_{1:T} \in dy_{1:T}] =\frac{1}{f_T(y_{1:T})}\Big [ \prod_{t=1}^T f_t(y_t|e_t) \Big ] \times \bb{P}_T[dx_{1:T}],
\end{equation}
with $f_T(\cdot)$ denoting the marginal pdf of $Y_{1:T}$.

A main contribution of the paper is to provide effective particle-based algorithms for the approximation of the target smoothing measure \eqref{eq:CD-conditional}, 
and of related distributions arising when working with CD-SSMs (e.g.~filtering distributions, posterior laws of static parameters). In particular, we aim to develop tailored Feynman-Kac models via careful transformations of the variates involved in (\ref{eq:CD-conditional}), so that important challenges arising due to the continuous-time nature of the signal can be overcome -- see challenges (C.i)-(C.iii) stated in the  Introduction. Once we appropriately define the reparameterised Feynman-Kac model, we will obtain access to the complete toolkit of state-of-the-art particle-based methods for carrying out statistical inference for SSMs. 

\subsection{Feynman-Kac Formulations}

%We define in more detail the kernels $P_t[x_{t-1},x_t]$. 
%For $t \in \mathcal{T}$, we consider the time-shifted process $X_t(s)=X(s+s_{t-1})$, with $s\in[0,\Delta_t]$. 
%Note that $X_t(s)$ is an diffusion process with 
% drift function $b_t(s, x) = b(s_{t-1} + s, x)$ and diffusion $\sigma_{t}(s, v) = \sigma(s_{t-1}+s,v)$. Thus, definition (\ref{eq:CD-latent})% refers to:
%
%\begin{align*}
%V_t = \{X_t(s)\}_{s\in[0,\Delta_t]}; \qquad E_t = X_t(\Delta_t) = \texttt{end}(V_t).
%\end{align*}
%
We begin by writing down the `forward decomposition' for $P_t[x_{t-1}, dx_t]$, $t\in \mathcal{T}$:
\begin{flalign}
\textrm{F-Decomposition:}\qquad \qquad 
%\label{eq:forward_decomposition0}
%P_1[dx_1] &= P_1[e_0, dv_1]\otimes \delta_{\textrm{end}(v_1)}(de_1);  \\
P_t[x_{t-1}, dx_t] &= P_t[e_{t-1}, dv_t]\otimes \delta_{\texttt{end}(v_t)}(de_t). &&
\label{eq:FD-P}
\end{flalign}
%
%Again, we make use of the convention $P_1[x_{0}, dx_1]\equiv P_1[dx_1]$.
%The law of an SDE is determined uniquely by the initial point and the drift and diffusion coefficients. 
The kernel $P_t[e_{t-1}, dv_t]$ corresponds to the law of the time-shifted process 
$V_t=\{V_t(s)\}_{s\in[0,\Delta_t]}$.
Note that $V_t(s)$ is an diffusion process with 
drift function $b_t(s, v) = b(s_{t-1} + s, v)$ and diffusion $\sigma_{t}(s, v) = \sigma(s_{t-1}+s,v)$.
I.e., $V_t$ is obtained as the solution of the following SDE on $[0,\Delta_t]$:
\begin{flalign}
\label{eq:P-SDE}
P_t[e_{t-1}, dv_t]: \qquad\quad  dV_t(s) = b_t(s, V_t(s))ds + \sigma_t(s, V_t(s))dB(s),\quad V_t(0) = e_{t-1}.&&
\end{flalign}
Expression \eqref{eq:FD-P} involves the distribution of the path $V_t$ given the starting point $E_{t-1}$, along with that of the ending point $E_t$, the latter being fully determined by $V_t$ in the construction shown so far. 

An alternative `backward decomposition' of $P_t[x_{t-1}, dx_t]$ can be constructed by instead considering first the distribution of the ending point $E_t$ given the starting point $E_{t-1}$ and then the distribution of $V_t|E_{t-1}, E_t$, that latter variate being a \emph{diffusion bridge} conditioned to start at $E_{t-1}$ and end at $E_t$. In a probabilistic language, this second construction provides the following expression for  $P_t[x_{t-1}, dx_t]$, $t\in \mathcal{T}$:
%
%\begin{align}
%P_1[dx_1] &= P_1[de_1]\otimes %P_1^\leftarrow[(e_0,e_{1}),dv_1] 
%\nonumber
%\\ 
%&\qquad \qquad  = %\big(p_1(e_1|e_{0})de_1\big) \otimes 
%P_1^\leftarrow[(e_0,e_1), dv_1],
%\label{eq:backward_decomposition0}
%\end{align}
%
%followed by: 
%
\begin{flalign}
\textrm{B-Decomposition:}\qquad
\label{eq:BD-P} 
P_t[x_{t-1}, dx_t] &= P_t[e_{t-1}, de_t]\otimes P_t[(e_{t-1},e_t), dv_t] \nonumber && \\ &\qquad \qquad = \big(p_t(e_t|e_{t-1})de_t\big)\otimes P_t[(e_{t-1},e_t), dv_t]. &&
\end{flalign}
Here, $p_t(\cdot|e_{t-1})$ is the (typically intractable) pdf of $V_t(\Delta_t)|V_t(0)=e_{t-1}$ and $P_t[(e_{t-1},e_t), dv_t]$ is the law of the bridge of the signal SDE (\ref{eq:P-SDE}) between times $0$ and $\Delta_t$, pinned at the starting point $e_{t-1}$ and the ending point $e_t$. Under regularity conditions, such a bridge is known  to correspond to the path-valued process defined as the solution of the SDE:
\begin{flalign}
 P_t[(e_{t-1},e_t),dv_t]: \qquad  dV_t(s) &= b_{t}(s, V_t(s); e_t)ds + \sigma_t(s, V_t(s))dB(s), \quad V_t(0) = e_{t-1}, &&
 \label{eq:P-bridge}
 \end{flalign}
for the following drift function ($\nabla_{\!v}$ denotes the vector of partial derivatives w.r.t.~argument $v$): 
 \begin{align}
    b_{t}(s,v; e_t) &= b_t(s,v) + \sigma_t(s,v)\sigma^\top_t(s,v)\nabla_{\!v} \log(p_{\Delta_t, e_t}(s,v)), 
    \label{eq:P-bridge-drift}
\end{align}
where  we have made use of the following function related with the SDE transition density:
\begin{align}
p_{\Delta_t, e_t}(s,v) = P[V_t(\Delta_t)\in de_t | V_t(s)=v]/\mathrm{Leb}(de_t), \qquad (s,v)\in [0,\Delta_t]\times \bb{R}^{d}.
\label{eq:P-td}
\end{align}
The bridge SDE determined via (\ref{eq:P-bridge})-(\ref{eq:P-bridge-drift}) can be obtained via use of Doob's $h$-transform, see e.g.~\cite{mide:21} for a recent exposition.  

\subsection{Bootstrap Particle Filter}
\label{subsec:boot}
%Under either the forward decomposition in (\ref{eq:forward_decomposition}) or the backward one in 
%(\ref{eq:backward_decomposition}),
Given the above sequence $P_t[x_{t-1},dx_t]$, $t\in\mathcal{T}$,
the targeted conditional law \eqref{eq:CD-conditional} 
of the CD-SSM has been written in the form of the Feynman-Kac formula (\ref{eq:FK-model}), 
with Markov kernels 
$M_t\equiv P_t$ %as implied by the definitions in the referenced equations,  
and weights given by $G_t(x_{t-1},x_t) \equiv f_t(y_t|e_t)$. 
%admits a bootstrap 
%particle filter implementation,   
%, as follows:
%
%\begin{align}
%    M_1[dx'_1] &= \mathbb{P}_1[dx'_1] & M_t[v_{t-1}, dx_t] = P_t[v_{t-1}, dx_t]\label{eq:bootstrap_formalism_1}\\
%    G_1(x'_1) &= f_t(y_t|v_t)  & G_t(v_{t-1}, x'_t) = f_t(y_t|x_t) \label{eq:bootstrap_formalism_2}
%\end{align}
%
For such a specification, the probability law $\bb{Q}_t[dx_{1:t}]$ in (\ref{eq:FK-model}) coincides with the conditional distribution of $X_{1:t}|Y_{1:t}$, as given in (\ref{eq:CD-conditional}) for $t=T$. 

We stress that, as discussed in the Introduction, the above construction can provide a non-degenerate particle filter that targets our CD-SSM via execution of Algorithm \ref{alg:FK-PF}. However, attempts to use this construct within FFBS in Algorithm \ref{alg:FK-FFBS}, and related smoothing-based algorithms, will lead to a degenerate methodology. 
%This construction of a Feynman-Kac formalism immediately implies the application of filtering ($X'_t | Y_{1:t}$), smoothing ($X'_{1:t}|Y_{1:t}$) and extended smoothing ($X'_t, \theta | Y_{1:t}$). However, the performance of algorithms based on this formalism is likely to suffer in problems where the $y_t$ is highly informative of the latent $v_t$, or in high dimensions. This occurs because the proposal kernels $M_t$ under the bootstrap formalism amount to simulating from the distribution of the signal, and thus do not take information from the observation $y_t$ into account. 
Even in the context of a particle filter, greater flexibility is desirable and can be made possible with the selection of `guided' proposal kernels. We present such a guided formalism in the next section, where we  target the conditional distribution $X_{1:t}| Y_{1:t}$, by making use of proposal kernels that are non-blind to the data.

\section{Guided Feynman-Kac Formulae for CD-SSMs}
\label{sec:guided}

For a guided Feynman-Kac formalism of a SSM (see e.g.~\cite{chop:20}) 
we can assume the selection of a Markov measure $\bb{M}_T[dx_{1:T}]$ in our Feynman-Kac formulation that does not coincide with the prior dynamics of the latent signal, $\bb{P}_T[dx_{1:T}]$.
That is, the Feynman-Kac expression will now involve the following weights:
\begin{align}
   % G_1(x_1) &= 
    %\frac{dP_1}{dM_1}(x_1) \times f_1(y_1|x(s_1))\label{eq:guided_formalism0};\\[0.1cm]
    G_t(x_{t-1}, x_t) = \frac{dP_t[x_{t-1}, \cdot]}{dM_t[x_{t-1}, \cdot]}(x_t)\times f_t(y_t|x(s_t)). \label{eq:CD-G}
\end{align}
Given that absolute continuity $P_t[x_{t-1}, dx_t] \ll M_t[x_{t-1}, dx_t]$ holds, the weights in (\ref{eq:CD-G}) will be well-defined.
%The bootstrap approach in Section \ref{subsec:boot} %(\ref{eq:forward_decomposition0})-(\ref{eq:forward_decomposition}) or in 
%(\ref{eq:backward_decomposition0})-(\ref{eq:backward_decomposition}) 
 %coincides with the guided one in the particular setting of $M_t \equiv P_t$, i.e.~when %one proposes from the signal dynamics.

We outline two different approaches for the selection of guided proposals that take the data point $y_t$ into account. 
%The two approaches build upon the forward and backward decompositions of $P_t$ given in (\ref{eq:forward_decomposition}) and (\ref{eq:backward_decomposition}) respectively.
We will refer to our first construct as the `forward proposal', as it can be thought as building upon the forward decomposition \eqref{eq:FD-P} of $P_t[x_{t-1},dx_t]$. We call the second the `backward proposal', as it is based on the corresponding backward decomposition \eqref{eq:BD-P}. 
Though seemingly similar, it will turn out that these two constructs have subtle differences. In particular, 
the backward proposal will allow for the derivation of a Feynman-Kac formula (and access to all accompanying particle-based algorithms) in the case of hypo-elliptic SDEs, when the diffusion coefficient is degenerate. That is not possible for the forward construct. The forward construct resembles several standard approaches followed in the literature in the context of MCMC algorithms and particle filters. 
The backward construct has not appeared before in the literature, to the best of our knowledge. 

%It turns out that as we will see later, the subtle difference between the forward and the backward proposals enables one to use particle-based smoothing algorithms on the important class of hypo-elliptic diffusions, when using a backward proposal. 

\begin{remark}
In the context of the forward proposal, and the corresponding reparameterised Feynman-Kac model that we will later on obtain via use of such a proposal, we will henceforth be assuming that $d_w\equiv d$, 
and that $\sigma_t(s,v)\in\mathbb{R}^{d\times d}$ is invertible for all $(s,v)\in [0,\Delta_t]\times \bb{R}^{d}$.  
That is, the forward approach will only be applicable to elliptic SDEs. In contrast, we allow full generality, with $d_w\le d$ in the setting of the backward proposal, so that the Feynman-Kac construct in this context will also cover the important class of hypo-elliptic SDEs.
\end{remark}

\subsection{Forward Proposal}
\label{subsec:FP}

\subsubsection{Construction of Forward Proposal}
We define the forward proposal $M_t^{\rightarrow}[x_{t-1}, dx_t]$ as follows:
\begin{flalign}
\label{eq:FD-M}
\textrm{F-Proposal:}\qquad \qquad\quad   M_t^{\rightarrow}[x_{t-1}, dx_t] = M_t^{\rightarrow}[e_{t-1}, dv_t]\otimes\delta_{\texttt{end}(v_t)}(de_t), &&
\end{flalign}
where we assume that 
$P_t[e_{t-1}, dv_t] \ll M_t^{\rightarrow}[e_{t-1}, dv_t]$.
%Extending both of the pathspace laws with the end point of the path, we therefore also have that $M_t^{\rightarrow}[v_{t-1}, dx'_t] >> P_t[v_{t-1}, dx'_t]$. 
The proposal kernel $M_t^{\rightarrow}[e_{t-1}, dv_t]$ corresponds to the solution of an SDE with the same diffusion coefficient as the target $P_t[e_{t-1}, dv_t]$ in 
(\ref{eq:P-SDE}), and  a drift function given as $b^\rightarrow_t(\cdot,\cdot; y_t):[0,\Delta_t]\times \bb{R}^{d}\rightarrow \bb{R}^{d}$. In particular, the drift function of the proposal SDE aims to take into account information from the observation $y_t$. We thus define the proposal SDE and its corresponding law as follows: 
%and may also use the starting point $v_{t-1}$.
\begin{flalign}
\label{eq:FD-proposal}
M_t^{\rightarrow}[e_{t-1}, dv_t]: \qquad\qquad dV_t(s) &= b^\rightarrow_t(s, V_t(s); y_t)ds + \sigma_t(s, V_t(s))dB(s),&& \\ \nonumber V_t(0) &= e_{t-1}. &&
\end{flalign}
We set
%
%\begin{align*}
$\Sigma_t = \sigma_t \sigma_t^\top\in \bb{R}^{d\times d}$. Recall that we assume
that $\Sigma_t=\Sigma_t(s,v)$ is invertible for all $(s,v)\in [0,\Delta_t]\times \bb{R}^{d}$.
%\end{align*}
Girsanov's theorem \citep{okse:13} states that, under regularity conditions,  we have the following  Radon-Nikodym derivative:
\begin{align}
&\frac{dP_t[e_{t-1}, \cdot]}{dM_t^{\rightarrow}[e_{t-1}, \cdot]}(v_t)   \nonumber\\
\nonumber
&\qquad\qquad = \exp \Bigg[\int_{0}^{\Delta_t} \big\{(b_t-b^\rightarrow_t)^\top\Sigma_t^{-1}\big\}(s, v_t(s);y_t)dv_t(s) \\ &\qquad \qquad \qquad\qquad\qquad - \tfrac{1}{2}\int_0^{\Delta_t} \big\{(b_t - b^\rightarrow_t)^\top\Sigma_t^{-1}(b_t + b^\rightarrow_t)\big\}(s, v_t(s);y_t)ds \Bigg].
\label{eq:FD-Girsanov}
\end{align}
In brief, sufficient regularity conditions involve global Lipschitz continuity and linear or slower growth for the drifts and diffusion coefficient, together with Novikov's condition \citep{okse:13}.
The likelihood \eqref{eq:FD-Girsanov} can be evaluated up-to time-discretisation. 

Upon choosing a tractable drift function $b^\rightarrow_t(s, v; y_t)$, one can sample from $M_t^{\rightarrow}[x_{t-1}, dx_t]$.
The corresponding potential $G_t^{\rightarrow}(x_{t-1}, x_t)$ of the generated Feynmac-Kac model writes as: 
\begin{flalign}
\textrm{F-Weight:}\qquad \qquad\qquad 
   % G_1(x_1) &= 
    %\frac{dP_1}{dM_1}(x_1) \times f_1(y_1|x(s_1))\label{eq:guided_formalism0};\\[0.1cm]
    G_t^{\rightarrow}(x_{t-1}, x_t) = \frac{dP_t[x_{t-1}, \cdot]}{dM_t^{\rightarrow}[x_{t-1}, \cdot]}(x_t)\times f_t(y_t|v_t(\Delta_t)), && 
    \label{eq:FD-G}
\end{flalign}
where the involved Radon-Nikodym derivative has been given in (\ref{eq:FD-Girsanov}).   
\begin{remark}
\label{rem:FB}
The proposal $M_t^{\rightarrow}[x_{t-1},dx_t]$ is defined via the forward decomposition (\ref{eq:FD-M}), in agreement with a similar expression for the signal kernel $P_t[x_{t-1},dx_t]$ in~(\ref{eq:FD-P}). Notice though that  $M_t^{\rightarrow}[x_{t-1},dx_t]$ also admits a backward-type decomposition, as one can write:
\begin{equation}
\label{eq:FD-BD}
    M_t^{\rightarrow}[x_{t-1}, dx_t] =M_t^{\rightarrow}[e_{t-1}, dx_t]  =  \big(m_t^{\rightarrow}(e_t|e_{t-1})de_t\big)\otimes M_t^{\rightarrow}[(e_{t-1},e_t), dv_t]. 
\end{equation}
In (\ref{eq:FD-BD}), $m_t^{\rightarrow}(\cdot|\cdot)$ is the, typically intractable, transition density from time $0$ to $\Delta_t$ of the proposal SDE (\ref{eq:FD-proposal}), and $M_t^{\rightarrow}[(e_{t-1},e_t), dv_t]$ is the distribution of the bridge of SDE (\ref{eq:FD-proposal}). Expression~(\ref{eq:FD-BD}) will be relevant when presenting a reparameterisation in the sequel. 
\end{remark}
%\begin{remark}
%Assuming that the diffusion coefficient $\sigma_t$ is invertible, using a bootstrap proposal $M_t=P_t$ can be considered to be a special case of a forward proposal.
%\end{remark}

\subsubsection{Choice of Forward Proposals}
\label{subsubsec:FP}

We give some guidance over the choice of the drift function $b^\rightarrow_t(s, v; y_t)$ in the proposal SDE \eqref{eq:FD-proposal} that takes the observation $y_t$ into account. \cite{sott:08} develop proposal SDEs for the case of state-independent diffusion coefficient by using expressions from a continuous-discrete extended Kalman Filter to form the drift function. %Such ideas have been considerably extended by \cite{mide:21}.
Alternative proposal SDEs can be obtained by adapting ideas for diffusion bridges (in settings of observations without noise) to the case of noisy observations \citep{dely:06, llop:18}. 

We present in more detail approaches based on the `optimal' proposal SDE and its approximation, following the recent developments in \cite{scha:17, mide:21} -- the relevant methodology provides a flexible framework that provides proposal SDEs for a broad class of diffusion models.
The optimal proposal minimises the variance of the incremental weights \citep{douc:00} and is obtained by conditioning the signal dynamics on the observation $y_t$. That is:
\begin{align*}
M_t^{\rightarrow,\textrm{opt}}[e_{t-1}, dv_t] \propto P_t[e_{t-1}, dv_t]\times f_t(y_t|e_t). 
\end{align*}
Under conditions, the distribution $M_t^{\rightarrow,\textrm{opt}}[e_{t-1}, dv_t]$ coincides with the law of an SDE with the following drift function (see \cite{mide:21} for a proof):
\begin{equation}
\label{eq:FD-optb}
b^{\rightarrow, \textrm{opt}}_t(s, v; y_t) = b_t(s, v) + \Sigma_t(s, v)\nabla_v(\log \rho_{\Delta_t, y_t}(s, v)),
\end{equation}
where we make use of the density function: 
\begin{equation}
\label{eq:P-td-y}
    \rho_{\Delta_t, y_t}(s,v) = \int_{\mathbb{R}^{d}} f_t(y_t|v')p_{\Delta_t,v'}(s,v)dv',\qquad  (s,v)\in [0,\Delta_t]\times \bb{R}^{d},
\end{equation}
for $p_{\Delta_t,v'}(s,v)$ as defined in (\ref{eq:P-td}).
%$y\mapsto \rho_{t, y}(s,x)$, for $(s,x)\in (0,\Delta_t)\times \bb{R}^{d}$, is the density function of $\bb{P}\,[\,Y_t\in dy\,|\,X(s_{t-1} + s) = x\,]$, given by:
%
%\begin{equation}\label{eq:y_t|v_s}
%    \rho_{t, y}(s,x) = \int_{\mathbb{R}^{d}} f_t(y_t|x')p_{\Delta_t,x'}(s,x)dx'.
%\end{equation}
%
%In the above expression $p^t_{s_1, s_2} = p_{s_{t-1} + s_1, s_{t-1} + s_2}$, with $0<s_1<s_2<\Delta_t$, denotes the transition density of the time-shifted SDE \eqref{eq:P_t_diffusion} from time $s_1$ to time $s_2$, so $p^t_{s, \Delta s} = p_{s_{t-1} + s, s_t}$, and $p^t_{0, \Delta_t} \equiv p_t$. 

The expression for $\rho_{\Delta_t, y_t}$ is typically intractable, as it involves an integral w.r.t.~the transition density of the signal SDE. Following the approach of \cite{mide:21}, a reasonable choice can be obtained by replacing $\rho_{\Delta_t, y_t}(s, v)$ in \eqref{eq:FD-optb} with an appropriate proxy $\tilde{\rho}_{\Delta_t, y_t}(s, v)$.  Thus, one makes use of the following drift function:
\begin{equation}
\label{eq:FD-b}
b^\rightarrow_t(s, v; y_t) = b_t(s, v) + \Sigma_t(s, v)\nabla_x(\log \tilde{\rho}_{t, y_t}(s, v)).
\end{equation}
Function $\tilde{\rho}_{\Delta_t, y_t}$ can  be obtained by replacing $p_{\Delta_t,v'}(s,v)$ in the integral (\ref{eq:P-td-y}) with a corresponding proxy $\tilde{p}_{\Delta_t,v'}(s,v)$. If $v'\mapsto \tilde{p}_{\Delta_t,v'}(s,v)$ is Gaussian and the observation operator for $y_t$ is linear (the latter being a common case in applications), 
then the integral (\ref{eq:P-td-y}) is tractable. 
In \cite{mide:21}, $\tilde{p}_{\Delta_t,v'}(s,v)$ is chosen via use of the transition density of a member within a class of linear SDEs.
\begin{remark}
\label{rem:solve}
Consider the following linear SDE:
\begin{equation}
\label{eq:linear-SDE}
 d\tilde{V}_t(s) = \big\{\tilde{b}_0(s) + \tilde{b}_1(s)\tilde{V}_t(s)\big\}ds + \tilde{\sigma}(s)dB(s), 
 %\qquad \tilde{X}_t(0) = e_{t-1},
\end{equation}
for $\tilde{b}_0: [0,\Delta_t] \rightarrow \bb{R}^{d}$, $\tilde{b}_1: [0,\Delta_t] \rightarrow \bb{R}^{d\times d}$ and $\tilde{\sigma}: [0,\Delta_t] \rightarrow \bb{R}^{{d\x d_{w}}}$, with $d_w\le d$.
Such a linear SDE has an analytical solution that writes as follows, for $s\in[0,\Delta_t]$ and $\tilde{V}_t(s)=v\in\bb{R}^{d}$:
%
%\begin{align*}
%\tilde{X}_t(s') = \Phi(s',s)\,x + 
%\int_{s}^{s'}\Phi(s',u)\tilde{A}(u)\,du + \int_{s}^{s'}\Phi(s',u)\tilde{\sigma}(u)\,dB(u),
%\end{align*}
%
\begin{align*}
\tilde{V}_t(\Delta_t) = \Phi(\Delta_t,s)\,v + 
\int_{s}^{\Delta_t}\Phi(\Delta_t,\tau)\tilde{b}_0(\tau)\,d\tau + \int_{s}^{\Delta_t}\Phi(\Delta_t,\tau)\tilde{\sigma}(\tau)\,dB(\tau),
\end{align*}
where, for $s\in[0,\Delta_t]$, we set $\Phi(s,u)=\Phi(s)\Phi(u)^{-1}$, with $\Phi(\cdot)$ obtained from the solution of the ODE specified as  $d\Phi(t)=\tilde{b}_1(t)\Phi(t)dt$, $\Phi(0)=\mathrm{I}_d$.
Thus, the conditional law of $\tilde{V}_t(\Delta_t)|\tilde{V}_t(s)=v$ corresponds to a Gaussian distribution $\mathcal{N}(\mu(v), \mathrm{V})$, with:
\begin{align*}
\mu(v) = \Phi(\Delta_t,s)v + \int_{s}^{\Delta_t}\Phi(\Delta_t,\tau)\tilde{b}_0(\tau)\,d\tau; \qquad  
\mathrm{V} = \int_{s}^{\Delta_t}\Phi(\Delta_t,\tau)
(\tilde{\sigma}\tilde{\sigma}^{\top})(\tau)\Phi(\Delta_t,\tau)^{\top}\,d\tau.
\end{align*}
\end{remark}
One could aim to choose $\tilde{b}_0(s)$, $\tilde{b}_1(s)$, and $\tilde{\sigma}(s)$ so that the linear SDE (\ref{eq:linear-SDE}) approximates the dynamics $P_t[e_{t-1}, dv_t]$ of the signal SDE. A strategy for specifying (\ref{eq:linear-SDE}) can involve linearisation of the drift of the signal SDE (\ref{eq:P-SDE}), or the following simpler approaches \citep{mide:21}: 
\begin{align*}
\tilde{b}_0(s) = b_t(0, e_{t-1}), \,\,\tilde{b}_1(s) = 0,\,\, \tilde{\sigma}(s) = \sigma_t(0, e_{t-1});\qquad 
\tilde{b}_0(s)=0, \,\,\tilde{b}_1(s)=0, \,\,\tilde{\sigma}(s) = \mathrm{I}_d.
\end{align*}%--------

%We outline some simple default choices, in which functions $A, B, C$ are chosen to be constants: 
%(i) $A(s)=0$, $B(s)=0$ and $C(s) = 1$ yields a proposal SDE commonly used in the literature; (ii) $A(s) = b_t(0, v_{t-1}), B(s) = 0, C(s) = \sigma_t(0, v_{t-1})$, a local linearisation of the drift $b_t$ in $v$, about the point $(0, v_{t-1})$.
    %\item $A(s) = b_t(0, v_{t-1}) - \frac{\partial}{\partial v}(b_t(0, v_{t-1}))v_{t-1}$, $B(s) = \frac{\partial}{\partial v}(b_t(0, v_{t-1}))$, $C(s) = \sigma_t(0, v_{t-1})$ a local linearisation of the drift $b_t$ in $v$, about the point $(0, v_{t-1})$.
%\begin{itemize}
    %\item $A(s)=0, B(s)=0, C(s) = 1$ approximation using the standard Brownian motion. May perform poorly in e.g a low noise regime. The term $\nabla_v(\log \tilde{\rho}_{t, y_t}(s, v))$ acts as a forcing term that pushes the process towards $y_t$. In a low noise regime, this term will not push the diffusion towards $y_t$ quickly enough.
    %\item $A(s)=0, B(s)=0, C(s) = \sigma_t(0, v_{t-1})$ approximation using a Brownian motion with diffusion that depends on the diffusion of the original SDE.
    %\item $A(s) = b_t(0, v_{t-1}), B(s) = 0, C(s) = \sigma_t(0, v_{t-1})$ 
    %\item $A(s) = b_t(0, v_{t-1}) - \frac{\partial}{\partial v}(b_t(0, v_{t-1}))v_{t-1}$, $B(s) = \frac{\partial}{\partial v}(b_t(0, v_{t-1}))$, $C(s) = \sigma_t(0, v_{t-1})$ a local linearisation of the drift $b_t$ in $v$, about the point $(0, v_{t-1})$.
%\end{itemize}

\subsection{Backward Proposal}
\label{subsec:BP}
\subsubsection{Construction of Backward Proposal}

Recall that we now allow for $d_w\le d$, as we cover both the elliptic and hypo-elliptic case for the signal SDE in (\ref{eq:CD-X}) (equivalently, in (\ref{eq:P-SDE})).

In this direction, we define the backward proposal $M^{\leftarrow}_t[x_{t-1}, dx_t]$ as follows:
\begin{flalign}
\label{eq:BD-M}
\textrm{B-Proposal:}\qquad \quad 
    M_t^{\leftarrow}[x_{t-1}, dx_t] = 
    \big(m_t^{\leftarrow}(e_t|e_{t-1})de_t\big)\otimes M^{\leftarrow}_t[(e_{t-1},e_t), dv_t]. &&
\end{flalign}
The density $m_t^{\leftarrow}(e_t|e_{t-1})$ is user-specified, under the requirements that one can evaluate it and sample from it.
Then, $M^{\leftarrow}_t[(e_{t-1},e_t), dv_t]$ is a proposal for the target bridge law $P_t[(e_{t-1},e_t),dv_t]$ in (\ref{eq:P-bridge}) connecting $e_{t-1}$ and $e_t$, chosen so that $P_t[(e_{t-1},e_t),dv_t]\ll M^{\leftarrow}_t[(e_{t-1},e_t), dv_t]$. 
%(`$\ll$' denotes absolute continuity w.r.t.~the dominating measure). %Simulation from the backward  proposal (\ref{eq:BD-M}) thus consists of first sampling $e_t$ at time $\Delta_t$ according to $m_t^{\leftarrow}(e_t|e_{t-1})$, and then generating a proposal bridge between times $s_{t-1}$ and $s_t$, starting at $e_{t-1}$ and ending at $e_t$. 
%This latter proposal $M^{\leftarrow}_t[(e_{t-1},e_t), dv_t]$ is chosen so that $P^{\leftarrow}_t[(e_{t-1},e_t),dv_t]\ll M^{\leftarrow}_t[(e_{t-1},e_t), dv_t]$. 
In such a case, the potential $G_t^{\leftarrow}(x_{t-1}, x_t)$ for the induced Feynman-Kac model is expressed in the form of the following product: 
%is the product of the ratio of densities for the end-point, of the Radon-Nikodym derivative between the  true and proposed diffusion bridges and of the likelihood, i.e.:
%
\begin{flalign}
\textrm{B-Weight:}\qquad 
\label{eq:BD-G}
\nonumber 
  G_t^{\leftarrow}(x_{t-1},x_t) &= \frac{dP_t[e_{t-1}, \cdot]}{dM_t^{\leftarrow}[e_{t-1}, \cdot]}(v_t) && \\[0.2cm] &\qquad \qquad = \frac{p_t(e_t|e_{t-1})}{m_t^{\leftarrow}(e_t|e_{t-1})} \times \frac{dP_t[(e_{t-1},e_t), \cdot]}{dM^{\leftarrow}_t[(e_{t-1},e_t), \cdot]}(v_t)\times f_t(y_t|e_t).&&
\end{flalign}%
%We briefly present some choices for a bridge proposal that are common in the literature. 

%\subsubsection{Choice of $m_t(e_t|e_{t-1})$}
%
\subsubsection{Choice of Backward Proposals}
\label{subsubsec:BP}
We provide some details on the specification of the backward proposal so that $y_t$ is taken into account. 

\medskip
\noindent \textbf{Choice of $m_t^{\leftarrow}(e_t|e_{t-1})$ -- Elliptic and Hypo-Elliptic Case:}
\medskip

The optimal proposal \citep{douc:00}, as also discussed in the forward setting of Section \ref{subsec:FP}, denoted here as
$m_t^{\leftarrow, \textrm{opt}}$, is obtained from the true conditional distribution of $(e_t,v_t)$ given $e_{t-1}$ and $y_t$.
Thus, in the current backward setting, we can write that:
%
%so that it conditions perfectly on the observed $y_t$, and to choose $M_t^{\leftarrow}[v_{t-1:t}, dx_t]$ to be the true diffusion bridge of the signal $P^{\leftarrow}_t[v_{t-1:t}, dx_t]$. 
%That is: 
%
\begin{equation}
\label{eq:BD-optM}
    M_t^{\leftarrow, \textrm{opt}}[e_{t-1}, dx_t] = \big(m_t^{\leftarrow,\textrm{opt}}(e_t|e_{t-1})de_t\big)\otimes P_t[(e_{t-1},e_t), dv_t],
\end{equation}
where the optimal proposal for the end point $e_t$, is given by:
\begin{equation}
\label{eq:BD-optm}
    m_t^{\leftarrow,\textrm{opt}}(e_t|e_{t-1}) \propto f_t(y_t|e_t)p_t(e_t|e_{t-1}).
\end{equation}
The expression for $m_t^{\leftarrow,\textrm{opt}}$ involves the typically intractable transition density of the SDE model for the latent signal. 
From this point onwards, one can follow an approach  similar to the one set out within Section \ref{subsec:FP}, where the true transition density needed within the drift of the optimal SDE proposal is replaced by a tractable proxy. Thus, in the current setting, 
to obtain an alternative to $m_t^{\leftarrow,\textrm{opt}}(e_t|e_{t-1})$ in (\ref{eq:BD-optm}), one can replace $p_t(e_t|e_{t-1})$ with a proxy, $\tilde{p}_t(e_t|e_{t-1})$, so that: 
\begin{equation}
\label{eq:BD-m}
        m_t^{\leftarrow}(e_t|e_{t-1}) \propto f_t(y_t|e_t)\tilde{p}_t(e_t|e_{t-1}).
\end{equation}
In the scenario when $e_t \mapsto \tilde{p}_t(e_t|e_{t-1})$ in a Gaussian pdf and the observation operator providing $y_t$ is linear in $e_t$, then one obtains a Gaussian density for $m^{\leftarrow}_t(e_t|e_{t-1})$.

Proceeding as in the discussion for the forward proposal in Section \ref{subsec:FP}, one can construct 
proxies $\tilde{p}_t(e_t|e_{t-1})$ that are Gaussian conditionally on $e_{t-1}$, by utilising the transition density of the linear SDE (\ref{eq:linear-SDE}).
We consider here in detail SDE (\ref{eq:linear-SDE}) as specified by a linearisation of the drift $b_t(s,v)$ of the signal SDE (\ref{eq:P-SDE}), as its transition density will be non-degenerate even for hypo-elliptic SDEs. 
That is, applying a first-order Taylor expansion around $e_{t-1}$ upon the $v$-component of the drift function $b_t(s,v)$, one obtains the following linear proxy of the true drift: 
\begin{align*}
b_t(0,e_{t-1}) + (D_v b_t)(0,e_{t-1})(v-e_{t-1}), 
\end{align*}
where $(D_v b_t)(s,v')\in\mathbb{R}^{d\times d}$ with   $(D_v b_t)(s,v')_{i,j}= (\partial_{v_j}b_{t}^{i})(s,v')$, $1\le i,j\le d$. Thus, in the notation of (\ref{eq:linear-SDE}), we have: 
\begin{align*}
\tilde{b}_0(s) = b_t(0,e_{t-1})-  (D_v b_t)(0,e_{t-1})e_{t-1}; 
\quad \tilde{b}_1(s)=(D_v b_t)(0,e_{t-1})v; \quad \tilde{\sigma}(s) = \sigma_{t}(0,e_{t-1}).
\end{align*}
The transition density $\tilde{p}_t(e_t|e_{t-1})$ can now be obtained via the formulae in Remark \ref{rem:solve}.
In combination with a linear observation operator, standard calculations will yield a Gaussian density for $m_{t}^{\leftarrow}(e_t|e_{t-1})$ in (\ref{eq:BD-m}).
%We present two approaches to obtaining a proxy $\tilde{p}_t$. 
%The first uses the transition density from a numerical scheme with size $\Delta_t$. That is, assuming the Euler-Maruyama scheme, one obtains:
%
%\begin{equation}
%\label{eq:em_end_point_proposal}
%    \tilde{p}_t(e_t|e_{t-1}) = \cl{N}(e_t; e_{t-1} + (\Delta s) b_t(0, e_{t-1}), (\Delta s)\Sigma_t(0, e_{t-1})).
%\end{equation}
%
%
We note that simpler choices for $\tilde{p}_t(e_t|e_{t-1})$  can also be considered. These can range from a standard Euler scheme in the elliptic case, to schemes tailored to particular classes of hypo-elliptic SDEs, see for example \cite{poke:09, ditl:19}.
%-------

%Another possible approach is to choose $\tilde{p}_t$ to be the transition density from a linear SDE. This then reduces the problem of selecting a proposal for the end point to choosing a Linear SDE $\tilde{P}_t[e_{t-1}, dv_t]$ with dynamics that are similar to the signal process $P_t[e_{t-1}, dv_t]$. 

%\subsubsection{Choice of $M^{\leftarrow}_t[(e_{t-1},e_t), dv_t]$}

\medskip
\noindent \textbf{Choice of $M^{\leftarrow}_t[(e_{t-1},e_t), dv_t]$:}
\medskip

We now proceed to the choice of the proposal $M_t^{\leftarrow}[(e_{t-1},e_t), dv_t]$ for the bridge. 
Recall that the true diffusion bridge is given by \eqref{eq:P-bridge}-\eqref{eq:P-bridge-drift}, with a drift that involves the intractable transition density of the signal SDE. 
%
%The Delyon-Hu bridge \eqref{eq:delyon_hu_bridge} is an option for a proposal. \cite{delyon2006simulation} provide the Radon-Nikodym derivative between the true diffusion bridge and the proposal, written in \eqref{eq:delyon_hu_likelihood}.
A general expression for a proposal bridge SDE can be as follows:
\begin{flalign}
\label{eq:BD-proposal}
  M_t^{\leftarrow}[(e_{t-1},e_t), dv_t]: \qquad\quad dV_t(s) &= b^{\leftarrow,br}_t(s, V_t(s); e_t)ds + \sigma_t(s, V_t(s))dB(s), && \\\nonumber  V_t(0) &= e_{t-1}. &&
\end{flalign}
The drift $b^{\leftarrow,br}_t(s,v;e_t)$ appearing above should have a particular structure, as it must force paths to terminate at point $e_t$ at time $\Delta_t$, while preserving absolute continuity w.r.t.~the law of the bridge of the latent SDE (\ref{eq:P-SDE}) in the CD-SSM.
%Under regularity conditions on the drift $b^{\leftarrow}_t$, 
The works in \cite{dely:06, scha:17, bier:20} provide detailed analytical studies into the theme of diffusion bridges. The latter of these   \citep{bier:20} covers both elliptic and hypo-elliptic SDEs. We briefly summarise some standard choices for SDE bridges that one can sample from, and for which analytical expressions of Radon-Nikodym derivatives w.r.t.~target SDE bridges have been obtained. We stress that further options can be directly integrated within the framework of our methodology. 

\medskip
\noindent \textbf{Choice of $M_t^{\leftarrow}[(e_{t-1},e_t), dv_t]$ -- Elliptic Case, \cite{dely:06}:}
\medskip

In the context of elliptic SDEs, with $d_w=d$ and $\sigma_t(s,x)\in\bb{R}^{d\times d}$ assumed positive-definite, a simple choice used in the literature is the Delyon-Hu bridge \citep{dely:06}, when: 
\begin{align*}
b^{\leftarrow,br}_t(s, v; e_t) = \frac{e_t - v}{\Delta_t - s},
\end{align*}
in which case the bridge SDE (\ref{eq:BD-proposal}) writes as follows:
%and assuming that the diffusion $\sigma_t$ is invertible, one can use the Delyon-Hu bridge \cite{delyon2006simulation},
%which is the solution of the SDE:
\begin{flalign}
\label{eq:BD-proposal-DH}
  M_t^{\leftarrow,\textrm{DH}}[(e_{t-1},e_{t}), dv_t]: \qquad  dV_t(s) = \frac{e_t - V_t(s)}{\Delta_t - s}ds + \sigma_t(s, V_t(s))dB(s), \quad V_t(0) = e_{t-1}. &&
\end{flalign}
Under regularity conditions, 
\cite{dely:06} obtain the following Radon-Nikodym derivative: 
%Assuming that $X_t$ is distributed according to \eqref{eq:delyon_hu_bridge}, This choice of proposal for the bridge gives continuous-time likelihood:
\begin{align}
\label{eq:BD-Girsanov-DH}
\nonumber 
    &\frac{dP_t[(e_{t-1},e_{t}), \cdot]}{dM^{\leftarrow,\textrm{DH}}_t[(e_{t-1},e_{t}), \cdot]}(v_t) \\[0.2cm] & \qquad \qquad\qquad = \frac{\cl{N}(e_t; e_{t-1}, \Sigma_t(0, e_{t-1}))}{p_t(e_t|e_{t-1})}\sqrt{\frac{\det(\Sigma_t(0, e_{t-1}))}{\det(\Sigma_t(\Delta_t, e_{t}))}}\times \exp\big\{\phi_t(v_t)\big\},
\end{align}
where we have defined the following functional: 
\begin{align*}
    \phi_t(v_t) &= \int_0^{\Delta_t} (b_t^\top\Sigma_t^{-1})(s, v_t(s))dv_t(s) \\
                            &\qquad - \tfrac{1}{2}\int_0^{  \Delta_t}(b_t^\top \Sigma_t^{-1}b_t)(s, v_t(s)) ds \\
                            &\qquad\qquad- \tfrac{1}{2} \int_0^{\Delta_t} \frac{e_t - v_t(s)}{\Delta_t  - s} \diamond d\big\{\Sigma_t^{-1}(s, v_t(s))(e_t - v_t(s))\big\}.
\end{align*}
In brief, the integral with `$\diamond$' is defined by taking sums on sub-intervals that partition $[0,\Delta_t]$ and replacing the integrand in each summand with its right-side value, before taking the limit for vanishing length of the sub-intervals  (whereas It\^o integrals use left-side values). 
We note that the normalising constant in the above Radon-Nikodym derivative is omitted in \cite{dely:06}, 
but can be obtained via use of Bayes' theorem \citep[Section 4.1]{meul:17}. 
Use of (\ref{eq:BD-Girsanov-DH}) within \eqref{eq:BD-G} provides an analytical expression of the potential $G_t^{\leftarrow}(x_{t-1}, x_t)$ for the induced Feynman-Kac model. The intractable transition density $p_t(e_t|e_{t-1})$ appearing in \eqref{eq:BD-G} will cancel out when obtaining $G^{\leftarrow}_t(x_{t-1}, x_t)$ as it also  appears as a denominator in the density expression for the diffusion bridges in (\ref{eq:BD-Girsanov-DH}). All remaining terms in the formula for $G_t(x_{t-1}, x_t)$ can be analytically evaluated, up-to time-discretisation. 
Note that the above choice of proposal bridge does not take into account the drift of the signal SDE. 

\medskip
\noindent \textbf{Choice of $M_t^{\leftarrow}[(e_{t-1},e_t), dv_t]$ -- Elliptic Case, \cite{scha:17}:}
\medskip

\cite{scha:17} develop a class of guided proposals for diffusion bridges, in the setting of elliptic SDEs, that are of the following form:
\begin{flalign}
\label{eq:BD-proposal-D}
  M_t^{\leftarrow, \textrm{G}}[(e_{t-1},e_t), dv_t]: \qquad\qquad   dV_t(s) &=  b_t^{\mathrm{G}}(s, V_t(s); e_t)ds + \sigma_t(s, V_t(s))dB(s), && \\\nonumber   V_t(0) &= e_{t-1}, &&
\end{flalign}
where we have defined: 
\begin{align}
\label{eq:BD-proposal-D-drift}
  b_{t}^{\mathrm{G}}(s, v; e_t) &= b_t(s,v) + \Sigma_t(s, v) \nabla_v \log(\tilde{p}_{\Delta_t, e_t}(s,v)),
\end{align}
for a transition density 
$\tilde{p}_{\Delta_t, e_t}(s,v)$, used as a proxy to the true $p_{\Delta_t, e_t}(s,v)$. In \cite{scha:17},  
$\tilde{p}_{\Delta_t, e_t}(s,v)$ is chosen to correspond to the transitions of a linear SDE as given in (\ref{eq:linear-SDE}), with constituent functions 
$\tilde{b}_0(s)$, $\tilde{b}_1(s)$ and $\tilde{\sigma}(s)$, and corresponding mean $\mu(v)$,  variance $V$ and matrix $\Phi(\Delta_t,s)$, involved in expression of the Gaussian transition density of the linear SDE, as specified in Remark \ref{rem:solve}.
We  define: 
\begin{align*}
\tilde{\Sigma}(s) &= \tilde{\sigma}(s)\tilde{\sigma}(s)^{\top};\\ 
\tilde{r}(s,v) &= \nabla_v\log \tilde{p}_{\Delta_t, e_t}(s,v) \equiv \Phi(\Delta_t,s)^{\top}V^{-1}(e_t
-\mu(v));\\
\tilde{H}(s) &= -(\nabla_v \nabla_v^{\top})\log \tilde{p}_{\Delta_t, e_t}(s,v) \equiv  \Phi(\Delta_t,s)^{\top}V^{-1}\Phi(\Delta_t,s). 
\end{align*}
%
%Let $\tilde{\phi}_{t}(e_t|e_{t-1})$ 
\cite{scha:17} show that, under the main `matching' requirement that: 
\begin{align}
\label{eq:BD-matching}
\tilde{\Sigma}(\Delta_t) \equiv \Sigma_t(\Delta_t, e_t),
\end{align}
and given that  additional regularity conditions are satisfied, then the law for the class of SDEs defined in (\ref{eq:BD-proposal-D})-(\ref{eq:BD-proposal-D-drift}), 
with $\tilde{p}_{\Delta_t, e_t}(s,v)$ obtained via the linear SDE (\ref{eq:linear-SDE}),
dominates the law of the diffusion bridge $P_t[(e_{t-1},e_t),dv_t]$, with a Radon-Nikodym derivative:
\begin{equation}
\label{eq:BD-Girsanov-D}
    \frac{dP_t[(e_{t-1},e_t), \cdot]}{dM_t^{\leftarrow,\mathrm{G}}[(e_{t-1},e_t), \cdot]}(v_t) = \frac{p^{\mathrm{G}}_t(e_t|e_{t-1})}{p_t(e_t|e_{t-1})}\exp\Big\{\int_0^{\Delta_t} \phi(s, v_t(s))ds \Big\},
\end{equation}
with $p^{\mathrm{G}}_t(de_t|e_{t-1})=\mathcal{N}(e_{t-1},\Delta_t \mathrm{I}_{d})$, where we have defined the function: 
\begin{align*}
\phi(s, v)= & (b_t(s, v)-\tilde{b}(s, v))^{\top} \tilde{r}(s, v) \nonumber \\
&\qquad  -\tfrac{1}{2} \operatorname{trace}\Big(\big(\Sigma_t(s, v)-\tilde{\Sigma}(s)\big)\big(\tilde{H}(s)-\tilde{r}(s, v) \tilde{r}(s, v)^{\top}\big)\Big) .
\end{align*}  
Specific choices of $\tilde{b}_0(s)$, $\tilde{b}_1(s)$, and $\tilde{\sigma}(s)$ are discussed in \cite{meul:17}. E.g., a simple linear SDE satisfying the matching condition (\ref{eq:BD-matching}) is deduced when: 
\begin{align*}
d\tilde{V}_t(s) = \sigma_t(\Delta_t,e_t)dB(s),\quad \tilde{V}_{t}(0)=e_{t-1}, 
\end{align*}
which yields
$(\Delta_t,e_t)\mapsto \tilde{p}_{\Delta_t,e_t}(s,u)$ corresponding to 
$\mathcal{N}(e_t;v,(\Delta_t-s) \Sigma_t(\Delta_t,e_t))$
 to be used in (\ref{eq:BD-proposal-D-drift}). Indeed, the drift of the guided proposal in this case writes as: 
\begin{align*}
  b_{t}^{\mathrm{G}}(s, x; e_t) &= b_t(s,x) + \Sigma_t(s, x)\Sigma_t(\Delta_t, e_t)^{-1}\frac{e_t-x}{\Delta_t - s}.
\end{align*}
%
%\begin{equation}
%\label{eq:DH_bridge}
%  M_t^{\leftarrow \textrm{DH}}[(e_{t-1},e_{t}), dv_t]: \qquad  dX_t(s) = \frac{e_t - X_t(s)}{\Delta s - s}ds + \sigma_t(s, X_t(s))dB(s), \quad X_t(0) = e_{t-1}.
%\end{equation}
%

\medskip
\noindent \textbf{Choice of $M_t^{\leftarrow}[(e_{t-1},e_t), dv_t]$ -- Hypo-Elliptic Case, \cite{bier:20}:}
\medskip

The methodology developed in \cite{bier:20} follows along the lines of the elliptic case. That is, a guided proposal is constructed as in (\ref{eq:BD-proposal-D})-(\ref{eq:BD-proposal-D-drift}), with a proxy $\tilde{p}_{\Delta_t, e_t}(s,v)$ of the true transition density $p_{\Delta_t, e_t}(s,v)$ used within the drift function. The tractable density $\tilde{p}_{\Delta_t, e_t}(s,v)$ derives from the dynamics of the linear SDE (\ref{eq:linear-SDE}), however in the hypo-elliptic setting extra care is required on the specification of the drift $\tilde{b}_0(s)+\tilde{b}_1(s) v$ to yield absolute continuity between the constructed proposal and the bridge of the signal SDE.
`Matching' restrictions must now be placed both upon $\tilde{\sigma}(s)$ and the linear drift $\tilde{b}_0(s)+\tilde{b}_1(s) v$.
Under conditions (some of which we briefly discuss below) one has absolute continuity between the target bridge $P[(e_{t-1},e_t),dv_t]$ and proposed one $M_t^{\leftarrow,\mathrm{G}}[(e_{t-1},e_t), dv_t]$, with a density given as in (\ref{eq:BD-Girsanov-D}), with terms as defined therein. Note that the Radon-Nikodym derivative (\ref{eq:BD-Girsanov-D}) is well-defined for coefficient matrices $\sigma_t(s,v), \tilde{\sigma}(s)\in \bb{R}^{d\times d_w}$, $d_w\le  d$, even if these are non-rectangular.

\begin{remark}
\label{rem:integrated}
\cite{bier:20} look carefully into the class of `integrated diffusions', whereby the model for the signal writes as (here $d=2$, $d_w=1$): 
\begin{align*}
dX_{1}(s) &= X_{2}(s)ds; \\
dX_{2}(s) &= \zeta(s,X(s))\,ds + \sigma\,dB(s).
\end{align*}
with $\zeta:[0,\Delta]\times \mathbb{R}^{2}$, $\Delta>0$,  bounded, globally Lipschitz and $\sigma>0$. In this case, the choice: 
\begin{align*}
\tilde{b}_0(s) = 0, \quad \tilde{b}_1(s) = \left(   
\begin{array}{cc} 0 & 1 \\ 0 & 0  
\end{array}
\right), \quad \tilde{\sigma}(s) = (0, \sigma)^{\top},
\end{align*}
is shown in \cite{bier:20} to satisfy both the drift and diffusion coefficient matching conditions so that target and proposed bridges and their density in (\ref{eq:BD-Girsanov-D}) are all well-defined. The hypo-elliptic SDE used in our numerics later on will lie in the class of integrated diffusions. Though this class is of relatively simple structure, the SDE we will use is of interest in applications and we believe that it will serve well to illustrate our methodology.  
\end{remark}

\section{Reparameterised Feynman-Kac Formulae}
\label{sec:transform}

Assume now that we have chosen a kernel $M_t[x_{t-1},dx_t]$ within the framework of the guided Feynman-Kac formula, either by using a forward proposal $M_t^{\rightarrow}[x_{t-1},dx_t]$ or a backward one $M_t^{\leftarrow}[x_{t-1},dx_t]$ as in  (\ref{eq:FD-M}) or~(\ref{eq:BD-M}), respectively. The corresponding potentials are then given by $G_{t}^{\rightarrow}(x_{t-1},x_t)$ in  (\ref{eq:FD-G})  or $G_{t}^{\leftarrow}(x_{t-1},x_t)$ in (\ref{eq:BD-G}), so that the induced Feynman-Kac model delivers the conditional distributions of the signal $X_{1:t}|Y_{1:t}$, $t \in \cl{T}$, for the CD-SSM of interest. One can immediately apply particle-based methods to approximate, e.g.: (i) the filtering distributions, using for instance~the standard version of a particle filter as described in Algorithm \ref{alg:FK-PF}; (ii) the smoothing distribution as in Algorithm \ref{alg:FK-GT}, albeit the method does not attempt to reselect ancestors in the backward pass and will suffer from path degeneracy. As already explained in the Introduction, due to the continuous-time nature of the signal, one cannot implement efficient smoothing methods that do aim to overcome path degeneracy 
with the derivations so far provided, as is the case for FFBS described in 
Algorithm~\ref{alg:FK-FFBS}. 
%based on the backward kernel  \eqref{eq:fk-backward-kernel}. 
We repeat here that in a probabilistic language this matter relates directly with the fact that the proposal kernel $M_t[x_{t-1},dx_t]$, as defined so far, does not have a density w.r.t.~a common reference measure that will not depend on the starting point $e_{t-1}$. The critical practical consequence of such a property is that the ancestor $x_{t-1}$ of a particle $x_t$ cannot be altered within any of the many algorithms that involve a smoothing approximation procedure and generate changes in the particle genealogy to overcome path degeneracy.  

To deal with the above issue we now construct new Feynman-Kac formulae via a transform of the models we have so far defined. That is,
we carefully set-up new proposal kernels $\bar{M}_t[z_{t-1}, dz_t]$ and potentials $\bar{G}_t(z_{t-1}, z_t)$, on corresponding measurable spaces $(\mathsf{Z}_t,\mathscr{Z}_t)$,
so that we overcome challenges (C.i)-(C.iii) for CD-SSMs stated in the Introduction.
%This formalism is defined on the same space as the guided formalism: $z'_t = (z_t, v_t) \in \cl{X}'$. 
Such constructions will lead to a new sequence of target distributions $\bar{\bb{Q}}_{t}[dz_{1:t}]$ on $\big(\mathsf{Z}_1\times \cdots \times \mathsf{Z}_t, \mathscr{Z}_1\otimes\cdots \otimes \mathscr{Z}_t  \big)$, through the standard Feynmac-Kac expression:
\begin{equation}
\label{eq:fk_guided_reparameterised}
    \bar{\bb{Q}}_t[dz_{1:t}] = \frac{1}{\bar{L}_t}\Big\{\bar{G}_1(z_1)\prod_{i=2}^t \bar{G}_i(z_{i-1}, z_i)\Big\}\times \bar{\bb{M}}_t[dz_{1:t}], \quad t \in \cl{T}.
\end{equation}
We will sometimes refer to the Feynman-Kac model given by \eqref{eq:fk_guided_reparameterised} as the `transformed Feynman-Kac formula/model' of a CD-SSM. The key elements of this construction (yet to be defined) are summarised as follows:
\begin{itemize}
    \item[i.] We can sample from the kernels $\bar{M}_t[z_{t-1}, dz_t]$, up-to time-discretisation.
    \item[ii.] The kernels $\bar{M}_t[z_{t-1}, dz_t]$ have a tractable density w.r.t.~a common reference measure that does not depend on the ancestor $z_{t-1}$. 
    The densities can be evaluated, up-to time-discretisation.
   % $\bb{W}[dz_t] \otimes dv_t$  
    %that can be evaluated.
    \item[iii.] The potentials $\bar{G}_t(z_{t-1},z_t)$ can be evaluated, up-to time-discretisation.
    %\item[4.] Each of the normalising constants $L_t = p_t(y_{1:t})$ are the marginal likelihoods of the CD-SSM.
    \item[iv.] 
    The transform involvess maps ${\mathsf{Z}}_t \mapsto {\mathsf{X}}_t$, $t\in\mathcal{T}$. 
    %For both $\overline{\bb{M}}_t$ and $\overline{\bb{Q}}_t$, there is a map $\mathfrak{F}_{t}:{\mathsf{Z}}^t \rightarrow {\mathsf{X}}^t$. 
    The image of 
    %$\overline{\bb{M}}_t[dz_{1:t}]$ and 
    $\bar{\bb{Q}}_t[dz_{1:t}]$ under such maps gives back  
    %$\bb{M}_t[dx_{1:t}]$ and 
    $\bb{Q}_t[dx_{1:t}]$ 
    %respectively, 
    from the original Feynman-Kac model.
\end{itemize}
Given such a transformed Feynman-Kac model, one will be able to implement the full spectrum of particle-based inference methods for the class of CD-SSMs. 
%that requires that the proposal kernels have densities with respect to a common reference measure. 
%For example, application of a given smoothing method will provide a Monte Carlo approximation of $\overline{\bb{Q}}_t[dz_{1:t}]$. Then, a Monte Carlo approximation of $\bb{Q}_t[dx_{1:t}]$, which under the original guided formalism is the filtering distribution of the signal of the CD-SSM at time $t$, can then immediately be recovered via application of the map $\mathfrak{F}_{t}$ upon the obtained particles.

We now provide the details for  the construction of the transformed model \eqref{eq:fk_guided_reparameterised}. 
Following the development of the forward and backward proposals in Section \ref{sec:guided}, we will also present two different definitions for $\bar{\bb{Q}}_t[dz_{1:t}]$.
%We place great focus on the method for constructing, based on the proposal kernel $M_t$, a new proposal kernel $\overline{M}_t$. 
%The new proposal $\overline{M}_t$ is defined implicitly through a map $F'_t: \cl{X}' \rightarrow \cl{X}'$. We present two possible constructions. 
We refer to the first as the `forward transform', as it is applicable to the case 
when the original Feynman-Kac model builds upon forward proposals, as described in Section \ref{subsec:FP}. This reparameterisation affords high flexibility in the choice of corresponding proposals, but will only be applicable when the diffusion coefficient $\sigma_t(s,v)$ is invertible. We refer to the second construct as  the `backward transform', as it is relevant for the Feynman-Kac models that make  use of the backward proposals in Section \ref{subsec:BP}. 
This direction does not require invertibility of $\sigma_t(s,v)$ and will be applicable to a broader class of signal SDEs for the CD-SSM, including hypo-elliptic SDEs.

%Having shown 2 approaches as to how to construct new proposal kernels $\overline{M}_t$, such that key elements 1. and 2. hold, we proceed to show how to construct the potentials $\overline{G}_t$ that, together with the proposal kernels, define a guided reparameterised Feynman-Kac formalism. We conclude by showing that key elements 3., 4. and 5. hold for the constructed formalism.

\subsection{Defining $\bar{M}^{\rightarrow}_t$: Forward Transform}

We consider the setting of the forward proposal  
$M_t^{\rightarrow}[x_{t-1}, dx_t]$ in Section \ref{subsec:FP}. Recall from (\ref{eq:FD-M}) that this kernel derives from the law of the proposal SDE in  (\ref{eq:FD-proposal}), extended with the end-point.
%and can be expressed in a backward decomposition as in \eqref{eq:M_t_forward_proposal_bd}. 
We will define a new kernel $\bar{M}^{\rightarrow}_t[z_{t-1}, dz_t]$, obtained from the forward proposal $M_t^{\rightarrow}[x_{t-1}, dx_t]$ via use of an invertible map,
the latter defined via an auxiliary bridge SDE.

The construction of this map will relate with the backward decomposition of $M_t^{\rightarrow}[x_{t-1}, dx_t]$ discussed in Remark \ref{rem:FB} and expressed in (\ref{eq:FD-BD}) in terms of the transition density $m_t^{\rightarrow}(e_t|e_{t-1})$ for the ending point and a connecting SDE bridge $M_t^{\rightarrow}[(e_{t-1},e_t),dv_t]$. Indeed, we define an auxiliary bridge SDE with distribution 
$R_t[(e_{t-1},e_t), dv_t]$ that dominates the law of the bridge of the forward proposal, 
that is, $M_t^{\rightarrow}[(e_{t-1},e_t), dv_t] \ll R_t[(e_{t-1},e_t), dv_t]$. The auxiliary bridge SDE writes as follows: 
\begin{flalign}
\label{eq:FD-Rbridge}
  R_t[(e_{t-1},e_t), dv_t]: \qquad  dV_t(s) = a_t^{br}(s, V_t(s); y_t) + \sigma_t(s, V_t(s))dB(s), \quad V_t(0) = e_{t-1}, &&
\end{flalign}
for a drift function $a_t^{br}(\cdot, \cdot; y_t):[0,\Delta_t]\times \bb{R}^{d}\to \bb{R}^{d}$. This drift must be such so that the solution of (\ref{eq:FD-Rbridge}) is forced to hit the ending point $e_t$. Following the earlier discussion about the Delyon-Hu bridge, a simple choice of drift is the following: 
\begin{equation}
\label{eq:FD-Rbridge-drift}
a_t^{br,\mathrm{DH}}(s, v; y_t) = \frac{e_t - v}{\Delta_t - s}.
\end{equation}
We denote the corresponding bridge distribution for this choice as $R_t^{\mathrm{DH}}[(e_{t-1},e_t), dv_t]$.
\begin{remark}
The bridge SDE law $R_t^{\mathrm{DH}}[(e_{t-1},e_t), dv_t]$ coincides with  $M^{\leftarrow,\textrm{DH}}_t[(e_{t-1},e_{t}), dv_t]$ defined in the context of setting up a backward proposal in Section \ref{subsec:BP}. We note though that this bridge SDE will now take up a very different role compared to the one in Section \ref{subsec:BP}. 
That is, here  bridge SDE (\ref{eq:FD-Rbridge}) will provide the means for a transform of paths of the proposal SDE (\ref{eq:FD-proposal}). In contrast, in Section \ref{subsec:BP}, $M^{\leftarrow,\textrm{DH}}_t[(e_{t-1},e_{t}), dv_t]$ formed part of the proposal.  
\end{remark}
\noindent Following \cite{dely:06}, we obtain a density function which is as the one in (\ref{eq:BD-Girsanov-DH}), with the difference that we now work with the bridge $M_t^{\rightarrow}[(e_{t-1},e_t),dv_t]$ of the proposal SDE (\ref{eq:FD-proposal}) rather than the bridge of the signal SDE, $P_t[(e_{t-1},e_{t}), dv_t]$. That is, under the drift choice (\ref{eq:FD-Rbridge-drift}), and regularity conditions, we have: 
%This SDE has a drift function $a_t^{br}$ that is tractable. This drift may or may not depend on the drift of the unconditioned diffusion $a_t$. A simple, drift-independent choice is the Delyon-Hu bridge \cite{delyon2006simulation}, which would be the special case:
%\begin{equation}\label{eq:R_t_dh_bridge}
%  R_t[v_{t-1:t}, dx_t]: \quad  dX_t(s) = \frac{v_t - X_t(s)}{\Delta s - s} + \sigma_t(s, X_t(s))dW(s) \qquad X_t(0) = v_{t-1}
%\end{equation}
%
%
\begin{equation}
\label{eq:FD-Girsanov-RDH}
\frac{dM_t^{\rightarrow}[(e_{t-1},e_t), \cdot]}{dR_t^{\mathrm{DH}}[(e_{t-1},e_t), \cdot]}(v_t) = \frac{\cl{N}(e_t; e_{t-1},  \Sigma_t(0, e_{t-1}))}{m_t^{\rightarrow}(e_t|e_{t-1})}\sqrt{\frac{\det(\Sigma_t(0, e_{t-1}))}{\det(\Sigma_t(\Delta_t, e_{t}))}}\times \exp\big\{\phi_t(v_t)\big\},
\end{equation}
for the functional: 
\begin{align*}
    \phi_t(v_t)= &\int_0^{\Delta_t} ((b_t^{\rightarrow})^\top\Sigma_t^{-1})(s, v_t(s))dv_t(s) \\
    &\qquad - \tfrac{1}{2}\int_0^{\Delta_t}((b_t^{\rightarrow})^{\top} \Sigma_t^{-1}b_t^{\rightarrow})(s, v_t(s)) ds \\[0.1cm]
    &\qquad\qquad- \tfrac{1}{2}\int_0^{\Delta_t} \frac{e_t - v_t(s)}{\Delta_t  - s} \diamond d\big\{\Sigma_t^{-1}(s, v_t(s))(e_t - v_t(s))\big\}.
\end{align*}
%
%\textcolor{blue}{
%The Backward Transform (BT) makes use of the bridge SDE (\ref{eq:SSM-bX}) to replace the path-valued variable $\tbX_t|\tbX_{t-1}(s_{t-1})\sim Q(d\tbX_t|\tbX_{t-1})$ with an 1--1 representation of it. As discussed earlier, the objective is to develop a method that will allow particle-based algorithms to be applicable in the
%DC-SSM setting.} 

It is now important to consider the auxiliary bridge SDE (\ref{eq:FD-Rbridge}) as the means for obtaining a map $\mathfrak{F}_t(\cdot;e_{t-1},e_t):\mathsf{U}_t\to \mathsf{X}_t(e_t)$ from the driving noise 
$B_t:=\{B(s)\}_{s\in[0,\Delta_t]}$ to the solution $V_t=\{V_t(s)\}_{s\in[0,\Delta_t]}$ of (\ref{eq:FD-Rbridge}), so that:
\begin{align}
\label{eq:F_define}
B_t \mapsto V_t = \mathfrak{F}_t(B_t;e_{t-1},e_{t}).
\end{align}
We have defined spaces $\mathsf{U}_{t}$ and $\mathsf{X}_t(e_t)$, so that the former corresponds to the set of realisations of a standard $d$-dimensional Brownian motion on $[0,\Delta_t]$ and 
the latter to realisations of the solution of (\ref{eq:FD-Rbridge}), thus emphasising that paths must terminate at $e_t$ in the case of $\mathsf{X}_t(e_t)$. Recall that in the setting of the forward approach we only consider $d_w=d$.

\begin{assumption}
\label{ass:invertible}
The diffusion coefficient matrix $\sigma_t(s,v)\in\mathbb{R}^{d\times d}$ is invertible for all $(s,v)\in[0,\Delta_t]\times \bb{R}^{d}$, $t\in\mathcal{T}$.
\end{assumption}
\noindent Under Assumption \ref{ass:invertible}, 
we can make use of the inverse of map $\mathfrak{F}_t$ to transform paths generated from the  proposal SDE~(\ref{eq:FD-proposal}). That is, for $v_t\sim M_t^{\rightarrow}[e_{t-1},dv_t]$ we set $e_t=\texttt{end}(v_t)$ and 
define the map $\mathfrak{F}_t^{-1}:\mathsf{X}_t(e_t)\to \mathsf{U}_t$, given $e_{t-1}$, $e_t$, as follows: 
\begin{align}
\label{eq:FD-Finv}
u_t := \mathfrak{F}_t^{-1}(v_t;e_{t-1},e_t).
\end{align}
Note that $u_t$ is, in practice, obtained via recursive application of  the following expression: 
\begin{align}
\label{eq:FD-Finv-SDE}
du_t(s) = \sigma_t(s, v_t(s))^{-1}\big\{dv_t(s) - a_t^{br}(s, v_t(s))ds\big\}.
\end{align}

We now have all necessary ingredients for the definition of the transformed proposal kernel $\bar{M}^{\rightarrow}_t[z_{t-1}, dz_t]$. 
In particular, we have that: 
\begin{align*}
z_t = (u_{t},e_t)\in \mathsf{Z}_{t} = \mathsf{U}_t\times \bb{R}^{d},\qquad t\in\mathcal{T}.
\end{align*}
Then, 
given $z_{t-1}=(u_{t-1},e_{t-1})$, we have that the variate $z_t = (u_{t},e_{t})\sim\bar{M}^{\rightarrow}_t[z_{t-1},dz_t]$ is obtained as follows:
\begin{itemize}
\item[i.] 
Sample $v_t\sim M_t^{\rightarrow}[e_{t-1}, dv_t]$, with the kernel defined via the proposal SDE  (\ref{eq:FD-proposal}) and set $e_t=\texttt{end}(v_t)$.
\item[ii.] Retrieve $u_t=\mathfrak{F}_t^{-1}(v_t;e_{t-1},e_t)$ via use of (\ref{eq:FD-Finv-SDE}).
\end{itemize}
%
%This kernel is obtained from the forward proposal kernel $M_t[v_{t-1}, dx'_t]$ as the image of the following map $(F'_t)^{-1}: \cl{X}' \rightarrow \cl{X}'$
%\begin{equation}
 %   (F'_t)^{-1}(v_{t-1}, \cdot): (X_t, V_t) \rightarrow (F_t^{-1}(v_{t-1}, V_t, X_t), V_t)
%\end{equation}
%We can describe this map in words as follows: keep the end point $V_t$ fixed, and apply the map $F_t^{-1}$ from the auxiliary bridge \eqref{eq:R_t_bridge} to the driving noise, to the path $X_t$ between $v_{t-1}$ and $V_t$. By defining our new kernel $\overline{M}_t$ as the image of a map from a kernel $M_t$ that it is possible to simulate from, this implies that it is also possible to simulate from $\overline{M}_t$: this is done by generating a sample from $M_t$, then applying the map $(F'_t)^{-1}$ to the sample. The sample from $M_t$ is generated by consideration of its forward decomposition, whilst the constructed map $(F'_t)^{-1}$ from $M_t$ to $\overline{M}_t$ was constructed by considering the backward decomposition of $M_t$. 
%------
%\\\\Note that since we have defined $\overline{M}_t$ as the image of $M_t$ under an invertible map $(F'_t)^{-1}$, we can also obtain the $M_t$ from the $\overline{M}_t$, by applying the map:
%\begin{equation}
 %   (F'_t)(v_{t-1}, \cdot): (Z_t, V_t) \rightarrow (F_t(v_{t-1}, V_t, Z_t), V_t)
%\end{equation}
From the above specification of the reparameterised kernel $\bar{M}^{\rightarrow}_t[z_{t-1},dz_t]$, one can obtain the following expression for its backward decomposition:
\begin{equation}
\label{eq:FD-M-transform}
    \bar{M}^{\rightarrow}_t[z_{t-1}, dz_t] = \big(m_t^{\rightarrow}(e_t|e_{t-1})de_t\big)\otimes \bar{M}^{\rightarrow}_t[(e_{t-1},e_t), du_t].
\end{equation}
Recall that $m_t^{\rightarrow}(e_t|e_{t-1})$ is the transition density of the forward proposal SDE (\ref{eq:FD-proposal}). Then, the kernel $\bar{M}^{\rightarrow}_t[(e_{t-1},e_t), du_t]$ is defined as the distribution 
of $u_t\sim \mathfrak{F}^{-1}_t(v_t;e_{t-1}, e_t)$, for the path-variable $v_t\sim M_t^{\rightarrow}[e_{t-1}, dv_t]$ and for $e_t=\texttt{end}(v_t)$.
This definition of $u_t$ has connections with the `innovation process' used in the context of MCMC algorithms, e.g.~in  \cite{chib:04, goli:08}. 

Let $\mathbb{W}_t$ denote the law of a standard $d$-dimensional Brownian motion on $[0,\Delta_t]$. Notice that application of 
the map $\mathfrak{F}_t(\cdot;e_{t-1},e_t)$ to a Brownian motion $B_t=\{B(s)\}_{s\in[0,\Delta_t]}$   (resp.~to $u_t\sim \bar{M}^{\rightarrow}_t[(e_{t-1},e_t), du_t]$) will yield a random path distributed according to the solution of the bridge SDE (\ref{eq:FD-Rbridge}) (resp.~distributed according to $M^{\rightarrow}_t[(e_{t-1},e_t), dv_t]$).
Thus, we immediately obtain the following density, due to the 1--1 transformation of the random variates involved:
%
%If we apply the map $F_t(v_{t-1:t}, \cdot)$ to a Brownian motion $W$, we recover a $X_t$ with distribution the auxiliary bridge $R_t[v_{t-1:t}, dx_t]$. If we apply this map to a $Z_t$ with distribution given by $\overline{M}_t[v_{t-1:t}, dz_t]$, we obtain the bridge of the proposal $M_t[v_{t-1:t}, dx_t]$. Since $R_t[v_{t-1:t}, dx_t] >> M_t[v_{t-1:t}, dx_t]$ with tractable likelihood ratio, we have that $\bb{W}[dz] >> \overline{M}_t[v_{t-1:t}, dz_t]$, with Radon-Nikodym derivative: 
\begin{equation}
\label{eq:FD-Girsanov-Mbar-bridge}
    \frac{d\bar{M}^{\rightarrow}_t[(e_{t-1},e_t), \cdot]}{d\bb{W}_t}(u_t) = \frac{dM_t^{\rightarrow}[(e_{t-1},e_t), \cdot]}{dR_t[(e_{t-1},e_t), \cdot]}\big(\mathfrak{F}_t(u_t;e_{t-1}, e_t)\big).
\end{equation}
For the case when the  Delyon-Hu drift function (\ref{eq:FD-Rbridge-drift}) is used in the derivations, the density on the right-hand-side of (\ref{eq:FD-Girsanov-Mbar-bridge}) is as provided in 
(\ref{eq:FD-Girsanov-RDH}).
Continuing from (\ref{eq:FD-Girsanov-Mbar-bridge}), the complete proposal kernel $\bar{M}^{\rightarrow}_t[z_{t-1},dz_t]$ admits the following Radon-Nikodym derivative (recall that $z_t=(u_t,e_t)$):
%Therefore, the proposal kernel that we have defined as \eqref{eq:proposal_kernel} is absolutely continuous with respect to reference measure $\bb{W} \otimes dv_t$, with Radon-Nikodym derivative:
\begin{align}
   \bar{M}^{\rightarrow}_t(z_t|z_{t-1})&= \frac{\bar{M}^{\rightarrow}_t[z_{t-1}, dz_t]}{\mathrm{Leb}(de_t)\otimes \bb{W}_t[du_t]}\nonumber  \\[0.2cm] &= m_t^{\rightarrow}(e_{t}|e_{t-1}) \x \frac{dM^{\rightarrow}_t[(e_{t-1},e_t), \cdot]}{dR_t[(e_{t-1},e_t), \cdot]}\big(\mathfrak{F}_t(u_t;e_{t-1},e_t)\big).
   \label{eq:FD-density}
\end{align}
%
%For an appropriately chosen auxiliary bridge \eqref{eq:R_t_bridge}, the continuous-time likelihood \eqref{eq:M_t_rn_derivative} is tractable. Indeed, assuming that we have chosen our auxiliary bridge to be the Delyon-Hu bridge \eqref{eq:R_t_dh_bridge}, 
Notice that the transition density $m_t^{\rightarrow}(e_t|e_{t-1})$ of the proposal SDE 
(\ref{eq:FD-proposal}) cancels out when taking the product in the right-hand-side of (\ref{eq:FD-density}), as $m_t^{\rightarrow}(e_t|e_{t-1})$ also appears in the probability measure 
$M^{\rightarrow}_t[(e_{t-1},e_t),du_t] = M^{\rightarrow}_t[e_{t-1},(du_t,de_t)]/ (m_t^{\rightarrow}(e_{t}|e_{t-1})de_t)$.
For the case $R_t[(e_{t-1},e_t),dv_t]=R_t^{\mathrm{DH}}[(e_{t-1},e_t),dv_t]$, we provide the analytically tractable expression for $\bar{M}_t^{\rightarrow}[z_{t-1},dz_t]$ via use of (\ref{eq:FD-Girsanov-RDH}).

\subsection{Defining $\bar{M}_t^{\leftarrow}$: Backward Transform}
We present a method for the construction of a transform 
$\bar{M}_t^{\leftarrow}[z_{t-1}, dz_t]$ of the 
backward proposal $M^{\leftarrow}_t[x_{t-1}, dx_t]$, the latter specified
 via the backward decomposition in (\ref{eq:BD-M}), consisted of the kernels $m_{t}^{\leftarrow}(e_t|e_{t-1})de_t$ and $M_t^{\leftarrow}[(e_{t-1},e_t),dv_t]$.

%Recall that a backward proposal can be expressed in backward decomposition, as:
%\begin{equation}\label{eq:M_t_backward_proposalR}
%    M_t[v_{t-1}, dx'_t] = [m_t(v_t|v_{t-1})dv_t]M_t[v_{t-1:t}, dx_t]
%\end{equation}

%
Nothing needs to be done about $m_{t}^{\leftarrow}(e_t|e_{t-1})$. 
Moving to kernel $M_t^{\leftarrow}[(e_{t-1},e_t),dv_t]$ corresponding to the proposal bridge SDE (\ref{eq:BD-proposal}), we define the map $\mathfrak{H}_t(\cdot;e_{t-1},e_t): \mathsf{U}_t \rightarrow \mathsf{X}_t(e_t)$ that projects the driving noise $B_t=\{B(s)\}_{s\in[0,\Delta_t]}$ of (\ref{eq:BD-proposal}) onto the solution 
 of (\ref{eq:BD-proposal}). We now denote the solution of the bridge SDE (\ref{eq:BD-proposal}) as
$V^{\leftarrow,br}_t=\{V(s)^{\leftarrow,br}\}_{s\in[0,\Delta_t]}$ for clarity.
That is, we have: 
\begin{align*}
%\label{eq:H_define}
B_t \mapsto V^{\leftarrow,br}_t = \mathfrak{H}_t(B_t;e_{t-1},e_{t}).
\end{align*}
The definition resembles the one of $\mathfrak{F}_{t}(\cdot;\cdot)$ in (\ref{eq:F_define}) used in the case of the forward transform. For $v_t\sim M_t^{\leftarrow}[(e_{t-1},e_t),dv_t]$, the backward reparameterisation considers instead the driving Wiener noise, $u_t$, in the construction of $v_t$ via $v_t=\mathfrak{H}(u_t;e_{t-1},e_t)$.
In this case: (i) one simply has that $u_t\sim \mathbb{W}_t$; (ii) no inversion of $u_t\mapsto \mathfrak{H}(u_t;e_{t-1},e_t)$ is required. This second characteristic of the backward approach is important, as it sidesteps the requirement for invertibility of the diffusion coefficient in the SDE (\ref{eq:CD-X}) for the latent signal, thus allowing application of the backward approach to hypo-elliptic models.
%
%\begin{equation}\label{eq:bp_bridge_forward_map}
  %  F_t(v_{t-1}, v_t, \cdot): W \rightarrow X_t \qquad \bb{W}%[dz_t] \rightarrow M_t[v_{t-1:t}, dx_t] 
%\end{equation}

Overall, we define $\bar{M}^{\leftarrow}_t[z_{t-1},dz_t]$ as follows, for $z_t=(u_t,e_t)$ given $z_{t-1}=(u_{t-1}, e_{t-1})$:
\begin{equation}
\label{eq:BD-Mbar}
    \bar{M}_t^{\leftarrow}[z_{t-1}, dz_t] = \big(m_t^{\leftarrow}(e_t|e_{t-1})de_t\big) \otimes \bb{W}_t[dz_t].
\end{equation}
The definition of $\bar{M}_t^{\leftarrow}[z_{t-1},dz_t]$ immediately implies that one 
can sample from this kernel, by independently generating a proposed Brownian motion path $z_t\sim \mathbb{W}_t$ and a proposed end-point  $e_t\sim m_t^{\leftarrow}(e_t|e_{t-1})$. Further, this proposal kernel has a tractable density given by:
\begin{equation}
\label{eq:BD-density}
  \bar{M}_t^{\leftarrow}(z_{t}|z_{t-1}) =   \frac{\bar{M}_t^{\leftarrow}[z_{t-1}, dz_t]}{\mathrm{Leb}(de_t) \otimes \bb{W}_t[du_t]} = m_t^{\leftarrow}(e_t|e_{t-1}).
\end{equation}

\begin{remark}
We note a main difference between 
backward and forward proposals, $M_t^{\rightarrow}[x_{t-1},dx_t]$ and $M_t^{\leftarrow}[x_{t-1},dx_t]$, that ultimately leads to the method based on $\bar{M}_t^{\leftarrow}[z_{t-1},dz_t]$ being applicable  also for a non-invertible diffusion coefficient $\sigma_t(s,v)$. 
Kernel $M^{\leftarrow}_t[x_{t-1}, dx_t]$ is constructed in (\ref{eq:BD-M}) via direct consideration of the backward decomposition of the target $P_t[x_{t-1},dx_t]$ in (\ref{eq:BD-P}). That is, one selects \emph{tractable} laws $m_t^{\leftarrow}(e_t|e_{t-1})$ and $M_t^{\leftarrow}[(e_{t-1},e_t), dv_t]$, both for the proposed ending point $e_t$ and the proposed bridge SDE. Then, the transform $(e_t,u_t)\mapsto (e_t,v_t)$, with $(e_t,u_t)$ admitting a density w.r.t.~the reference measure $\mathrm{Leb}(de_t) \otimes \bb{W}_t[du_t]$, is obtained by working with the noise used to sample $v_t\sim M_t^{\leftarrow}[(e_{t-1},e_t),dv_t]$.
%A common choice here is that of the   Delyon-Hu bridge. 
Kernel $M^{\rightarrow}_t[x_{t-1}, dx_t]$ in (\ref{eq:FD-M}) relates to the forward decomposition of $P_t[x_{t-1},dx_t]$ in (\ref{eq:FD-P}), thus one must now work with $v_t\sim M_t^{\rightarrow}[e_{t-1},dv_t]$. However, use of the driving noise for $v_t$ cannot lead to non-degenerate particle-based smoothing methods. Instead, one must rely to the  auxiliary bridge SDE, of law $R_t[(e_{t-1},e_t), dv_t]$, which serves to transform the signal $(v_t, e_t)\mapsto (u_t, e_t)$, via a map that indeed requires the inverse of $\sigma_t(s,v)$.
\end{remark}

\subsection{Potentials $\bar{G}_t^{\rightarrow}$, $\bar{G}_t^{\leftarrow}$}
We have now defined two classes of tractable kernels  
$\bar{M}_t^{\rightarrow}[z_{t-1}, dz_t]$, $\bar{M}^{\leftarrow}_t[z_{t-1}, dz_t]$,  that we can sample from, 
and which have analytical densities, given in (\ref{eq:FD-density}) and \eqref{eq:BD-density} w.r.t.~the reference measure $\mathrm{Leb}(de_t)\otimes\bb{W}_t[du_t]$. 
Critically, the reference measure does not involve the 
end-point $e_{t-1}$. 
We now complete our construction of the transformed Feynman-Kac models, by determining the corresponding potentials, denoted $\bar{G}_t^{\rightarrow}(z_{t-1},z_t)$ and $\bar{G}_t^{\leftarrow}(z_{t-1},z_t)$. 
\subsubsection{Forward Potential $\bar{G}_t^{\rightarrow}$}
%
%
%We have thus re-defined the hidden variable of the original SSM. 
%%
%\begin{remark}
Recall that we work with the dynamics of the proposal SDE (\ref{eq:FD-proposal}) to obtain the path-valued variate  $v_t\sim M_t^{\rightarrow}[e_{t-1}, dv_t]$. Upon taking $u_t = \mathfrak{F}_{t}^{-1}(v_t;e_{t-1},e_t)$ with $e_t=\texttt{end}(v_t)$, one obtains $z_t=(v_t,e_t)\sim \bar{M}_{t}^{\rightarrow}[z_{t-1},dz_t]$.
The transform:
\begin{align*}
x_t=(v_t,e_t)\mapsto z_t = \big(u_t, e_t);\qquad u_t = \mathfrak{F}_{t}^{-1}(v_t;e_{t-1},e_t),
\end{align*}
when applied on the target signal variate 
$x_t\sim P_t[x_{t-1},dx_t]$ itself will similarly give rise to a probability law $\bar{P}_t^{\rightarrow}[z_{t-1},dz_t]$ on the $z$-space. 
Thus, the potential $\bar{G}_t^{\rightarrow}(z_{t-1},z_t)$ on the $z$-space is equal to: 
\begin{align}
\label{eq:FD-Gbar}
\bar{G}_t^{\rightarrow}(z_{t-1},z_t) = 
\frac{d\bar{P}_t^{\rightarrow}[z_{t-1},\cdot]}{d\bar{M}^{\rightarrow}_t[z_{t-1},\cdot]}(z_t) \times  f_t(y_t|e_t),
\end{align}
for the Radon-Nikodym derivative: 
\begin{align}
\nonumber
\frac{d\bar{P}_t^{\rightarrow}[z_{t-1},\cdot]}{d\bar{M}_t^{\rightarrow}[z_{t-1},\cdot]}(z_t) &= \frac{p_t(e_t|e_{t-1})}{m_{t}^{\rightarrow}(e_t|e_{t-1})}\times \frac{d\bar{P}_t[(e_{t-1},e_t),\cdot]}{d\bar{M}^{\rightarrow}_t[(e_{t-1},e_t),\cdot]}
(u_t) \\[0.2cm]
\nonumber
 &= \frac{p_t(e_t|e_{t-1})}{m_{t}^{\rightarrow}(e_t|e_{t-1})}\times \frac{dP_t[(e_{t-1},e_t),\cdot]}{dM^{\rightarrow}_t[(e_{t-1},e_t),\cdot]}
\big(\mathfrak{F}_t(u_t;e_{t-1},e_t)\big) \\[0.2cm] 
&\equiv 
\frac{dP_t[e_{t-1},\cdot]}{dM^{\rightarrow}_t[e_{t-1},\cdot]}
\big(\mathfrak{F}_t(u_t;e_{t-1},e_t)\big).
\label{eq:FD-Gbar2}
\end{align}
The expression in the last line of (\ref{eq:FD-Gbar2}) can be analytically evaluated up-to time-discretisation via use of the Girsanov density in (\ref{eq:FD-Girsanov}).

\subsubsection{Backward Potential $\bar{G}_t^{\leftarrow}$}

In this case, we work with the transform:
\begin{align*}
z_t = \big(u_t, e_t) \mapsto x_t=(v_t, e_t); \qquad v_t = \mathfrak{H}_{t}(u_t;e_{t-1},e_t).
\end{align*}
When considering the target signal path $x_t\sim P_t[x_{t-1},dx_t]$, we have that $x_t=(v_t,e_t)$ with laws $e_t\sim p_t(e_t|e_{t-1})de_t$ and $v_t\sim P_t[(e_{t-1},e_t),dv_t]$. Then, $u_t\sim \bar{P}_t^{\leftarrow}[(e_{t-1},e_t),du_t]$, with this latter distribution being defined via the requirement $v_t = \mathfrak{H}_{t}(u_t;e_{t-1},e_t)$ must follow $P_t[(e_{t-1},e_t),dv_t]$.

Thus, the potential on the $z$-space writes as follows: 
\begin{align}
\nonumber
\bar{G}_{t}^{\leftarrow}(z_{t-1},z_t)&= \frac{d\bar{P}_t^{\leftarrow}[z_{t-1},\cdot]}{d\bar{M}^{\leftarrow}_t[z_{t-1},\cdot]}(z_t)\times f_t(y_t|e_t)   \\[0.2cm]
\nonumber
&= \frac{p_t(e_t|e_{t-1})}{m_{t}^{\leftarrow}(e_t|e_{t-1})}\times \frac{d\bar{P}_t^{\leftarrow}[(e_{t-1},e_t),\cdot]}{d\bar{M}^{\leftarrow}_t[(e_{t-1},e_t),\cdot]}
(u_t)\times f_t(y_t|e_t)    \\[0.2cm] &= \frac{p_t(e_t|e_{t-1})}{m_{t}^{\leftarrow}(e_t|e_{t-1})}\times \frac{dP_t[(e_{t-1},e_t),\cdot]}{dM^{\leftarrow}_t[(e_{t-1},e_t),\cdot]}
\big(\mathfrak{H}_t(u_t;e_{t-1},e_t)\big)\times f_t(y_t|e_t).  
\label{eq:BD-Gbar}
\end{align}
The Radon-Nikodym derivative appearing on the right-hand-side of the last line in the above expression is given in (\ref{eq:BD-Girsanov-DH}) for proposal $M^{\leftarrow}_t[(e_{t-1},e_t),dv_t] = M^{\leftarrow,\mathrm{DH}}_t[(e_{t-1},e_t),dv_t]$, and in (\ref{eq:BD-Girsanov-D}) for the case when 
$M^{\leftarrow}_t[(e_{t-1},e_t),dv_t] =M_t^{\leftarrow,\mathrm{G}}[(e_{t-1},e_t), dv_t]$.

\subsection{Final Forward and Backward Feynman-Kac Formulae}

The developed proposal kernels and corresponding potentials in this section fully specify our forward and backward transformed Feynman-Kac formulae. The proposal kernels are defined in a way so that one can sample from these. They also satisfy the important property that is required for implementation of improved particle-based algorithms that reselect ancestors: they admit a tractable density w.r.t.~a common $\sigma$-finite dominating measure that does not depend on $e_{t-1}$, namely $\mathrm{Leb}(de_t) \otimes \bb{W}_t[du_t]$. These transformed Feynman-Kac models for CD-SSMs may then be used as inputs to the particle-based algorithms presented in Section \ref{sec:FK} -- and, effectively, to the full toolkit of particle-based methods that involve altering ancestors. To demonstrate this, we provide in Algorithms \ref{alg:FD-PF} and \ref{alg:FD-FFBS} pseudo-codes for the particle filter and FFBS based on the forward reparameterised Feynman-Kac model. Then, Algorithms \ref{alg:BD-PF} and \ref{alg:BD-FFBS} provide similar pseudo-codes for the corresponding backward one.

%The potential is defined in such a way that from the target distribution under reparameterisation $\overline{\bb{Q}}_t$, one can construct a map $F'_{1:t}$ to recover the original target distribution $\bb{Q}_t$ which is the distribution of the signal of the CD-SSM, given the noise. Each potential is defined to be the composition of applying the map $F'_t$ and $G_t$: 
%
%\begin{equation}
%\overline{G}_t(e_{t-1}, z_t) = G_t(e_{t-1}, F'_t(z'_t)) = G_t(v_{t-1}, (F_t(v_{t-1}, v_t, z_t), v_t))    
%\end{equation}
%
%So, when evaluating the potential in a particle filter, one takes a sample from $\overline{M}_t$, applies the map $F_t$ to recover a sample from $M_t$, then evaluates the potential $G_t$ from the guided formalism. The map $F_t$ is known, and the potential $G_t$ can be evaluated, therefore the potential $\overline{G}_t$ can be evaluated. We have now completed our construction of the guided, reparameterised formalism: our choices of reparameterised proposal kernels and potentials define a new sequence of target distributions $\overline{\bb{Q}}_t[dz'_{1:t}]: t \in \cl{T}$. For each t, we have by construction: 

\begin{algorithm}[!h] 
    \caption{Particle Filter for CD-SSM in (\ref{eq:CD-X})-(\ref{eq:CD-Y}) -- under forward methodology} %\citep{chopin2020introduction}}
    \label{alg:FD-PF}
    \SetKwInOut{Input}{Input}
    \SetKwInOut{Output}{Output}
    \KwIn{$\bar{M}_1^{\rightarrow}[dz_1]$, $\bar{M}_t^{\rightarrow}[z_{t-1},dz_{t}]$ and  $\bar{G}_1^{\rightarrow}(z_1)$, $\bar{G}_t^{\rightarrow}(z_{t-1},z_t)$, $t=2, \ldots, T$ (use (\ref{eq:FD-M-transform}), (\ref{eq:FD-Gbar})).}
    \KwOut{Particles $z^{1:N}_{1:T}=(u_{t}^{1:N},
    e_t^{1:N})_{t\in[1:T]}$, $x^{1:N}_{1:T}=(v_{t}^{1:N}, 
    e_t^{1:N})_{t\in[1:T]}$, ancestors $A^{1:N}_{2:T}$, weights $W_{1:T}^{1:N}$.}
    %Sample $v^j_{1}$ from proposal SDE (\ref{eq:FD-proposal}); set $e_0^{j}=\texttt{start}(v_1^{j})$ and $e_1^{j}=\texttt{end}(v_1^{j})$\;
    %Apply backward reparameterisation: %$u^{j}_{1}=\mathfrak{F}^{-1}_{1}\big(v_1^j;e_0^{j},e_1^{j}\big)$\;
    %Collect particles: $z^{j}_1 = (e_1^{j}, %u^{j}_{1})$\;
    %Assign weights: $w^j_1 \gets %\bar{G}_1^{\rightarrow}(z^{j}_1)$\;
    %Normalise weights: $W^j_1 \gets \frac{w^j_1}{\sum_{k=1}^{N}w^k_1}$\;
    \For{$t=1,...,T$}{
        If $t=1$, set $A_t^{j}\gets j$; else   $A_t^{j}\sim \texttt{Resample}(W^{1:N}_{t-1})$\; %User defined Resampling Scheme 
        %$\mathsf{X}_t^j \sim Q(dx_t|\mathsf{X}_{t-1}^{A_t^j})$\;
    Sample $v^j_{t}$ from the proposal SDE (\ref{eq:FD-proposal}) with starting point $e_{t-1}^{A_t^j}$ \; 
    Set $e_{t-1}^{j}=\texttt{start}(v_t^{j})$ and 
    $e_t^{j}=\texttt{end}(v_t^{j})$\;
    Apply  transform: 
    $u^{j}_{t}=\mathfrak{F}_t^{-1}(v^j_{t};e_{t-1}^{j},e_t^{j})$\;
    Collect particles: $z^{j}_t = (u^{j}_{t}, e_t^{j})$, $x^{j}_t = (v^{j}_{t}, e_t^{j})$\;
        Assign weights: $w_t^j = \bar{G}_t^{\rightarrow}(z_{t-1}^{A_t^j}, z_t^{j})$ using the expressions in (\ref{eq:FD-Gbar})-(\ref{eq:FD-Gbar2}) for $\bar{G}^{\rightarrow}_t(\cdot,\cdot)$\;
        Normalise weights: $W_t^j = \frac{w_t^j}{\sum_{k=1}^N w_t^k}$;
    }
\end{algorithm}
\begin{algorithm}[!h] 
    \caption{FFBS for CD-SSM in (\ref{eq:CD-X})-(\ref{eq:CD-Y}) -- under forward methodology} %\cite{godsill2004monte}}
    \label{alg:FD-FFBS}
    \SetKwInOut{Input}{Input}
    \SetKwInOut{Output}{Output}
    \KwIn{Output $z^{1:N}_{1:T}=(u_{t}^{1:N}, e_t^{1:N})_{t\in[1:T]}$, $A_{2:T}^{1:N}$, $W_{1:T}^{1:N}$, from Algorithm \ref{alg:FD-PF}.}
    \KwOut{Approximate sample $\big(x_1^{B_1}=(v_1, e_1)^{B_1}, \ldots, x_{T}^{B_T}=( v_T, e_T)^{B_T}\big)$ from the smoothing distribution $P\,[\,X_{1:T} \in dx_{1:T}\,|\, Y_{1:T} = y_{1:T}\,]$.}
    Sample $B_T \sim \mathcal{M}(W_T^{1:N})$\;
    %Return $z_{T}^{B_T}=(e_T, u_T)^{B_T}$ \;
    \For{$t=T,...,2$}{
        Assign weights: $\hat{w}_{t-1}^j \gets 
        \bar{M}_t^{\rightarrow}(z_{t}^{B_{t}}|z_{t-1}^j)\,\bar{G}^{\rightarrow}_t(z_{t-1}^{j},z_t^{B_t}) \cdot W_{t-1}^j $ using expression (\ref{eq:FD-density}) for $\bar{M}^{\rightarrow}_t(\cdot|\cdot)$\;
        Normalise weights: $\hat{W}_{t-1}^j = \frac{\hat{w}_{t-1}^j}{\sum_{k=1}^N \hat{w}_{t-1}^k}$\;
        $B_{t-1} \sim \mathcal{M}(\hat{W}_{t-1}^{1:N})$;
        }
     Return $x_1^{B_1} = \big(\mathfrak{F}_1(u^{B_1}_{1};e_{0},e_1^{B_1}),e_1^{B_1}\big)$, $x_2^{B_2} = \big( \mathfrak{F}_2(u^{B_2}_{2};e_{1}^{B_1},e_2^{B_2}), e_2^{B_2}\big)$, \ldots, $x_T^{B_T} = \big(\mathfrak{F}_T(u^{B_T}_{T};e_{T-1}^{B_{T-1}},e_T^{B_T}), e_T^{B_T}\big)$.   
\end{algorithm}
%This update is not possible without the assumption of \eqref{eq:smoothing_density}.
%
\begin{algorithm}[!h] 
    \caption{Particle Filter for CD-SSM in (\ref{eq:CD-X})-(\ref{eq:CD-Y}) -- under backward methodology} %\citep{chopin2020introduction}}
    \label{alg:BD-PF}
    \SetKwInOut{Input}{Input}
    \SetKwInOut{Output}{Output}
    \KwIn{$\bar{M}_1^{\leftarrow}[dz_1]$, $\bar{M}_t^{\leftarrow}[z_{t-1},dz_{t}]$ and  $\bar{G}_1^{\leftarrow}(z_1)$, $\bar{G}_t^{\leftarrow}(z_{t-1},z_t)$, $t=2, \ldots, T$ (use (\ref{eq:BD-Mbar}), (\ref{eq:BD-Gbar})).}
    \KwOut{Particles $z^{1:N}_{1:T}=(u_{t}^{1:N},
    e_t^{1:N})_{t\in[1:T]}$, $x^{1:N}_{1:T}=(v_{t}^{1:N},
    e_t^{1:N})_{t\in[1:T]}$, ancestors $A^{1:N}_{2:T}$, weights $W_{1:T}^{1:N}$.}
%    Sample $v^j_{1}$ from proposal SDE (\ref{eq:FD-proposal}); set $e_0^{j}=\texttt{start}(v_1^{j})$ and $e_1^{j}=\texttt{end}(v_1^{j})$\;
    %Apply backward reparameterisation: $u^{j}_{1}=\mathfrak{F}^{-1}_{1}\big(v_1^j;e_0^{j},e_1^{j}\big)$\;
    %Collect particles: $z^{j}_1 = (e_1^{j}, u^{j}_{1})$\;
    %Assign weights: $w^j_1 \gets \bar{G}_1^{\rightarrow}(z^{j}_1)$\;
    %Normalise weights: $W^j_1 \gets \frac{w^j_1}{\sum_{k=1}^{N}w^k_1}$\;
    \For{$t=1,...,T$}{
        If $t=1$ set $A_t^{j}\gets j$; else  $A_t^{j}\sim \texttt{Resample}(W^{1:N}_{t-1})$\; %User defined Resampling Scheme 
        %$\mathsf{X}_t^j \sim Q(dx_t|\mathsf{X}_{t-1}^{A_t^j})$\;
    Sample $e^j_{t}\sim m^{\leftarrow}_t(e_{t}|e_{t-1}^{A_t^j})$ and $u_t^{j}\sim \bb{W}_t[du_t]$
      \; 
    Obtain the path: $v_t^{j}\gets \mathfrak{h}_t(u_t^{j};e_{t-1}^{A_t^j},e_t^{j})$ \;
    Collect particles: $z^{j}_t = (u^{j}_{t},e_t^{j})$, $x^{j}_t = ( v^{j}_{t},e_t^{j})$\;
        Assign weights: $w_t^j = \bar{G}_t^{\leftarrow}(z_{t-1}^{A_t^j}, z_t^{j})$ using the expression in (\ref{eq:BD-Gbar}) for $\bar{G}^{\leftarrow}_t(\cdot,\cdot)$\;
        Normalise weights: $W_t^j = \frac{w_t^j}{\sum_{k=1}^N w_t^k}$;
    }
\end{algorithm}
\begin{algorithm}[!h] 
    \caption{FFBS for CD-SSM in (\ref{eq:CD-X})-(\ref{eq:CD-Y}) -- under backward methodology} %\cite{godsill2004monte}}
    \label{alg:BD-FFBS}
    \SetKwInOut{Input}{Input}
    \SetKwInOut{Output}{Output}
    \KwIn{Output, $z^{1:N}_{1:T}=(u_{t}^{1:N},e_t^{1:N})_{t\in[1:T]}$, $A_{2:T}^{1:N}$, $W_{1:T}^{1:N}$, from Algorithm \ref{alg:BD-PF}.}
    \KwOut{Approximate sample $\big(x_1^{B_1}=(v_1,e_1)^{B_1}, \ldots, x_T^{B_T}=(v_T, e_T)^{B_T}\big)$ from the smoothing distribution $P\,[\,X_{1:T} \in dx_{1:T}\,|\, Y_{1:T} = y_{1:T}\,]$.}
    Sample $B_T \sim \mathcal{M}(W_T^{1:N})$\;
    \For{$t=T,...,2$}{
        Assign weights: $\hat{w}_{t-1}^j \gets 
        \bar{M}_t^{\leftarrow}(z_{t}^{B_{t}}|z_{t-1}^j)\, \bar{G}^{\leftarrow}_t(z_{t-1}^{j},z_t^{B_t}) \cdot W_{t-1}^j $ using the expression in (\ref{eq:BD-density}) for $\bar{M}^{\leftarrow}_t(\cdot|\cdot)$\;
        Normalise weights: $\hat{W}_{t-1}^j = \frac{\hat{w}_{t-1}^j}{\sum_{k=1}^N \hat{w}_{t-1}^k}$\;
        $B_{t-1} \sim \mathcal{M}(\hat{W}_t^{1:N})$;
        }
        Return $x_1^{B_1} = \big( \mathfrak{H}_1(u^{B_1}_{1};e_{0},e_1^{B_1}),e_1^{B_1}\big)$, $x_2^{B_2} = \big( \mathfrak{H}_2(u^{B_2}_{2};e_{1}^{B_1},e_2^{B_2}),e_2^{B_2}\big)$, \ldots, $x_T^{B_T} = \big( \mathfrak{H}_T(u^{B_T}_{T};e_{T-1}^{B_{T-1}},e_T^{B_T}), e_T^{B_T}\big)$.
\end{algorithm}
%This update is not possible without the assumption of \eqref{eq:smoothing_density}.

\begin{remark}
A key point in the implementation of Algorithm \ref{alg:BD-FFBS} is the calculation of the backward weights $\hat{w}_{t-1}^{j}$ for the current particle $z_t^{B_t} = (u_{t}^{B_t}, e_t^{B_t})$. Implementation of the transformed  formulae has allowed for all particles 
$z_{t-1}^j =(u_{t-1}^{j}, e_{t-1}^{j})$ to now be given positive backward weights. 
\end{remark}

\section{Numerical Experiments}
\label{sec:numerics}
We present a collection of experiments that demonstrate the utility of the new framework for particle-based methods for CD-SSMs outlined in this work. We present one experiment that demonstrates the filtering performance of the forward and backward proposal constructions outlined in Section \ref{sec:guided}, with such constructs addressing challenge (C.i) in the Introduction for the online filtering problem. Another two experiments demonstrate the performance of different smoothing algorithms that reselect ancestors to overcome path degeneracy and which overcome challenges (C.i)-(C.iii) due to use of the transformed Feynman-Kac formulae developed in Section \ref{sec:transform}. We implement our experiments on two choices of model. For the first two experiments we use a linear SDE for which the ground truth is known. For the last experiment we use a non-linear SDE with inter-observation length $\Delta_t$ such that the observation regime classifies as a low-frequency one (i.e.~avoiding data augmentation would lead to high biases). 
%is sufficiently large so that approximating the model with a one-step transition density would introduce bias, which is the intended setting for application of our methods. 
In all experiments, when representation of a continuous-time path is required, a time-discretisation with  $M=50$ imputed points in-between observations is used via means of  an Euler-Maruyama scheme. For resampling operations in particle filters, we use systematic resampling and we resample adaptively when $\textrm{ESS} \leq 0.5\cdot N$ (`ESS' stands for Effective Sample Size).

\subsection{Ornstein-Uhlenbeck CD-SSM: Elliptic and Hypo-elliptic Cases}
\label{subsec:ou}
We consider a CD-SSM with $d=2$ driven by the latent Ornstein-Uhlenbeck (OU) signal:
\begin{equation*}
%\label{eq:numerics_ou}
    dX(s) = AX(s)ds + \phi dB(s),  
\end{equation*}
with both components observed via $Y_t| X(s_t) = x \sim \mathcal{N}_2(x, \sigma_y^2 I_2)$ at equidistant times \mbox{($\Delta_t=1$)}. We benchmark our methods against the truth, since the filtering law $X(s_t)|Y_{1:t}$, the likelihood $p(y_{1:t})$ and the smoothing law $X(s_1),\dots, X(s_T)|Y_{1:T}$ are tractable. We consider both an elliptic and a hypo-elliptic  latent signal with corresponding specifications:
\begin{equation*}
(d_w, A, \phi) = (2, -I_2, I_2), \qquad (d_w, A, \phi) = (1, (\begin{smallmatrix} 0 & 1 \\ 0 & -1 \end{smallmatrix}), (\begin{smallmatrix} 0 \\ 1 \end{smallmatrix})),
\end{equation*}
under different noise regimes $\sigma_y \in \{0.05, 0.1, 0.2, 0.5, 1.0\}$, thus parameterising $10$ distinct CD-SSMs. We evaluate both filtering and smoothing performance. We simulate data from each of the $10$ models, of length $T=100$, and use the same data in both filtering and smoothing experiments for the same CD-SSM.

\subsubsection{Experiment 1: Filtering}
\label{subsubsec:ou_filtering_exp}
\begin{figure}[h]
  \centering
  \includegraphics[width=\linewidth]{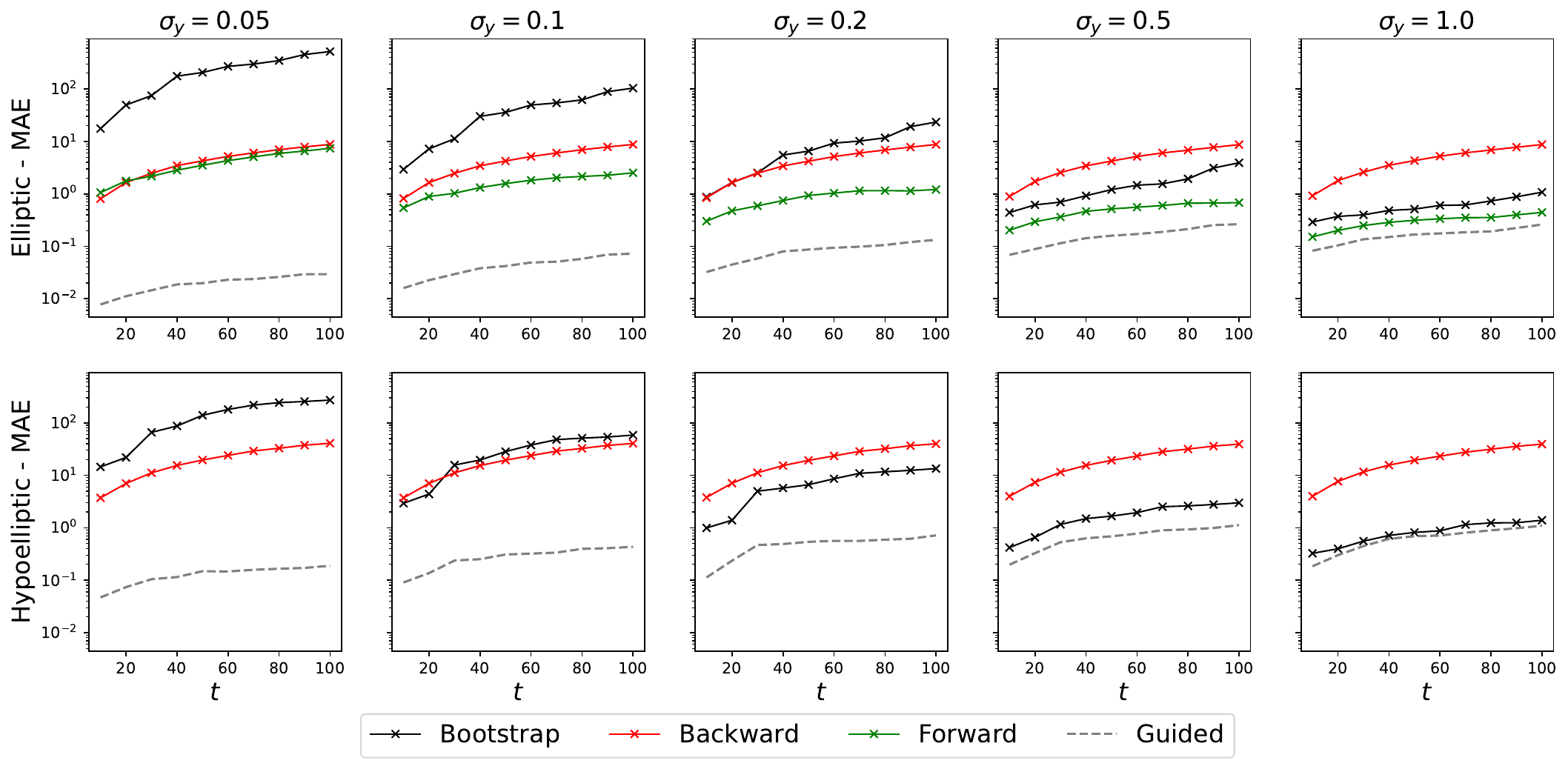}
  \caption{Results for Experiment 1. Top Row: Elliptic OU. Bottom Row: Hypo-elliptic OU. MAE for estimators of $\log p(y_{1:t})$ under different choices of proposal. (`Guided' here refers to the ideal PF in discrete-time that uses the optimal proposal.)}
   \label{fig:filtering_results}
\end{figure}

We investigate the performance of the data-informed proposals described in Section \ref{sec:guided} for filtering, using the selection methods %based on the construction of proxy linear SDEs
outlined in Subsections \ref{subsubsec:FP} and \ref{subsubsec:BP}. For the elliptic model, we consider both a forward and a backward proposal. For the hypo-elliptic case, we only look (necessarily) at a backward proposal. For each choice of proposal, we run $K=96$ replications of the particle filter (PF, Algorithm \ref{alg:FK-PF}) using the relevant guided Feynman-Kac formula. In Figure \ref{fig:filtering_results} we show the mean absolute error (MAE) $\mathrm{E}|\hat{\psi}_t-\psi_t|$ of the estimate of the log-likelihood of the data, as calculated via the $K=96$ replications, where $\psi_t=\log(p(y_{1:t}))$ and $\hat{\psi}_t$ is the standard estimator provided by the particle filter across times $t\in\mathcal{T}$ for each of the $5$ values of $\sigma_y$. We define our proposals for the experiments as follows: 

\begin{itemize}
    \item \textbf{Forward Proposal $M_t^\rightarrow[x_{t-1}, dx_t]$ (elliptic OU)}: We select a forward proposal SDE as in the approach outlined in Subsection \ref{subsubsec:FP}, with its diffusion coefficient matching that of the signal SDE. Recall that the drift of the forward proposal is obtained by replacing the term $\rho_{\Delta_t, y_t}(s, v)$ that appears in expression for the drift of the optimal proposal $b_t^{\rightarrow, \textrm{opt}}(s, v; y_t)$ with a proxy $\tilde{\rho}_{\Delta_t, y_t}(s, v)$.  This proxy is obtained from choosing a linear SDE in (\ref{eq:linear-SDE}) that resembles the true signal dynamics $P_t[e_{t-1}, dv_t]$. We select the linear SDE via the simple choices $\tilde{b}_0(s) = \tilde{b}_1(s) = 0$, $\tilde{\sigma}(s)=\sigma_t(0, e_{t-1})=I_2$. Via routine calculations, noting that $b_t(s, v)=-v$, $\Sigma_t(s, v)=I_2$ for the elliptic OU, we obtain a forward proposal $M_t^\rightarrow[x_{t-1}, dx_t]$ with drift function:
    \begin{equation*}
        b_t^{\rightarrow}(s, v; y_t) = -v + \tfrac{y_t - v}{\Delta_t - s + \sigma_y^2}. 
    \end{equation*}
    \item \textbf{Backward Proposal $M_t^\leftarrow[x_{t-1}, dx_t]$ (elliptic OU)}:  Recall that selection of the kernel $M_t^\leftarrow[x_{t-1}, dx_t]$ involves choosing a proposal for the end point $m^\leftarrow_t(e_t|e_{t-1})$ and a proposal for a bridge SDE $M_t^\leftarrow[(e_{t-1}, e_t), dv_t]$. We consider first the proposal for the ending point -- as the signal process is a linear SDE, we have access to $p_t(e_t|e_{t-1})$, which is Gaussian. The observation operator is linear in $e_t$, therefore one can find the optimal proposal $m_t^{\leftarrow, \textrm{opt}}(e_t|e_{t-1})\equiv p_t(e_t|e_{t-1},y_t)$, which we use here. For the bridge $M_t^\leftarrow[(e_{t-1}, e_t), dv_t]$, we use a guided proposal as in (\ref{eq:BD-proposal-D}). Recall that the drift of such a proposal is obtained by replacing $p_{\Delta_t, e_t}(s, v)$ that appears in the drift of the true bridge, with a proxy from a chosen linear SDE (\ref{eq:linear-SDE}). For the linear SDE, we make the simple choice $\tilde{b}_0(s) = \tilde{b}_1(s) = 0$, $\tilde{\sigma}(s)=\sigma_t(\Delta_t, e_t)=I_2$, the latter selection satisfying the required `matching condition' in the diffusion coefficient. Since $b_t(s, v)=-v$, $\Sigma_t(s, v)=I_2$, we end up with a bridge proposal $M_t^\leftarrow[(e_{t-1}, e_t), dv_t]$ with a drift function given by:
    \begin{equation*}
        b_t^G(s, v; e_t) = -v + \tfrac{e_t - v}{\Delta_t - s}.
    \end{equation*}
    \item \textbf{Backward Proposal $M_t^\leftarrow[x_{t-1}, dx_t]$ (hypo-elliptic OU):} As in the elliptic case, we have access to the Gaussian $p_t(e_t|e_{t-1})$ and the observation operator is linear in $e_t$, so we use $m_t^{\leftarrow, \textrm{opt}}(e_t|e_{t-1})$ for the ending point. To choose the bridge $M_t^\leftarrow[(e_{t-1}, e_t), dv_t]$ we use a guided proposal as considered in \cite{bier:20}, where again one replaces $p_{\Delta_t, e_t}(s, v)$ in the drift of the true bridge with a proxy $\tilde{p}_{\Delta_t, e_t}(s, v)$ that comes from a linear SDE. Further `matching conditions' are required, now on both the drift $\tilde{b}_0(s) + \tilde{b}_1(s)v$ and the diffusion coefficient $\sigma(s)$ of the linear SDE $\{\tilde{V}_t(s)\}_{s \in [0, \Delta_t]}$. Considering Remark \ref{rem:integrated} we make the choice $\tilde{b}_0(s)= 0$, $\tilde{b}_1(s)= (\begin{smallmatrix} 0 & 1 \\ 0 & 0 \end{smallmatrix})$, $\tilde{\sigma}(s) = (\begin{smallmatrix} 0 \\ 1 \end{smallmatrix})$. Noting that $b_t(s, v) = (\begin{smallmatrix} 0 & 1 \\ 0 & -1 \end{smallmatrix})v$, $\Sigma_t(s, v) = (\begin{smallmatrix} 0 & 0 \\ 0 & 1 \end{smallmatrix})$ we end up with a bridge proposal $M_t^\leftarrow[(e_{t-1}, e_t), dv_t]$  with drift function:
    \begin{equation*}
        b_t^G(s, v; e_t) = \begin{pmatrix} 0 & 1 \\ 0 & -1 \end{pmatrix} v + \begin{pmatrix} 0 & 0 \\ -\frac{6}{(\Delta_t - s)^2} & \frac{4}{(\Delta_t - s)} \end{pmatrix}(e_t - v).
    \end{equation*}
\end{itemize}
We also run a bootstrap PF 
%(algorithm \ref{alg:FK-PF} using the Feynman-Kac model of section \ref{subsec:boot} proposing particles from the signal dynamics that are blind to the observations, 
and an `ideal' guided PF. The ideal algorithm proposes from the law  
$X(s_t)\sim X(s_t)|X(s_{t-1}),y_t$ which is tractable in the linear setting,  without using  continuous-time paths. 
%where the proposal is obtained by conditioning the true signal dynamics $p(x(s_t)|x(s_{t-1}))$ on observation $y_t$ \citep{douc:00, chop:20}. 
%For the general, non-linear SDE, both the signal dynamics and the signal conditioned on the observation $y_t$ are intractable (recall (C.i) from the Introduction). 
We observe from Figure \ref{fig:filtering_results} that:
\begin{itemize}
    \item When focusing on the ideal guided PF and the bootstrap PF, the former is more effective for decreasing $\sigma_y$ (as expected). 
    %This numerically evidences that guiding most materially improves performance when observations are informative.   
    \item In the low noise regimes $\sigma_y \in \{0.05, 0.1\}$ where guiding materially impacts algorithmic performance, both forward and backward methods outperform the bootstrap PF. In the lowest noise regime ($\sigma_y = 0.05$), the difference in performance  is of at least an order of magnitude.
    % \item Under higher noise regimes ($\sigma_y \in \{0.5, 1.0\}$), despite a reasonable choice of backward proposal being used, the performance for estimating the log-likelihood is worse than the bootstrap, due to the mismatch between the proposed bridge and the true bridge. 
\end{itemize}

\begin{remark}
More complex choices are possible in both the forward and backward cases. Re-running the above experiments for more sophisticated proposals produces the expected outcome of improved performance of one forward/backward proposal against another but does not significantly improve performance relative to the bootstrap and the ideal guided PFs. Note that in the case of backward proposals, using the 1st order expansion to define $\tilde{b}_0$ and $\tilde{b}_1$ would correspond to using the true bridge $P_t[e_{t-1:t}, dv_t]$ as proposal, and from reviewing the backward weight expression (\ref{eq:BD-G}), one can see that the resulting algorithm then reduces to the ideal guided PF. 
\end{remark}

\subsubsection{Experiment 2: Smoothing via FFBS}
\label{subsubsec:ou_smoothing_exp}
\begin{figure}[h]
  \centering
  \includegraphics[width=\linewidth]{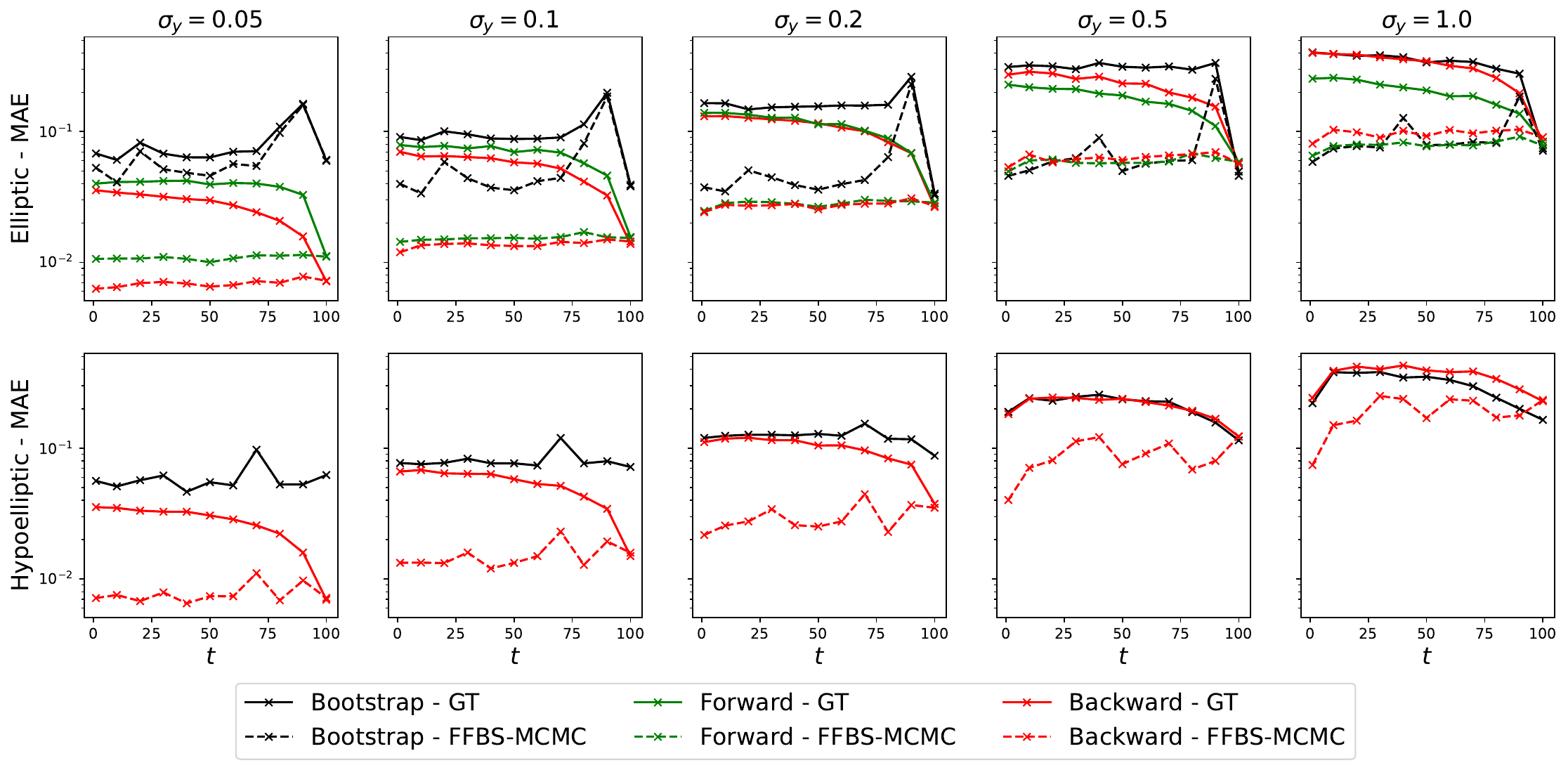}
  \caption{Results for Experiment 2. Top Row: Elliptic OU. Bottom Row: Hypo-elliptic OU. MAE of the estimator of the first component of the marginal first moments of the smoothing distribution across time.}
  \label{fig:smoothing_results}
\end{figure}

We apply smoothing algorithms that reselect ancestors for the CD-SSMs at hand, with such an approach made possible via the contributions in this work. We use the transformed Feynman-Kac formulae of Section \ref{sec:transform} and the deduced FFBS algorithm against a benchmark SMC algorithm that cannot alter ancestors. We compare the performance of the genealogy tracking (GT) method (Algorithm \ref{alg:FK-GT}) versus the MCMC variant of the FFBS algorithm \citep{bunc:13} (FFBS-MCMC). 
FFBS-MCMC is described in Algorithm \ref{alg:FK-FFBS-MCMC} in Appendix \ref{sec:further_methods}. In contrast to FFBS that involves $\mathcal{O}(N^2)$ computations, the cost of FFBS-MCMC is $\mathcal{O}(N)$.
%We compare with this algorithm instead of the standard FFBS algorithm (as presented in algorithm \ref{alg:FK-FFBS} of Appendix \ref{sec:further_methods}) as the cost of FFBS-MCMC is linear as opposed to quadratic in the number of particles (assuming that we use the same number of backward samples as number of particles), 
%and has been shown to have consistency and stability properties \citep{dau:22} thus ensuring direct comparability with GT in terms of compute cost. 
Further details about FFBS-MCMC are given in Appendix \ref{sec:further_methods}. The reduction in cost from quadratic to linear is achieved by replacing the normalisation of the weights with an Independent Metropolis proposal. The overall performance of a particle-based smoothing method will depend on both the choice of smoothing algorithm and the performance of the proposals used in the filtering steps. To ensure clarity in demonstrating that improved performance comes from using a better algorithm as opposed to the use of better proposals, where possible we run both algorithms for each of the choices of path proposal used in the previous experiment. It is only not possible to run FFBS-MCMC when we use a bootstrap (path) proposal for the hypo-elliptic OU, as there is no forward construct in this case. Note that a bootstrap proposal is a special case of a forward proposal, so to use a smoothing algorithm that alters ancestors for a hypo-elliptic model, we must introduce a proposal in the filtering steps. We run each algorithm using $N=100$ particles for $K=960$ replicates and record the MAE of the estimator of mean of the first coordinate of $X(s_t)$ (which is the smooth component in the case of the hypo-elliptic model) given all data, i.e.~we track $\mathrm{E}|\hat{\psi}_t - \psi_t|$ with known $\psi_t = \mathrm{E}[X_1(s_t)|Y_{1:T}=y_{1:T}]$ and $\hat{\psi}_t$ the estimator of $\psi_t$ from the smoothing algorithm, across $t \in \mathcal{T}$. The results are presented in Figure \ref{fig:smoothing_results}. For each proposal, solid lines indicate GT (Geneology Tracking), dashed lines indicate FFBS-MCMC. We observe from the results that:
\begin{itemize}
    \item For any given choice of model and proposal, the FFBS-MCMC algorithm made possible from our reparameterised constructions always outperforms GT for $t<T$. For $t<<T$, the outperformance is more pronounced.
    \item One can observe that smoothing performance depends on both algorithmic efficiency and choice of proposal. For example, for the elliptic OU with $
    \sigma_y = 0.05$, the bootstrap proposal with FFBS-MCMC performs worse than the forward proposal with GT and the backward proposal with GT.
    \item For $t=T=100$ we can observe results for estimation of the filtering mean: the performance difference between the backward and bootstrap proposals for estimation of the log-likelihood in higher noise regimes ($\sigma_y \in \{0.5, 1.0\}$) appears to impact less materially impact the performance in estimation of the filtering mean.
    \item For the hypo-elliptic OU with high noise ($\sigma_y \in \{0.2, 0.5, 1.0\}$), we note that in order to use improved particle-based methods that alter ancestors (resulting in improved performance in the smoothing steps), that can be seen empirically to result in improved performance, we need to introduce a proposal into the filtering steps that worsens performance of estimation of the log-likelihood of the data, relative to the bootstrap.
\end{itemize}

\subsection{Fitzhugh-Nagumo CD-SSM}
\label{subsec:fhn}
We consider a highly non-linear CD-SSM where the signal is  determined via the hypo-elliptic Fitzhugh-Nagumo (FHN) model (here, $d=2$, $d_w=1$):  
%for the joint evolution of the action potential and conductance dynamics of a neuron, as considered in \cite{ditl:19, sams25}:
%
\begin{align*}
dX_1(s) &= \tfrac{1}{\epsilon} (X_1(s) - X_1(s)^3 - X_2(s)) ds;  \\
dX_2(s) &=(\gamma X_1(s) - X_2(s) + \beta) ds + \sigma dB(s). 
\end{align*}
%
%The FHN model describes the  joint evolution of the action potential and conductance dynamics of a neuron, see e.g.~\citep{ditl:19, sams25}.
The smooth component $X_1(s)$ models the evolution of the membrane potential of a neuron  and $X_2(s)$ describes the neuron's conductance dynamics. This model has been studied in numerous works, see e.g.~\cite{ditl:19, sams:25}.
We assume that only the potential $X_1(s)$ is observed (as is typical in practice), at equidistant times $\Delta_t=0.05$, with additive measurement error $Y_t|X(s_t)=x \sim \mathcal{N}_1(x, \sigma_y^2)$, where the noise parameter is fixed to $\sigma_y = 0.01$. We choose the parameters in the latent signal to be in accordance with the ones used in \cite{sams:25}, i.e.~$(\epsilon, \gamma, \beta, \sigma) = (0.1, 1.5, 0.8,0.3)$. 
%
%Instead of inferring the membrane potential and the conductance dynamics $X(s) = (X_1(s), X_2(s))$, 
%
We work with the 1--1 transform of the original process defined as $\bar{X}(s) = (X_1(s), \bar{X}_2(s))$ that contains $X_1(s)$ and its derivative $\bar{X}_2(s)$, that is  $\bar{X}_2(s) := \tfrac{1}{\epsilon}(X_1(s) - X_1(s)^3 - X_2(s))$. 
We thus move into the setting of integrated hypo-elliptic SDEs where absolute continuity conditions between target and proposal bridges are easier to verify, see Remark \ref{rem:integrated}.
A routine application of It\^o's lemma gives the integrated bivariate SDE:
\begin{align}
dX_1(s) &= \bar{X}_2(s) ds; \nonumber  \\
d\bar{X}_2(s) &=\tfrac{1}{\epsilon}\big(1-\epsilon - 3X_1(s)^2\big)\bar{X}_2(s)\,ds  \nonumber \\ &\qquad \qquad + \tfrac{1}{\epsilon}\big((1-\gamma)X_1(s) - X_1(s)^3  - (s+\beta)\big)ds + \tfrac{\sigma}{\epsilon}\,dB(s).
\label{eq:t-FHN}
\end{align}
The initial value of the SDE is fixed at $(X_1(0), \bar{X}_2(0))=(0, 0)$.
Figure \ref{fig:fhn_simulation} (left panel) shows the simulated observations of the FHN CD-SSM we will consider in the numerics, with $T=100$ steps, corresponding to a time period $s \in [0, 5]$.

\subsubsection{Bias Evaluation for FHN Model with $\Delta_t=0.05$}
As stated in the Introduction, the methodology of this work is intended for the low-frequency observation setting where 
%(assuming that we only take interest in inferring the latent signal at the observation times (i.e in the case of the offline setup $|Y_{1:T}=y_{1:T}$), 
$\Delta_t$ is sufficiently large that if one were to replace 
the intractable transition density of the FHN model (\ref{eq:t-FHN}) over a period of length $\Delta_t$ with a proxy obtained via a single $\Delta_t$-step of a numerical scheme, then a large amount of bias would be introduced in the model. 
%$p_t(e_t|e_{t-1})$ in the prior for $e_{1:T}$ with a single $\Delta_t$-step of an approximation $\tilde{p}_t(e_t|e_{t-1})$ 
%then this would introduce significant bias into the model, thus altering the target (marginal) posterior distribution $E_{1:T}|Y_{1:T}=y_{1:T}$. 
%This happens when density $\tilde{p}_t(\cdot|e_{t-1})$ is far away from the true 
%density $p_t(\cdot|e_{t-1})$ that it intends to approximate for some values of $t \in \mathcal{T}$.
We show here that this is the case for the FHN model under consideration with $\Delta_t=0.05$. We refer by $p_t(e_t|e_{t-1})$ 
and $\tilde{p}_t(e_t|e_{t-1})$ to the true and (an) approximate  transition transition density, respectively, of the FHN model. 
Typically, $\tilde{p}_t(e_t|e_{t-1})$ may be chosen as the (Gaussian) transition density of a linear SDE as previously defined in Section \ref{subsubsec:BP},
or via a numerical scheme. E.g., the Euler-Maruyama scheme is the standard choice for an elliptic model, but it does not admit a density for hypo-elliptic SDEs.
Alternative schemes have been developed for the hypo-elliptic setting -- see e.g.~\cite{poke:09, ditl:19}. For our purposes we take $\tilde{p}_t(\cdot|e_{t-1})$ to be the recently developed and well-studied `locally Gaussian' scheme \citep{glot:21} to obtain the proxy $\tilde{p}_t(e_t|e_{t-1})$ of the FHN transition density. An explicit expressions for this scheme in the case of the FHN model is given in Appendix \ref{sec:loc_gauss}.

\begin{figure}[h]
      \centering
  \includegraphics[width=\linewidth]{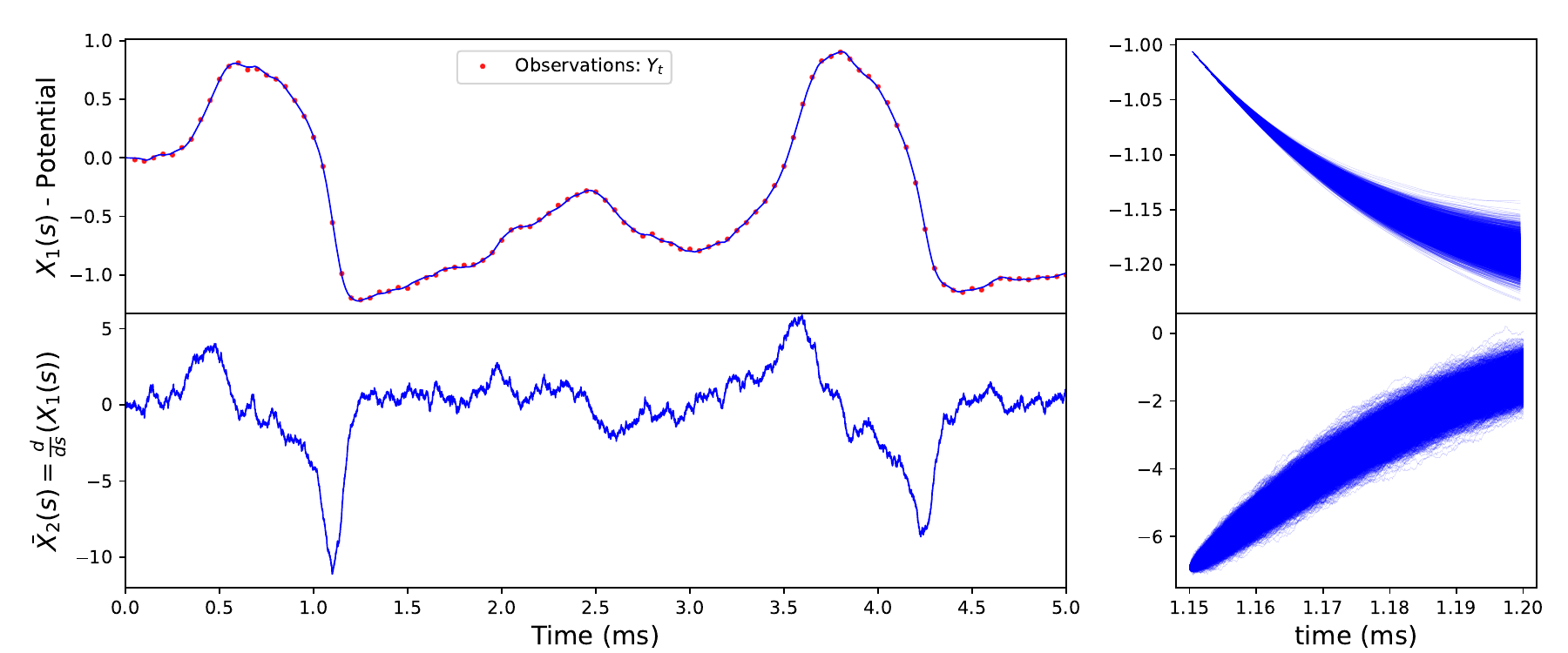}
  \caption{Left: Simulation of the FHN CD-SSM used in the numerical experiment. Right: Simulation of paths 
  $X_1([s_{t-1}, s_t])$,  $\bar{X}_2([s_{t-1}, s_t])$, for $s_{t-1}=1.15$, $x(s_{t-1}) = (-1.01, -6.92)^{\top}$ via the locally Gaussian scheme with $1,000$ imputation steps along $[s_{t-1},s_t]$.}
  \label{fig:fhn_simulation}
\end{figure}

 Approximation of $p_t(e_t|e_{t-1})$ via the locally Gaussian scheme (along with other possible alternatives) relies on a linear approximation of the signal dynamics over the inter-observation time interval $\Delta_t$. 
Thus, $p_t(\cdot|e_{t-1})$ will be badly approximated by $\tilde{p}(\cdot | e_{t-1})$ in cases when the signal exhibits strong non-linear behaviour in the interval $s \in [s_{t-1}, s_t]$.
Considering the simulation of the signal process given in Figure \ref{fig:fhn_simulation} (left panel), there are several values of $t \in \mathcal{T}$ when the signal exhibits non-linear behaviour within the interval $[s_{t-1}, s_t]$. 
Indicatively, we consider the step $t=24$, whence $s_{t-1}=s_{23}=1.15$, $s_t=s_{24}=1.20$ and the initial signal value is $e_{t-1}=x(s_{t-1})=(-1.01, -6.92)^\top$. 
In Figure~\ref{fig:fhn_simulation} (right panel), we show $5,000$ simulations from the signal process starting at $s_{23}=1.15$ with $1,000$ imputation steps along $[s_{t-1},s_t]$ using the locally Gaussian scheme, from which one can clearly observe non-linear signal dynamics. 
Let $\tilde{q}_\Delta(x'|x)$ be the transition density of the locally Gaussian scheme for a given time-step $\Delta>0$, thus $\tilde{p}_t(e_t|e_{t-1}) = \tilde{q}_{\Delta_t}(e_t|e_{t-1})$. Let $\tilde{p}_t^{(k)}$ be the proxy of $p_t$ when iterating the locally Gaussian scheme along $k\ge 1$ imputation steps, i.e:
\begin{equation*}
    \tilde{p}_{t}^{(k)}(e_t|e_{t-1}) = \int_{\mathbb{R}^{d\times (k-1)}} \Bigg[\prod_{i=1}^k q_{\frac{\Delta_t}{k}}(x_i|x_{i-1}) \Bigg]dx_{1:{k-1}},
\end{equation*}
with $x_0\equiv e_{t-1}$, $x_{k}\equiv e_t$. 
%We drop the superscript when $k=1$. 
For $k>1$ we can simulate  from $\tilde{p}^{(k)}(\cdot|e_{t-1})$ and produce a (Gaussian) kernel density estimator  (KDE) for its values (we use 5,000 such samples in the results below), whereas the above integral is intractable. We obtain the `true' $p_t(\cdot|e_{t-1})$ by using the $5,000$ simulated points in Figure~\ref{fig:fhn_simulation} (right) at time $s_{24}=1.2$ and utilising a KDE. Figure~\ref{fig:fhn_bias_results} presents contour plots of $p_t(\cdot | e_{t-1})$ and for $p_t^{(k)}(\cdot|e_{t-1})$, $k \in \{1, 2, 5, 10\}$  for $t=24$. We also report in Figure~\ref{fig:fhn_bias_results} the (tractable, when utilising the KDEs) $\mathcal{L}_2$-distance between each of the 4 approximations and the true $p_t(\cdot|e_{t-1})$. It can be seen from the results that:
\begin{figure}[h]
      \centering
  \includegraphics[width=\linewidth]{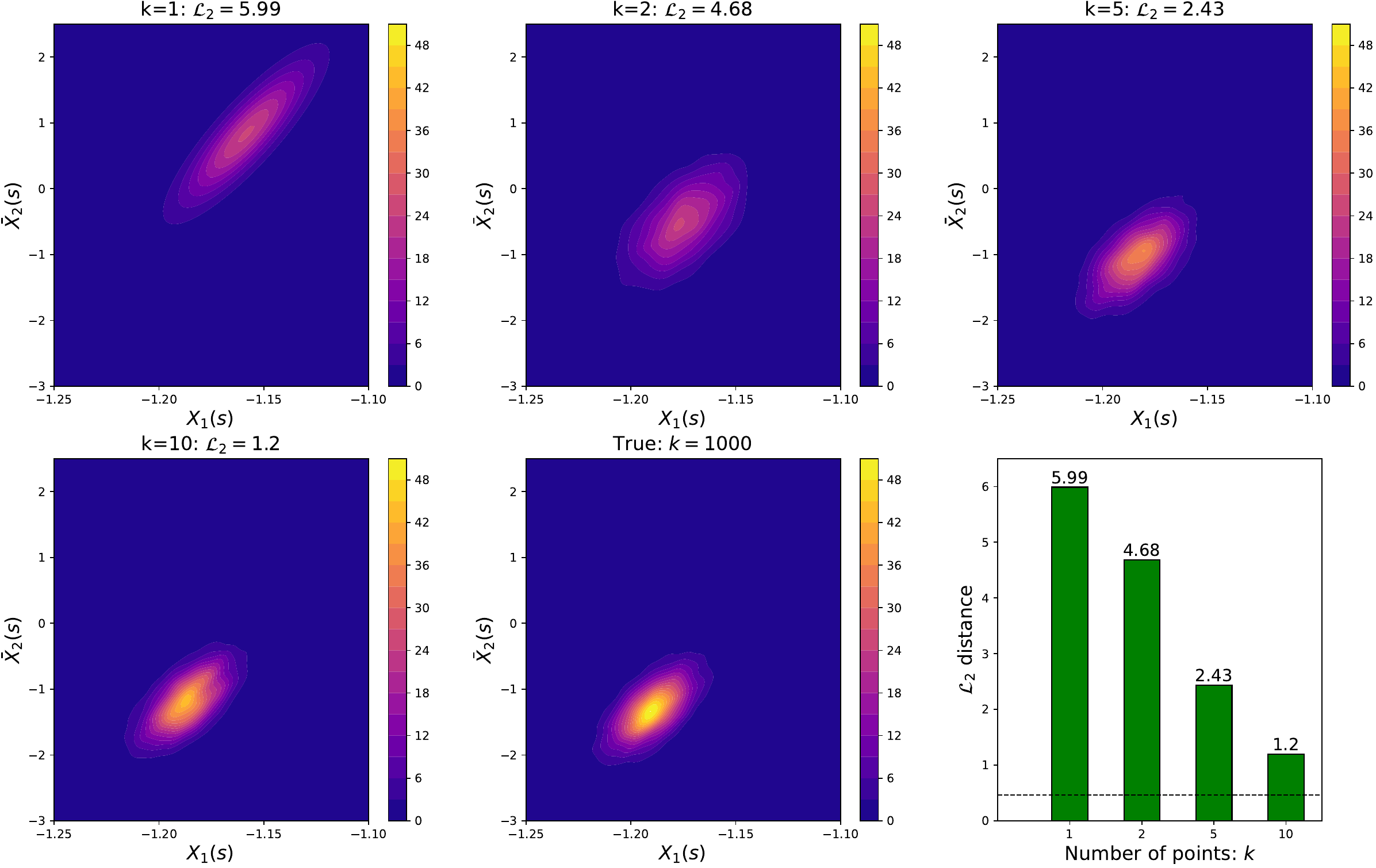}
  \caption{Contour plots of the transition density approximations $\tilde{p}_t^{(k)}(\cdot|e_{t-1})$, $k \in \{1, 2, 5, 10\}$ and of 
  the `true' density
  $p_t(\cdot|e_{t-1})\equiv \tilde{p}^{(1000)}(\cdot|e_{t-1})$, with $e_{t-1}=(-1.01, -6.92)^\top$. When $k>1$, the density is obtained via KDE with $5,000$ samples. Bottom right: Barplot of $\mathcal{L}_2$-distances between the true transition density and the approximations for increasing $k$. The dashed line shows the $95$-percentile of simulated $\mathcal{L}_2$-distances for $k=1,000$ between the baseline sample used in the bottom-centre plot and a number of independent samples of the same size ($5,000$).}
  \label{fig:fhn_bias_results}
\end{figure}
\begin{itemize}
    \item Comparing the contour plot of $\tilde{p}_t^{(1)}(\cdot|e_{t-1})$ (top, left) with $p_t(\cdot|e_{t-1})$ (bottom, centre), it is clear that a single $\Delta_t$-step of the locally Gaussian scheme delivers a very crude approximation of the true model transitions over $\Delta_t$. The approximation is particularly poor in the second component, $\bar{X}_2(s)$, which involves the highly non-linear drift. Thus,  data augmentation is needed to avoid large biases when carrying out statistical inference.
    \item As expected, for increasing $k$, $\tilde{p}_t^{(k)}(\cdot|e_{t-1})$ gets closer to $p_t(\cdot|e_{t-1})$. This can be observed both visually and quantitatively via the the reduction of the $\mathcal{L}_2$-distance.
    \item Contrasting the shading of contours at the plots for $\tilde{p}_t^{(k)}$,  $k=10$, (bottom, left) vs the truth $p_t(\cdot|e_{t-1})$ (bottom, centre) we notice that even when $k=10$, one still does not accurately approximate $p_t(\cdot|e_{t-1})$. Thus, taking a reasonably large number of imputed points between observations times (e.g.~we use $M=50$ in our experiments in the sequel) is necessary to avoid introducing high biases into the model.
\end{itemize}

\subsubsection{Experiment 3: Smoothing via Particle MCMC}
\label{section:experiment3}
We evaluate the performance of the iterated conditional SMC (iCSMC) algorithm \citep{andr:10}. This particle-based MCMC algorithm builds upon the particle filter, so that for iCSMC the particles are auxiliary variables within an MCMC update. 
It is well-understood that inclusion of a `backward step' (iCSMC-BS), originally proposed in \cite{whit:10}, that reselects particle ancestors improves algorithmic performance both empirically \citep{lind:13} and theoretically \citep{chop:15, karj:23}. 
The reparameterised Feynman-Kac formulae of Section \ref{sec:transform} enables one to use iCSMC-BS on the class of CD-SSMs.
For completeness, we provide pseudocode for iCSMC  in Algorithm \ref{alg:FK-iCSMC} of Appendix~\ref{sec:further_methods}.
Execution of the backward step requires one to generate a single trajectory via Algorithm \ref{alg:FK-FFBS}, which will lead to degenerate algorithms under the Feynman-Kac formulations of Section \ref{sec:CD-SSM} and \ref{sec:guided} -- an issue resolved under the reparameterised Feynman-Kac formulae of Section \ref{sec:transform}.
To implement iCSMC-BS in the case when the signal is a hypo-elliptic SDE, one must introduce a backward proposal (cf.~Section \ref{subsec:BP}) 
which involves a user-specified density for the end-point $m_t^{\leftarrow}(e_t|e_{t-1})$ and a bridge proposal $M_t[(e_{t-1}, e_t), dv_t]$ for the signal process. 

%the filtering steps - it is for this reason that we work with the reparameterisation of the FHN model $\bar{X}(s) = (X_1(s), \bar{X}_2(s))$ instead of the standard one $X(s) = (X_1(s), X_2(s))$: 
%this reparameterisation ensures that the resulting diffusion is within the class of integrated diffusions \cite{bier:20} for which a continuous-time likelihood exists and can be evaluated pointwise. 

\begin{figure}[h]
  \centering
  \includegraphics[width=\linewidth]{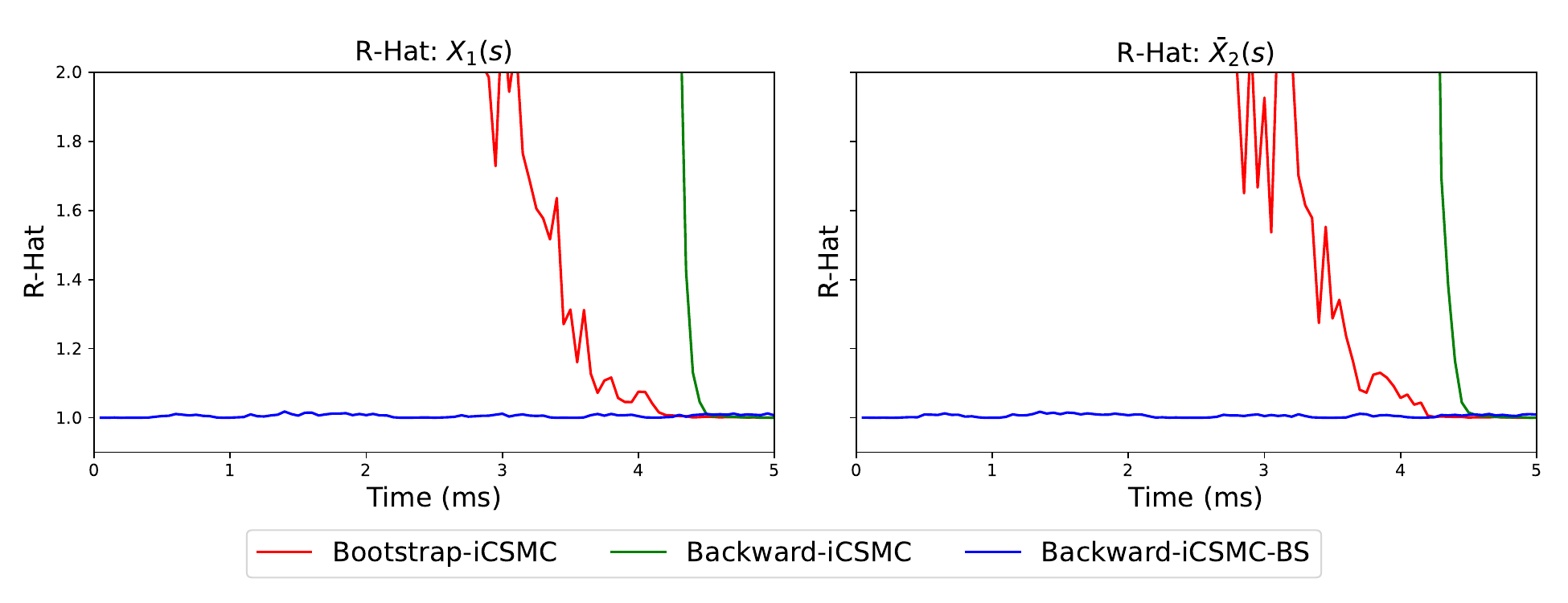}
  \caption{Results for Experiment 3: $R$-hat diagnostic for each of the FHN components across observations.}
  \label{fig:fhn_r_hat_results}
\end{figure}

We choose a guided proposal, $M_t^{\leftarrow, G}[(e_{t-1}, e_t), dv_t]$ for the bridge, %In our hypo-elliptic setting this requires one to choose an underlying linear SDE that satisfies matching conditions on both the drift, $\tilde{b}_0(s) + \tilde{b}_1(s)v$, and the diffusion coefficient, $\tilde{\sigma}(s)$.
%To this cause, via Remark \ref{rem:integrated}, 
%by considering the 
by selecting  
the following coefficients for the corresponding proxy underlying linear SDE (see Remark \ref{rem:integrated}): 
\begin{align}
\tilde{b}_0(s) = \begin{pmatrix}
  0 \\
  0 
\end{pmatrix}, \qquad \tilde{b}_1(s) =
\begin{pmatrix}
  0 & 1 \\
  0 & 0
\end{pmatrix},
\qquad  \tilde{\sigma}(s) = \begin{pmatrix}
  0 \\
   \sigma
\end{pmatrix}.
\label{eq:llinear}
\end{align}
%
%\alex{Under this choice (\ref{eq:llinear}), the matching conditions are satisfied.}
%Recall that these matching conditions imply the well-posedness of a Radon-Nikodym derivative between the bridge proposal $M_t[(e_{t-1}, e_t),dv_t]$ and the true bridge $P_t[(e_{t-1}, e_t),dv_t]$. 
We also utilise the above linear SDE in the specification of the 
end-point proposal $m_t^{\leftarrow}(e_t|e_{t-1})$. The transition density $\tilde{p}(e_t|e_{t-1})$ of this linear SDE over $\Delta_t$ is then conditioned upon the observation $y_t$ to provide $m_t^{\leftarrow}(e_t|e_{t-1})$. 
For this choice of proposal, we run the iCSMC algorithm both with the backward step (Backward-iCSMC--BS) and without it (Backward-iCSMC) -- we also run the iCSMC algorithm with a bootstrap proposal (Bootstrap--iCSMC), under which it is not possible to use a backward step. 
For each algorithm we run 8 MCMC chains for $L=1,000$ iterations, using $N=50$ particles within the particle filters. 
Each chain is initialised by running the corresponding standard particle filter (i.e.~without consideration of an CSMC step) and tracing a single ancestral path. 
\begin{figure}[h]
  \centering
  \includegraphics[width=\linewidth]{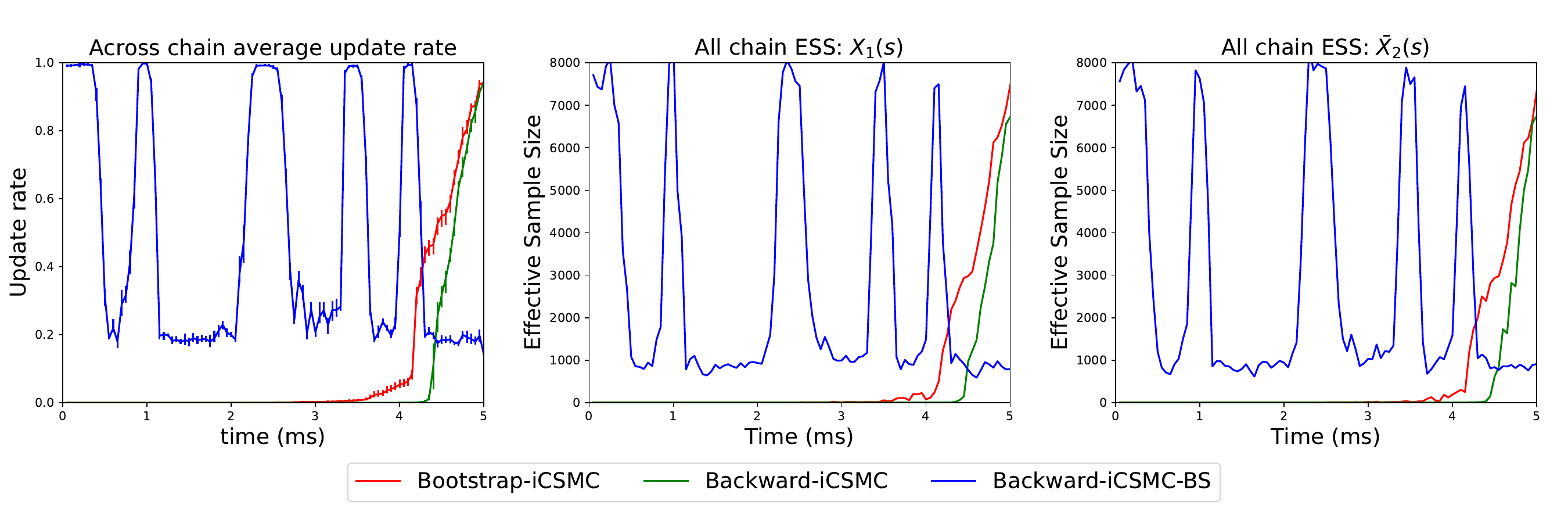}
  \caption{Results for Experiment 3: Left: Mean (across 8 chains) update rate across observation times -- error bars indicate min/max update rates from the 8 chains. Centre/Right: ESS for the first $X_1(s)$ and second $\bar{X}_2(s)$ components across all times for the 3 MCMC chains.}
  \label{fig:fhn_results}
\end{figure}
Figure~\ref{fig:fhn_results} (left) presents the average `update rate'  of the end-points $X(s_t)|Y_{1:T}=y_{1:T}$, across the 8 chains, for each $t \in \mathcal{T}$, for each of the 3 algorithms -- error bars indicate the maximum and minimum update rates across the 8 chains. 
Figure~\ref{fig:fhn_results} (centre, right) presents the ESS of the chain corresponding to the two components of the end-points $X_1(s_t),\bar{X}_2(s_t)|Y_{1:T}=y_{1:T}$ across observation times $t \in \mathcal{T}$. We observe that:
\begin{itemize}
    \item For iCSMC run without the backward step (with both the bootstrap and the informed proposals), the algorithm does not update the positions of $X_1(s_t),\bar{X}_2(s_t)|Y_{1:T}=y_{1:T}$ for (say) $t \leq 70$, corresponding to $s \in [0, 3.5]$. Upon inclusion of the backward step, update rates remain consistently above 15\% across times $t \in \mathcal{T}$. Thus, inclusion of the backward step successfully overcomes path degeneracy. %that reduces the representation of the smoothing distribution for times $t << T$ to a single particle.
    \item As expected, the deterioration in ESS in both signal components for $t<T$ in the iCSMC algorithm without the backward step matches the deterioration in the update rate.
    \item  Contrasting Figure~\ref{fig:fhn_results} with Figure~\ref{fig:fhn_simulation} (left panel), mixing of co-ordinates at particular times improves considerably under iCSMC with the backward step when the (unobserved) latent signal for the membrane potential $X_1(s)$ exhibits strong non-linear behaviour at corresponding inter-observation times $s \in [s_{t-1}, s_t]$.
\end{itemize}
We also evaluate chain convergence via the (improved) $R$-hat \citep{veht:21} diagnostic shown at Figure~\ref{fig:fhn_r_hat_results}, which indicates that the iCSMC-BS chains have converged across all time-steps $t \in \mathcal{T}$, whereas the chains do not converge for iCSMC without the backward step when most of times $t$ with $t < T$.
\section{Conclusions}
\label{sec:conc}
We have developed a new, comprehensive toolbox of particle-based algorithms 
for carrying out statistical inference for CD-SSMs. Presenting our methodological developments through the underlying Feynman-Kac formulae delivers two primary benefits:
\begin{itemize}
    \item[(i)] The Feynman-Kac formulae produced in this work 
    %in the context of CD-SSMs 
    immediately allow for application of state-of-the-art particle-based Monte Carlo methods upon CD-SSMs. Recent 
    (e.g.~\cite{core:25}) and future developments for particle-based  algorithms based on Feynman-Kac approximations can be readily incorporated within our framework. 
    \item[(ii)] Feynman-Kac formulae serve as useful abstractions in the context of probabilistic programming. We demonstrate this by plugging our Feynman-Kac constructions into the `\texttt{particles}' package \citep{chop:20}, a popular probabilistic programming framework that utilises the Feynman-Kac abstraction. The resulting package `\texttt{particles-cdssm}' is used in the implementation of all experiments in 
    Section~\ref{sec:numerics}.
\end{itemize}
Beyond the aforementioned access for CD-SSMs to additional particle-based methods proposed in the literature, the development of our path-reparameterised Feynman-Kac formulae leads also to other future research directions. Indicatively:
\begin{itemize}
    \item[(i)] Whilst we demonstrate that the forward and backward guided proposal constructions outlined in Section \ref{sec:guided} improve performance, these constructs are based on approximating intractable score function terms in the drift of ideal proposal SDEs with ones deduced from relatively simple linearisation approaches. Improvements in this area would result in performance benefits on the full spectrum of particle-based methods captured under our framework. A recent promising strand of literature has looked at learning the intractable score function in the optimal proposal SDEs via neural networks: see \cite{chop:23} in the context of the forward proposal, and \cite{heng:25, bake:24, yang:25} in the context of diffusion bridges (which form a major component for our backward proposal). \cite{yang:25} also take a variational approach to learn the optimal proxy `linear SDE' $\{\tilde{V}_t(s)\}_{s\in [0, \Delta_t]}$. 
    %Only the approach of \cite{chop:23} is directly applicable in the particle-based context (i.e can be plugged directly into our framework) and only when the $\Delta_t$ are constant. 
    %Empirical results shown in these works appear to show promise - thus further progress in this direction has the potential to greatly improve the performance of particle-based methods for CD-SSMs.
    \item[(ii)] Particle-based methods provide an apparent powerful direction in a \emph{sequential} inference setting 
    for the wide class of CD-SSMs.
    We provide both theoretical and heuristic arguments in the Introduction as to the reasons for which such approaches are many times preferred also in the \emph{offline/batch} setup where standard MCMC can offer an alternative. We briefly iterate here that: (i) popular derivative-driven MCMC algorithms (e.g.~HMC, NUTS) can break down for target posteriors of complex curvature that often arise for SSMs (e.g., in low-noise scenarios); (ii) particle-based algorithms have been shown analytically to have superior performance for SSMs of increased time length $T$.  
    Further work is required in this direction to 
    develop clearer guidelines for contrasts in performance of standard MCMC and particle-based algorithms when targeting SSMs (and CD-SSMs). 
    %Research that provides further clarification both theoretically and empirically on when a particle MCMC approach outperforms HMC would help build consensus over which method to use, and aid practitioners in selecting the most reliable and performant inference algorithm. One would expect that progress in the direction outlined in (i) would dramatically improve the performance of Particle MCMC approaches in the context of high-dimensions, informative observation regimes, highly non-linear signal processes or large $\Delta_t$.
    
    \item[(iii)] We believe that the Feynman-Kac constructions developed in this work 
    %that involve careful representation of the particles as a decomposition of the ending point and the driving noise of a bridge, 
    can be extended to other classes of CD-SSMs. Indicatively, one of the authors has been involved in \cite{yone:22} where elliptic SDEs \emph{with jumps} are considered and a particle-based algorithm for online parameter inference is proposed. Though that work lacks the general scope of the present one, it does serve to illustrate the potential for the full machinery of particle-based methods to be made available for general continuous-time signals with discontinuities. Discrete state-space Markov jump processes could also be considered, with the results of \cite{cors:25} giving an indicative direction for the development of path proposals and bridge constructions that can be relevant within this context. We note that derivative-driven MCMC algorithms are impractical and/or inapplicable for such discontinuous signal processes.
\end{itemize}
%\section{Previous version of the Conclusion}
% Previous version of the conclusion:
%We have presented a comprehensive methodology over particle-based algorithms for the Monte  the smoothing distribution of CD-SSMs. 
%By allowing for careful alternation of ancestors within algorithmic executions, our particle-based methods for CD-SSMs overcome path degeneracy characterising standard approaches. Further advances in the development of guided proposals for the filtering problem for CD-SSMs can be directly incorporated in our framework, see e.g.~\citep{mide:21, chop:23, heng:25} for some recent contributions in this direction.
%We have focused on continuous-time stochastic signals %defined via SDEs, and have allowed the latter to be members of the elliptic or hypo-elliptic class of diffusion processes. \chris{Feynman-Kac formulae may be used as abstractions in the development of probabilistic programs for particle-based methods. We have developed a new probabilistic programming package: `particles\_cdssm' that implements the spectrum of particle-based methods on the presented class of CD-SSMs, by embedding the outlined Feynman-Kac formulae into the `particles' package \citep{chop:20}. These constructions also imply the applicability of more current \citep{fink:23, core:25} and future developments in the literature of particle-based methods to this class of models.} In future work, we aim to extend our approach to signals determined by more general classes of latent continuous-time processes, including SDEs with jumps and discrete-valued Markov jump processes.
\section*{Acknowledgements}
CS was supported by the Engineering and Physical Sciences Research Council (EPSRC) through a Doctoral Training Partnership (DTP) studentship.

\section*{Code availability}
All numerical experiments were implemented using the \texttt{particles-cdssm} package, which extends the \texttt{particles} library to the class of CD-SSMs using as an abstraction the Feynman–Kac constructions developed in this work. The package is available, along with code to reproduce all numerical experiments at \url{https://github.com/chris-stanton50/particles-cdssm}.

\bibliographystyle{apalike}
\bibliography{reference}

\newpage

\appendix

\section{Further Particle-Based Methods}
\label{sec:further_methods}
FFBS (Algorithm \ref{alg:FK-FFBS}) is presented in Section \ref{sec:FK} as representative of particle-based methods that alter ancestors to tackle path degeneracy. Such methods require Feynman-Kac formulae with Markov kernels 
that $M_t[x_{t-1}, dx_t] \ll \nu(dx_t)$, for a reference measure $\nu(dx_t)$ that does not depend on $x_{t-1}$. We have seen in the main text that careful work in needed to enforce such a requirement for the class of CD-SSMs.
%with density $M_t(x_t|x_{t-1})$ that can be evaluated pointwise. 
We outline below two additional algorithms  made applicable for CD-SSMs under our contributions, and which are used in numerical examples of Section \ref{sec:numerics}. Similar to Section \ref{sec:FK}, we present the algorithms as they are applied to the generic Feynman-Kac model in \eqref{eq:FK-model}.
We provide pseudo-code for the FFBS-MCMC algorithm \citep{bunc:13} used in Experiment 2 (Subsection \ref{subsubsec:ou_smoothing_exp}) in  Algorithm \ref{alg:FK-FFBS-MCMC}, the latter leading to generation of one (approximate) sample from the smoothing distribution (thus, to, say, $N_S\ge 1$ such samples after repetition).
The cost of FFBS-MCMC is $\mathcal{O}(T\cdot N) + \mathcal{O}(N_S \cdot T\cdot K)$, which will typically compare favourably with the 
$\mathcal{O}(N_S \cdot T\cdot N)$ cost of FFBS (see \cite{bunc:13} for details).  
%Assuming that the number of samples generated from the smoothing distribution is equal to the number of particles, this algorithm has linear cost in the number of particles.
\\
\medskip
\begin{algorithm}[H] 
    \caption{FFBS-MCMC for Feynman-Kac model in (\ref{eq:FK-model}).} %\cite{godsill2004monte}}
    \label{alg:FK-FFBS-MCMC}
    \SetKwInOut{Input}{Input}
    \SetKwInOut{Output}{Output}
    \KwIn{Full particle output, $X_{1:T}^{1:N}$, $A_{2:T}^{1:N}$, $W_{1:T}^{1:N}$, from Algorithm \ref{alg:FK-PF}; no of steps $K$ of Independent Metropolis-Hastings MCMC.}
    \KwOut{Approximate sample $(X_1^{B_1}, \dots, X_T^{B_T})$ from $\bb{Q}_T[dx_{1:T}]$.}
    Sample: $B_T \sim \mathcal{M}(W_T^{1:N})$\;
    \For{$t=T,...,2$}{
        $B_{t-1} \gets A_{t}^{B_t}$\;
        \For{$k=1, \dots, K$}{
        Sample: $B_{t-1}^* \sim \mathcal{M}(W_{t-1}^{1:N})$\;
        Sample: $U \sim \mathcal{U}([0, 1])$\;
        Assign: $a \gets \frac{M_t(X_t^{B_t}|X_{t-1}^{B_{t-1}^*})G_t(X_{t-1}^{B_{t-1}^*}, X_t^{B_t})}{M_t(X_t^{B_{t}}|X_{t-1}^{B_{t-1}})G_t(X_{t-1}^{B_{t-1}}, X_t^{B_t})}$\; 
        \If{$U<a$}{
        Assign: $B_{t-1} \gets B_{t-1}^*$\;
        }}
        }
\end{algorithm}
\medskip

Algorithm \ref{alg:FK-iCSMC} presents the CSMC (`Conditional SMC') kernel applied to a generic Feynman-Kac model. We use standard multinomial resampling as in our numerical experiment in Subsection \ref{section:experiment3} -- other resampling schemes \citep{chop:15} can be trivially incorporated.

\begin{algorithm}[H] 
    \caption{CSMC kernel for Feynman-Kac model in (\ref{eq:FK-model}).} %\citep{chopin2020introduction}}
    \label{alg:FK-iCSMC}
    \SetKwInOut{Input}{Input}
    \SetKwInOut{Output}{Output}
    \KwIn{$x_{1:T}^*$: Current signal value.}
    \KwOut{Variate $x_{1:T}$, so that the transition  $x_{1:T}^*\to x_{1:T}$ preserves  $\mathbb{Q}_T[dx_{1:T}]$.}
    $X_{1:T}^1 \gets x_{1:T}^*$\;
        Sample (for $j = 2,\dots, N$): $X^{j}_1\sim M_1[dx_1]$\;
    Assign weights: $w^j_1 \gets G_1(X^j_1)$\;
    Normalise weights: $W^j_1 \gets \frac{w^j_1}
        {\sum_{k=1}^{N}w^k_1}$\;
    \For{$t=2,...,T$}{
        Set: $A_t^1 \gets 1$\;
        Sample (for $j = 2,\dots, N$): $A_t^{j} \sim \mathcal{M}(W^{1:N}_{t-1})$\; %User defined Resampling Scheme 
        Sample (for $j = 2,\dots, N$): $X_t^{j} \sim M_t[X_{t-1}^{A_t^j},dx_t]$\;
        Assign weights: $w_t^j \gets G_t(X_{t-1}^{A_t^j}, X_t^{j})$\;
        Normalise weights: $W_t^j \gets \frac{w_t^j}{\sum_{k=1}^N w_t^k}$;
    }
    $x_{1:T} \gets$ output of Algorithm \ref{alg:FK-GT} or \ref{alg:FK-FFBS} (backward step) with input  $X_{1:T}^{1:T}, A_{2:T}^{1:N}, W_{1:T}^{1:N}$;
 \end{algorithm}

\section{Locally Gaussian Scheme for FHN Model}
\label{sec:loc_gauss}
The locally Gaussian scheme \citep{glot:21} applied to the (integrated form of the) FHN SDE, $(X_1(s), \bar{X}_2(s))$, over an arbitrary time-step $\Delta>0$ has a density:
\begin{equation*}
    q_\Delta(x'|x) \sim \mathcal{N}_2(x'; \mu_{LG}(\Delta , x), \Sigma_{LG}(\Delta)).
\end{equation*}
The mean and covariance functions $\mu_{LG}: [0, \infty) \times \mathbb{R}^2 \rightarrow \mathbb{R}^2$, $\Sigma_{LG}: [0, \infty) \rightarrow \mathbb{R}^{2\times 2}$ are given by:
\begin{gather*}
    \mu_{LG}(\Delta, x) = \begin{pmatrix} x_1  \\ x_2 \end{pmatrix} + \begin{pmatrix} x_2  \\[0.2cm] b_{\,\mathrm{FHN}, 2}(x_1, x_2) \end{pmatrix} \Delta  + \begin{pmatrix} b_{\,\mathrm{FHN}, 2}(x_1, x_2)  \\ 0 \end{pmatrix} \tfrac{\Delta^2}{2},\\ \Sigma_{LG}(\Delta) = \begin{pmatrix} \frac{\Delta^3}{3} & \frac{\Delta^2}{2} \\ \frac{\Delta^2}{2} & \Delta\end{pmatrix}.
\end{gather*}
Here, $b_{\,\mathrm{FHN}, 2}: [0, \infty) \times \mathbb{R}^2 \rightarrow \mathbb{R}$ is the second component of the drift of the (integrated) FHN SDE in (\ref{eq:t-FHN}).

% Draft of argument to derive locally Gaussian scheme in terms of the linear SDE.

% Considering Remark \ref{rem:taylor}, this corresponds to setting $D_x(b)_{21}=D_x(b)_{22}=0$ then obtaining $\tilde{b}_0, \tilde{b}_1$ from equation \ref{eq:linear_coefs}. We thus obtain the co-efficients:
% \begin{equation}
% \label{eq:coefs-ldl-1}
%     \tilde{b}_{0, LDL-1} = \begin{pmatrix} 
%                     0 \\ 
%                     b_{2, FHN}(x(s_{t-1}))                
%                     \end{pmatrix} 
%                     \quad 
%     \tilde{b}_{1, LDL-1} = \begin{pmatrix}
%                     0 & 1\\
%                     0 & 0
%                   \end{pmatrix}
% \end{equation}
% The resulting linear SDE is then an integrated Brownian motion, for which we have the transition density:

% \begin{align}
% \label{eq:ldl-1-linear-sde}
% d\tilde{X}_1(s) &= \tilde{X}_2(s) ds  & \tilde{X}_1(0) =x_1(s_{t-1})\\
% d\tilde{X}_2(s) &= b_{2, FHN}(x_1(s_{t-1}), x_2(s_{t-1})) ds + \frac{\sigma}{\epsilon} dB(s)  & \tilde{X}_2(0) =x_2(s_{t-1})
% \end{align}

% \begin{align}
% \label{eq:ldl-1-trans-density}
% \tilde{p}_{LDL-1}(x(s_t) | x(s_{t-1})) &= \mathcal{N}_2(x(s_t); \mu_{LDL-1}(x(s_{t-1})), \Sigma_{LDL-1})
% \end{align}
% For mean function $\mu(\Delta, x)$ and covariance $\Sigma(\Delta)$ 
\section{Challenge (C.iii): Particle Gibbs for $\theta$}
We argue in the Introduction that the Feynman-Kac formulations of this work `decouple' the strong  dependence between a parameter vector $\theta\in \Theta\subseteq \mathbb{R}^{p}$ that is involved in the diffusion coefficient of the model SDE and the signal $\{X(s)\}_{s\in[0,s_T]}$ \citep{robe:01}, thus 
enabling non-degenerate MCMC updates of $\theta|\{X(s)\}_{s\in[0,s_T]}$ in the context of joint parameter and signal smoothing via Particle Gibbs \citep{andr:10}. For completeness, we
present below a clarifying argument about the above.
\subsection{Joint Feynman-Kac Model}
Extending the  definitions of the Feynman-Kac model presented in Section \ref{sec:FK} to include the parameter $\theta$, we define the initial law $M_1^\theta[dx_1]$, the Markov probability kernels $M_t^\theta[x_{t-1}, dx_t]$, $t\ge 2$, and the potential functions
$G_1^\theta: \Theta \times \mathsf{X}_1 \rightarrow [0, \infty)$, $G_t^\theta: \Theta \times \mathsf{X}_{t-1}\times \mathsf{X}_{t} \rightarrow [0, \infty)$, $t \ge 2$. %to depend on $\theta$. 
We also introduce the pdf $\psi:\mathbb{R}^{p} \rightarrow [0, \infty)$ and a corresponding ($\sigma$-finite) dominating measure $d\theta$. 
We define the joint Feynman-Kac model as the sequence of probability measures on $\big(\Theta \times \mathsf{X}_1\times \cdots \times \mathsf{X}_t, \mathcal{B}(\Theta) \otimes \mathscr{X}_1\otimes\cdots \otimes \mathscr{X}_t  \big)$ defined through a change of measure as follows:
\begin{align}
   \mathbb{Q}_t[d\theta, dx_{1:t}] 
   &= \Big[ \tfrac{1}{L_t} \cdot L_t(\theta)\psi(\theta) d\theta \Big] \otimes \mathbb{Q}_t^{\theta}[dx_{1:t}]
   \nonumber\\ 
   &= \Big[ \tfrac{1}{L_t} \cdot L_t(\theta)\psi(\theta) d\theta\Big] \otimes 
   \Bigg[ \frac{1}{L_t(\theta)} 
   \Big\{G^\theta_1(x_1)\prod_{i=2}^t G^\theta_t(x_{t-1}, x_t)\Big\}
   \times \mathbb{M}^\theta_{t}[dx_{1:t}] \Bigg].
   \label{eq:joint-fk}
\end{align}
As in Section \ref{sec:FK}, we have set: 
\begin{align*}
\mathbb{M}_{t}^\theta[dx_{1:t}] = M_1^\theta[dx_1]\bigotimes_{i=2}^{t} M_{i}^\theta[x_{i-1},dx_i]. %\qquad t \in \mathcal{T}.
\end{align*}
We assume that the normalising constants $L_t(\theta)$ and $L_t$ are well-defined, that is:
\begin{equation*}
    L_t(\theta) := \bb{E}_{\bb{M}_t^\theta}\Big[G_1^\theta(X_1)\prod_{i=2}^t G_i^\theta(X_{i-1}, X_i)\Big] < \infty, \qquad L_t := \bb{E}_{\psi}\big[ L_t(\theta)\big] < \infty.
\end{equation*}
For $(\theta, X_{1:t})\sim\bb{Q}_t[d\theta, dx_{1:t}]$, $t \in \mathcal{T}$, the conditional law $X_{1:t} | \theta$ is a standard Feynman-Kac model, normalised by $L_t(\theta)$, in particular such a conditional law writes as:
\begin{equation}
    \label{eq:param-fk}
    \mathbb{Q}_t^{\theta}[dx_{1:t}] = \frac{1}{L_t(\theta)} \Big\{G^\theta_1(x_1)\prod_{i=2}^t G^\theta_t(x_{t-1}, x_t)\Big\}\times \mathbb{M}^\theta_{t}[dx_{1:t}]
\end{equation}
It is of interest to approximate integrals w.r.t.~$\mathbb{Q}_t[d\theta, dx_{1:t}]$. For an SSM, 
the latter law corresponds to its \emph{joint smoothing distribution} at step $t$ under the prior $\psi(\theta)d\theta$ set for the parameter.

Particle Gibbs (PG) is a particle-based MCMC algorithm that retains invariance of an extended target distribution that admits the joint Feynman-Kac law (\ref{eq:joint-fk}) as its marginal.
The PG kernel iterates between updating a sample of the signal given $\theta$ via the CSMC iterate presented in Algorithm \ref{alg:FK-iCSMC}, 
 followed by a Gibbs update of $\theta$ given the signal from
 $\mathbb{Q}_T[d\theta|X_{1:T}=x_{1:T}]$,
 %$\mathbb{Q}_T[d\theta|X_{1:T}=x_{1:T}]=P[\theta \in d\theta|X_{1:T}=x_{1:T}, Y_{1:T}=y_{1:T}]$:
 the latter being 
 the law of $\theta | X_{1:T}$ when $(\theta, X_{1:T}) \sim \mathbb{Q}_T[d\theta, dx_{1:T}]$. A pseudocode for PG is given in Algorithm \ref{alg:FK-PG}.

\medskip
\begin{algorithm}[H] 
    \caption{PG Kernel targeting joint Feynman-Kac model (\ref{eq:joint-fk})} %\cite{godsill2004monte}}
    \label{alg:FK-PG}
    \SetKwInOut{Input}{Input}
    \SetKwInOut{Output}{Output}
    \KwIn{Current parameter $\theta^*$ and current signal  $x_{1:T}^*$.}
    \KwOut{Variates $(\theta$, $x_{1:T})$, so that the transition $(\theta^*, x_{1:T}^*) \rightarrow (\theta, x_{1:T})$ preserves $\bb{Q}_T[d\theta, dx_{1:T}]$.}
    $x_{1:T} \gets$ Output of Algorithm \ref{alg:FK-iCSMC} with input $x_{1:T}^*$, via update that preserves $\mathbb{Q}_T^{\theta^*}[dx_{1:T}]$\; %either with (PGBS) or Algorithm \ref{alg:FK-GT} (PG) with input ($\tilde{X}_{1:T}^{1:N}, \tilde{A}_{2:T}^{1:N}, \tilde{W}_{1:T}^{1:N}$) \;
    $\theta\gets$ Output of Markov kernel, providing update that preserves  $\bb{Q}_T[d\theta | X_{1:T} = x_{1:T}]$.
\end{algorithm}
\medskip
%Without further assumptions on the target $\mathbb{Q
%}_T[d\theta, dx_{1:T}]$, implementation of Algorithm \ref{alg:FK-PG} is not possible, as one cannot sample from $\bb{Q}_T[d\theta | X_{1:T} = x_{1:T}]$.
\noindent  Assume  that the law $M_t^\theta[x_{t-1}, dx_t]$ has a density, say $M_t^\theta(x_t | x_{t-1})$, 
w.r.t.~a reference $\sigma$-finite measure $\nu(dx_t)$ that does not depend on $(x_{t-1},\theta)$, for all $x_{t-1} \in \mathsf{X}_{t-1}$, $\theta \in \Theta$. 
We can thus apply Bayes' formula to obtain the following expression:
\begin{equation}
    \label{eq:FK-conditional-theta}
    \bb{Q}_T[d\theta | X_{1:T}=x_{1:T}] \propto \psi(\theta) \cdot \prod_{t=1}^T \Big[ G^\theta_t(x_{t-1},x_t)M^\theta_t(x_t|x_{t-1}) \Big] d\theta.
\end{equation}
This expression enables implementation of Metropolis-within-Gibbs-type of updates for $\theta$.  
%sampling from $\bb{Q}_T[d\theta | X_{1:T} = \tilde{X}_{1:T}]$ via a Metropolis within Gibbs update.
%
\subsection{Particle Gibbs for CD-SSMs}

We now consider the application of PG (Algorithm \ref{alg:FK-PG}) in the context of  the joint Feynman-Kac formulation arising for CD-SSMs. 
In this direction, we consider the model (\ref{eq:CD-X})-(\ref{eq:CD-Y}) with unknown parameter $\theta$, which may be involved in the drift, $b_\theta(s, x)$], and/or diffusion $\sigma_\theta(s, x)$ coefficient of the signal SDE (\ref{eq:CD-X}) and the observation density $f_t^\theta(y_t|x_t)$ (\ref{eq:CD-Y}).
We then use the same latent state definition as  in Section \ref{sec:CD-SSM}: $X_t = (V_t, E_t)$ on space $\mathsf{X}_t = (C[0, \Delta_t], \mathbb{R}^d) \times \mathbb{R}^d$, 
involving the time-shifted path segments $V_t = \{V(s)\}_{s \in [0, \Delta_t]}$, with $V_t(s) = X(s_{t-1} + s)$ and ending point $E_t=V_t(\Delta_t)\equiv\texttt{end}(V_t)$.
One can then define the law of $X_{1:T}$ as the Markov measure $\mathbb{P}_T^\theta[dx_{1:T}]$ consisting of kernels $P_t^\theta[x_{t-1}, dx_t]\equiv P_t^\theta[e_{t-1}, dx_t]$
in the same way as in Section \ref{sec:CD-SSM}, but now with a parameter involved. 
We set a prior $\varphi(\theta)d\theta$ on the parameter. Then, $(\theta, X_{1:T}, Y_{1:T})$ have the joint distribution:
\begin{equation*}
\label{eq:ssm-joint}
P\,\big[\,\theta \in d\theta, X_{1:T} \in dx_{1:T}, Y_{1:T} \in dy_{1:T}\,\big] = 
\big[ \varphi(\theta) d\theta \big] \otimes \mathbb{P}^\theta_T[dx_{1:T}] \otimes \Big[ \prod_{t=1}^T f^\theta_t(y_t|x_t) \Big]\ dy_{1:T}.
\end{equation*}
Conditioning on the observations, we obtain the \emph{joint smoothing distribution}:
\begin{align}
%\label{eq:ssm-joint-smoothing-1}
\nonumber 
&P\,[\,\theta \in d\theta, X_{1:T} \in dx_{1:T}\,| \,Y_{1:T} = y_{1:T}\,]=\Big[\tfrac{1}{f_T(y_{1:T})} 
 \big\{\prod_{t=1}^T f^\theta_t(y_t | x_t) \big\}\,\varphi(\theta) \Big] \big[ d\theta \otimes \mathbb{P}^\theta_T[dx_{1:T}] \big] \\
\label{eq:ssm-joint-smoothing-2}
&\qquad\qquad\qquad =\Big[\tfrac{1}{f_T(y_{1:T})} f^\theta_T(y_{1:T})\varphi(\theta) \Big] d\theta \otimes 
    \Big[\tfrac{1}{f^\theta_T(y_{1:T})} \big\{ \prod_{t=1}^T f^\theta_t(y_t | x_t) \big\}  \Big]\times \mathbb{P}^\theta_T[dx_{1:T}].
\end{align}
In (\ref{eq:ssm-joint-smoothing-2}) we express the initial distribution as the product of the marginal $\theta | Y_{1:T}$ and the conditional $X_{1:T}|\theta, Y_{1:T}$. 
In this context, $f_T(y_{1:T})$ is now the marginal pdf of $Y_{1:T}$, whilst $f^\theta_T(y_{1:T})$ is the conditional pdf of $Y_{1:T} | \theta$.

Comparing the expressions in (\ref{eq:joint-fk}) and (\ref{eq:ssm-joint-smoothing-2}), one can see that the CD-SSM joint smoothing law $\mathbb{P}_T\,[\,\theta \in d\theta, X_{1:T} \in dx_{1:T}\,|\,Y_{1:T}=y_{1:T}\,]$ writes as a joint Feynman-Kac model (\ref{eq:joint-fk}) -- we set the priors to match $\psi=\varphi$. Then, we can define a (joint) bootstrap formalism as in Subsection \ref{subsec:boot}: $M_t^\theta[x_{t-1}, dx_t] \equiv P_t^\theta[x_{t-1}, dx_t]$ and $G_t^\theta(x_{t-1}, x_t)\equiv f_t^\theta(y_t|x_t)$. 

Returning to a possible implementation of PG (Algorithm \ref{alg:FK-PG}), under the current setup of the joint Feynman-Kac model for the CD-SSM (one yet to consider any reparameterisation), 
the conditional $\mathbb{Q}_T\,[\,d\theta\,|\,X_{1:T}=x_{1:T}\,]\equiv\mathbb{P}\,[\,\theta \in d\theta\,|\,X_{1:T}=x_{1:T}, Y_{1:T}=y_{1:T}\,]$ does not admit an expression of the form (\ref{eq:FK-conditional-theta}): kernels $M_t^\theta[x_{t-1}, dx_t] \equiv P_t^\theta[x_{t-1}, dx_t]$ do not admit a density w.r.t.~a common $\sigma$-finite dominating measure that does not depend on $x_{t-1}$. Marginalising this law so that only parameters that are in the diffusion coefficient are involved will yield a Dirac measure \citep{robe:01}. 
Thus, even if it were possible to simulate from $\mathbb{P}\,[\,\theta \in d\theta\,|X_{1:T}=x_{1:T}, Y_{1:T}=y_{1:T}\,]$, the PG kernel would be reducible. 

Following the arguments in Section \ref{sec:guided}, but now involving the parameter $\theta$, one can alternatively write (\ref{eq:ssm-joint-smoothing-2}) in the form (\ref{eq:joint-fk}) with guided formalisms of the joint Feynman-Kac model. 
In brief, in the case of the forward proposal, one can follow the arguments of Subsection \ref{subsec:FP} to construct kernels $M_t^{\theta, \rightarrow}[x_{t-1}, dx_t]$ and corresponding potentials $G_t^{\theta, \rightarrow}(x_{t-1}, x_t)$ that take the observations $y_t$ into account. 
The arguments do not change when parameter $\theta$ is included. The same holds for the backward proposal, yielding kernels $M_t^{\theta, \leftarrow}[x_{t-1}, dx_t]$ and potentials $G_t^{\theta, \leftarrow}(x_{t-1}, x_t)$ and updates  that preserve the target joint smoothing law (\ref{eq:ssm-joint-smoothing-2}).
Under both of these approaches, we still cannot implement PG, as introducing the path proposals alone does not change the conditional 
$\mathbb{Q}_T\,[\,d\theta\,|\,X_{1:T}=x_{1:T}\,]\equiv\mathbb{P}\,[\,\theta \in d\theta\,|\,X_{1:T}=x_{1:T}, Y_{1:T}=y_{1:T}\,]$.

Following the developments of Section \ref{sec:transform}, we can set-up, for each of the (joint) guided Feynman-Kac formulae defined above, a corresponding transformed Feynman-Kac formula. This leads to a new sequence of target distributions $\bar{\bb{Q}}_{t}[d\theta, dz_{1:t}]$ on $\big(\Theta \times \mathsf{Z}_1\times \cdots \times \mathsf{Z}_t, \mathcal{B}(\Theta) \times \mathscr{Z}_1\otimes\cdots \otimes \mathscr{Z}_t  \big)$ of the form:
\begin{align}
   \bar{\mathbb{Q}}_t[d\theta, dz_{1:t}] 
   &= \Big[ \tfrac{1}{\bar{L}_t} \cdot \bar{L}_t(\theta)\psi(\theta) d\theta \Big] \otimes \bar{\mathbb{Q}}_t^{\theta}[dz_{1:t}]
   \nonumber\\ 
   &= \Big[ \tfrac{1}{\bar{L}_t} \cdot \bar{L}_t(\theta)\psi(\theta) d\theta\Big] \otimes 
   \Big[ \tfrac{1}{\bar{L}_t(\theta)} 
   \big\{\bar{G}^\theta_1(z_1)\prod_{i=2}^t \bar{G}^\theta_t(z_{t-1}, z_t)\big\}
   \times \bar{\mathbb{M}}^\theta_{t}[dz_{1:t}] \Big].
\label{eq:FK-joint-reparameterised}
\end{align}
This expression involves new kernels $\bar{M}_t^\theta[z_{t-1}, dz_t]$ and potentials $\bar{G}_t^{\theta}(z_{t-1}, z_t)$ that can be defined in the same way as in Section \ref{sec:transform}, for both the forward and backward cases. The arguments still hold -- the maps $\mathfrak{F}_t^\theta$ and $\mathfrak{H}_t^\theta$ that appear when defining the kernels $M_t^{\theta, \rightarrow}[z_{t-1}, dz_t]$ and $M_t^{\theta, \leftarrow}[z_{t-1}, dz_t]$ for the forward and backward approach respectively will now depend on $\theta$.

Through applying the maps $\mathsf{Z}_t \rightarrow \mathsf{X}_t$ to this transformed model and keeping the parameter fixed, the original Feynman-Kac model $\mathbb{Q}_t[d\theta, dx_{1:t}]$ is recovered. Thus, joint latent state and parameter inference for CD-SSMs can proceed via application of a (particle) MCMC approach that preserves the joint Feynman-Kac model (\ref{eq:FK-joint-reparameterised}). Considering application of PG in this context, when updating the parameter, this would involve simulation from (a Markov kernel that leaves invariant) the conditional law of $\theta|Z_{1:T}=z_{1:T}$ for variate $(\theta, Z_{1:T})\sim\bar{\mathbb{Q}}_T[d\theta, dz_{1:T}]$. This conditional admits an expression of the form (\ref{eq:FK-conditional-theta}), as under the transformed Feynman-Kac model, proposal kernels $\bar{M}_t^{\theta}[z_{t-1}, dz_t]$ have densities $\bar{M}_t^\theta(z_t|z_{t-1})$ w.r.t.~a common dominating measure $\textrm{Leb}(de_t)\otimes\mathbb{W}_t(du_t)$ that does not depend on $(\theta,z_{t-1})$. The PG parameter update is therefore \emph{possible} via Metropolis-within-Gibbs. An expression of the form (\ref{eq:FK-conditional-theta}) also implies that under the transform, the law of components of $\theta$ involved in the diffusion coefficient given the transformed signal have a density w.r.t.~the Lebesgue measure -- thus the transform successfully `decouples' the strong dependence between parameters involved in the diffusion coefficient and the signal. In the subsequent PG step when the latent signal is being updated given the parameter, via iCSMC, the transform also enables the use of backward steps, as discussed in (C.iii) in the Introduction.

\end{document}